\documentclass[preprint,prd,showpacs,showkeys,superscriptaddress]{revtex4}
\usepackage{amsmath,graphicx,epsfig,amssymb}
\allowdisplaybreaks

\newcommand{\lQ}{\Lambda_{\text{QCD}}}

\newcommand{\be}{\begin{equation}}
\newcommand{\ee}{\end{equation}}
\newcommand{\bea}{\begin{eqnarray}}
\newcommand{\eea}{\end{eqnarray}}
\newcommand{\efti}{\text{EFT}_\text{I}}
\newcommand{\eftii}{\text{EFT}_\text{II}}

\def\siml{{\ \lower-1.2pt\vbox{\hbox{\rlap{$<$}\lower6pt\vbox{\hbox{$\sim$}}}}\ }}
\def\simg{{\ \lower-1.2pt\vbox{\hbox{\rlap{$>$}\lower6pt\vbox{\hbox{$\sim$}}}}\ }}

\def\slashchar#1{\setbox0=\hbox{$#1$}\dimen0=\wd0\setbox1=\hbox{/}\dimen1=\wd1\ifdim\dimen0>\dimen1
                 \rlap{\hbox to \dimen0{\hfil/\hfil}} #1 \else \rlap{\hbox to
           \dimen1{\hfil$#1$\hfil}} /  \fi}

\graphicspath{{../}}

\begin{document}

\title{Exclusive decays of $\chi_{bJ}$ and $\eta_b$ into two charmed mesons}

\author{Regina S. Azevedo}
\email{razevedo@physics.arizona.edu}
\affiliation{Department of Physics, University of Arizona, Tucson, AZ 85721, USA}
\author{Bingwei Long}
\email{long@ect.it}
\affiliation{European Centre for Theoretical Studies in Nuclear
  Physics and Related Areas (ECT*), I-38100 Villazzano (TN), Italy}
\affiliation{Department of Physics, University of Arizona, Tucson, AZ
  85721, USA}
\author{Emanuele Mereghetti}
\email{emanuele@physics.arizona.edu}
\affiliation{Department of Physics, University of Arizona, Tucson, AZ 85721, USA}

\begin{abstract}

We develop a framework to study the exclusive two-body decays of
bottomonium into two charmed mesons and
apply it to study the decays of the $C$-even bottomonia. 
Using a sequence of effective field theories, we take advantage of the
separation between the scales contributing to the decay processes,
$2m_b \gg m_c \gg \lQ$. We prove that, at leading order in the EFT
power counting, the decay rate factorizes into the convolution of two
perturbative matching coefficients and three non-perturbative matrix
elements, one for each hadron.
We calculate the relations between the decay rate and non-perturbative
bottomonium and $D$-meson matrix elements at leading order, with
next-to-leading log resummation.
The phenomenological implications of these relations are discussed.
\end{abstract}

\pacs{12.39.Hg, 13.25.Gv}
\keywords{quarkonium decay; non-relativistic QCD; soft-collinear
  effective theory}

\preprint{ECT*-09-09}

\maketitle
\section{Introduction}\label{Intro}
The exclusive two-body decays of  heavy quarkonium into light hadrons have
been studied in the framework of perturbative QCD by many authors (for
reviews, see \cite{Chernyak:1983ej} \cite{Brodsky:1989pv}). These
processes exhibit a large hierarchy between the heavy quark mass, 
which sets the scale
for annihilation processes, 
and the scales that determine the
dynamical structure of the particles in the initial and final
states. The large energy released in the annihilation of the heavy
quark-antiquark pair and the
kinematics of the decay ---
with
the products flying away from the decay point in two
back-to-back, almost light-like directions--- allow
for rigorously
deriving a factorization formula for the decay rate at leading twist
(for an up-to-date review of the theoretical and experimental status of
the exclusive decays into light hadrons, see \cite{Brambilla:2004wf}).

For the bottomonium system, a particularly interesting class of
two-body final states is the
ones containing two charmed mesons. In
these cases the picture is complicated by the appearance of an
additional intermediate scale, the charm mass $m_c$,
which is much smaller than
the bottom mass $m_b$ but is large enough to be perturbative. 
These decays differ significantly from
those involving only light quarks. The creation of mesons that are
made up of purely light quarks involves 
creating two quark-antiquark pairs,
with the energy shared between the
quark and antiquark in each pair. In the production of two $D$
mesons, however, almost all
the energy of the bottomonium is carried away by the
heavy $c$
and $\bar c$, while the light quark and antiquark, which
bind to the $\bar c$ and $c$ respectively, carry away
(boosted) residual energies.

The existence of well-separated scales in the system and
the intuitive picture of the decay process suggest to tackle the
problem using a sequence of effective field theories (EFTs) that are obtained by
subsequently integrating out the dynamics
relevant to the perturbative
scales $m_b$ and $m_c$.

In the first step, we integrate out the scale $m_b$
by describing the $b$ and
$\bar b$ with Non-Relativistic
QCD (NRQCD) \cite{Bodwin:1994jh},
and the highly energetic $c$ and $\bar c$
with two copies of Soft-Collinear Effective Theory
(SCET)
\cite{Bauer:2000ew} \cite{Bauer:2000yr} \cite{Bauer:2001yt}
\cite{Bauer:2002nz} \cite{Leibovich:2003jd} in opposite light-cone
directions. In the second step, we integrate out the dynamics
manifested
at scales of
order $m_c$
by treating the quarkonium with potential
NRQCD (pNRQCD) \cite{Pineda:1997bj} \cite{Brambilla:1999xf}
\cite{Brambilla:2004jw}, and the $D$ mesons with a boosted version of
Heavy-Quark Effective Theory
(HQET) \cite{Grinstein:1990mj}
\cite{Eichten:1989zv} \cite{Georgi:1990um} \cite{Beneke:2003xh}
\cite{Beneke:2004km} \cite{Fleming:2007qr} \cite{Fleming:2007xt}.
The detailed explanation of why the aforementioned EFTs are employed
is offered in Sec.~\ref{Eft}.
We will
prove that, at leading order in the EFT
expansion, the decay rate factors into a convolution of two
perturbative matching coefficients
and three (one for each hadron) non-perturbative matrix
elements. The
non-perturbative matrix elements are 
process-independent and encode
information on both the initial and final
states.

For simplicity, in this paper we focus on the decays of the $C$-even
quarkonia $\chi_{bJ}$ and $\eta_b$
that, at leading order in the
strong coupling $\alpha_s$, proceed via the 
emission
of two virtual gluons.
The same method can be generalized to the
decays of $C$-odd states
$\Upsilon$ and $h_b$,
which require an additional virtual gluon.
We also refrain from processes that have vanishing contributions at
leading order in the EFT power counting.
So the specific processes
studied in this paper are $\chi_{b0, 2} \rightarrow D D$,
$\chi_{b0,\, 2} \rightarrow D^* D^*$, and $\eta_b \rightarrow D D^* +
\text{c.c.}$
However, the EFT approach
developed in this paper enables one to systematically include
power-suppressed effects, making it possible
to go beyond the leading-twist approximation.

The study of the inclusive and exclusive charm
production in bottomonium
decays and of
the role played by the
charm mass $m_c$ in such processes
have recently drawn renewed
attention \cite{Bodwin:2007zf} \cite{Kang:2007uv}
\cite{Braguta:2009df} \cite{Braguta:2009xu}, in 
connection with the experimental advances spurred in the
past few years by the abundance of bottomonium data produced at
facilities like BABAR, BELLE, and CLEO. 
The most notable result was the observation of the bottomonium ground
state $\eta_b$, 
recently reported by the BABAR collaboration \cite{:2008vj}.
Furthermore,
the CLEO collaboration published the first results for several exclusive decays of $\chi_b$
into light hadrons \cite{:2008sx} and
for the inclusive decay of $\chi_b$ into open charm \cite{Briere:2008cv}.  
In particular, they measured the branching ratio $\mathcal B(\chi_{bJ}
\rightarrow D^0 X)$, where $J$ is the total angular momentum of the
$\chi_b$ state, and conclusively showed that for $J=1 $ the production of open charm
is substantial: $\mathcal B(\chi_{b1}(1P) \rightarrow D^0 X)  = 12.59
\pm 1.94 \%$. For the $J=0,2$ states the data
are weaker, but
the production of open charm still appears to be relevant. 
The measurements of the CLEO collaboration are in good
agreement with the prediction of Bodwin
\textit{et al.} \cite{Bodwin:2007zf}, where
EFT techniques (in particular NRQCD)
were for the first time applied to
study the production of charm in bottomonium decays.
 
The double-charm decay channels analyzed here have not yet been
observed, so one of our aims is to see if they may be observable given the
current data. Unfortunately, the poor knowledge of the $D$-meson matrix
elements prevents us from providing definitive
predictions for the decay rates $\Gamma(\chi_{b J} \rightarrow D D)$, $\Gamma(\chi_{bJ}
\rightarrow D^* D^*)$, and $\Gamma(\eta_b  \rightarrow D D^* + \textrm{c.c.})$. As we
will show, these rates are
indeed strongly dependent on the parameters of the $D$- and
$D^*$-meson distribution amplitudes, in particular on their first
inverse moments $\lambda_D$ and $\lambda_{D^*}$:
the rates vary by an order of magnitude in the accepted
ranges for $\lambda_D$ and $\lambda_{D^*}$. 
On the other hand, the factorization formula implies that these
channels, if measured with sufficient accuracy, could constrain the
form of the $D$-meson distribution amplitude and the value of its
first inverse moment. In turn, the details of the $D$-meson structure
are relevant to other $D$-meson observables,
which are crucial for a model-independent determination of the CKM
matrix elements $|V_{cd}|$ and $|V_{cs}|$ \cite{Hill:2005ju}.

This paper is organized as follows.
In Sec.~\ref{Eft} we discuss the
degrees of freedom and the EFTs we use. In Sec.~\ref{SCETmatch} we match
QCD onto NRQCD and SCET at the scale $2m_b$. The renormalization-group
equation (RGE) for the matching coefficient
is derived and solved in Sec.~\ref{SCETrun}. In  Sec.~\ref{bHQETmatch} the scale $m_c$ is
integrated out by matching NRQCD and SCET onto pNRQCD and
bHQET. The renormalization of the low-energy EFT operators is
performed in Sec.~\ref{bHQETrun}, with some technical details left to
App.~\ref{AppA}. The decay rates are calculated in Sec.~\ref{rate}
using two model distribution amplitudes. In Sec.~\ref{concl} we draw
our conclusions.

\section{Degrees of freedom and the Effective Field
  Theories}\label{Eft}

Several well-separated scales are involved in the
decays of the $C$-even bottomonia $\eta_b$ and $\chi_{bJ}$
into two $D$ mesons,
making them ideal processes for the application of
EFT techniques. The distinctive structures of the bottomonium
(a heavy quark-antiquark pair) and the $D$ meson (a bound state of a heavy quark and a
light quark) suggest that one needs different EFTs to describe the
initial and final states.

We first look at the initial state.
The $\eta_b$ is the ground state of the bottomonium system. It is a
pseudoscalar particle, with spin $S = 0$, orbital angular momentum
$L=0$, and total angular momentum $J =0$. In what follows we will often
use the spectroscopic notation $^{2S+1}L_J$, in which the $\eta_b$ is
denoted by $^1S_0$.
The $\chi_{bJ}$ is a triplet of states with quantum numbers
$^3P_J$. The $\eta_b$ and $\chi_{bJ}$
are non-relativistic bound states of a $b$ quark and a $\bar b$ antiquark. 
The scales in
the system are the $b$ quark mass $m_b$,
the relative momentum of the $b \, \bar b$ pair  $m_b w$, the binding
energy $m_b w^2$, and $\lQ$, the scale where QCD becomes strongly
coupled. $w$ is the relative velocity of the
quark-antiquark pair in the meson, and from the bottomonium spectrum it can be inferred that
$w^2 \sim 0.1$. 
Since $m_b \gg \lQ$,
$m_b$ can be integrated out in
perturbation theory and the bottomonium can be described in
NRQCD. The degrees of freedom of NRQCD are non-relativistic heavy
quarks and antiquarks, with energy and momentum $(E, |\vec p\,|)$ of order
$(m_b w^2, m_b w)$, light quarks and gluons. In NRQCD, the gluons can
be soft $(m_b w, m_b w)$, potential $(m_b w^2, m_b w)$, and ultrasoft (usoft)
$(m_b w^2, m_b w^2)$.
The NRQCD Lagrangian is constructed as a systematic expansion in
$1/m_b$
whose first few terms are 
\begin{equation*}
 \mathcal L_{\rm{NRQCD}} = \psi^{\dagger} \left( i D_0 + \frac{\vec
     D^2}{2 m_b} + \frac{\vec\sigma \cdot g \vec B}{2m_b}+
   \ldots\right) \psi +
\chi^{\dagger} \left(  i D_0 - \frac{\vec D^2}{2 m_b} -
  \frac{\vec\sigma \cdot g \vec B}{2m_b} + \ldots \right) \chi \; ,
\end{equation*}
where $\psi$ and $\chi^{\dagger}$ annihilate a $b$ quark and a $\bar
b$ antiquark respectively, and $\cdots$ denotes higher-order
contributions in $1/m_b$. In NRQCD several
mass scales are still dynamical and different assumptions on the
hierarchy of these scales may
lead to
different power countings for operators of higher
dimensionality. However, as long as $w \ll 1$, higher-dimension
operators are suppressed
by powers of $w$ (for a critical discussion on
the different power countings we refer to \cite{Brambilla:2004jw}).

NRQCD still contains interactions that can excite the heavy quarkonium far
from
its mass shell, for example, through the interaction of a non-relativistic quark
with a soft gluon.
In the case
$m_b w  \gg \lQ$, we can integrate out these
fluctuations, matching perturbatively NRQCD onto a low-energy effective
theory, pNRQCD. We are then left with a theory of non-relativistic
quarks and ultrasoft gluons, with non-local potentials
induced by the integration over soft- and potential-gluon modes.  
The
interactions of the heavy quark with ultrasoft gluons
are still described
by the NRQCD Lagrangian, with the constraint that all the gluons are
ultrasoft. In the weak coupling regime  $m_b w \gg \lQ$, the
potentials are organized by an expansion in $\alpha_s(m_b w)$,
$1/m_b$, and $r$, where $r$ is the distance between the quark and
antiquark in the quarkonium, 
$r \sim 1/m_b w$.  If we assume $m_b w^2
\sim \lQ$, each term in the expansion has a definite power counting in
$w$ and the leading potential is Coulombic $V \sim \alpha_s(m_b w)/r$.  

An alternative approach,
which does not require a two-step matching,
has been developed in the effective theory vNRQCD \cite{Luke:1996hj}
\cite{Luke:1999kz} \cite{Manohar:1999xd} \cite{Hoang:2002yy}. In the
vNRQCD approach there is only one EFT below $m_b$,
which is obtained by
integrating out all the off-shell fluctuations at the hard scale $m_b$
and introducing different fields for
various propagating degrees of freedom (non-relativistic quarks and
soft and ultrasoft gluons). In spite of the differences between the
two formalisms, pNRQCD and vNRQCD give
equivalent final answers in all the known examples in which both theories can be applied.  

We now turn to the structure of the $D$ meson.
The most relevant features of the $D$ meson are captured by a
description in
HQET. In HQET, in order to integrate out the inert scale $m_c$, the
momentum of the heavy quark is generically written as~\cite{Georgi:1990um}
\be
p = m_c v + k\; ,
\ee
where $v$ is the four-velocity label, satisfying $v^2 = 1$, and $k$ is the
residual momentum. If one chooses $v$ to be the center-of-mass 
velocity of the $D$ meson, $k$ scales as $k \sim v \lQ$.
Introducing the light-cone vectors $n^{\mu} = (1,0,0,1)$ and $\bar
n^{\mu} = (1,0,0,-1)$, one can express the
residual
momentum in light-cone
coordinates, $k^{\mu} = \bar n \cdot k\, n^{\mu}/2 + n \cdot k \, \bar
n^{\mu}/2 + k^{\mu}_{\perp}$ or simply $k = (n \cdot k,\bar n \cdot k, \vec
k_{\perp})$. There are two relevant frames. One is the $D$-meson rest
frame, in which $v$ is conveniently chosen as $v_0 = (1,0,0,0)$, and the
other is the bottomonium rest frame, in which the $D$ mesons are
highly boosted in opposite directions, with $v$ chosen as $v = v_D$,
the four-velocity of one of the $D$ mesons. By a simple
consideration of kinematics and the scaling $k \sim v\lQ$, one can work out the
scalings for $k$ in the two frames. In the $D$-meson rest frame, $k
\sim \Lambda_{\text{QCD}}(1,1,1)$,
and in the bottomonium rest frame (supposing the $D$ meson moving in
the positive $z$-direction),
\begin{equation}\label{eq:2.1}
k  \sim \Lambda_{\text{QCD}} \left(n \cdot v_D, \bar n \cdot v_D, 1 \right) \sim
\Lambda_{\text{QCD}} \bar n \cdot v_D \left(\lambda^2, 1,
  \lambda\right)\; ,
\end{equation}
where
$\bar n \cdot v_D \sim 2 m_b / m_c$
and $\lambda = m_c / 2 m_b \ll 1 $.
It is convenient for the calculation in this
paper to use the bottomonium rest frame, so we drop the subscript in
$v_D$ and we assume $v = v_D$ in the rest of this paper.
The momentum scaling in Eq. \eqref{eq:2.1} is called
ultracollinear (ucollinear), and boosted HQET (bHQET) is the theory that describes heavy
quarks with ultracollinear residual 
momenta and light degrees of freedom (including gluons and light
quarks) with the same momentum scaling.

The bHQET Lagrangian is organized as a series in powers of $\lQ/m_c$
and, for residual momentum ultracollinear in the $n$-direction, the
leading term is \cite{Fleming:2007qr}
\begin{equation}
\mathcal L_{\rm{bHQET}} = \bar h_{n} i v \cdot D h_n \; ,
\end{equation}
where the field $h_n$ annihilates a heavy quark 
and the covariant derivative $D$ contains ultracollinear and ultrasoft gluons,
 \begin{equation}
i D^{\mu} = \frac{n^{\mu}}{2} \left(i \bar n \cdot \partial  + g \bar n \cdot A_n\right) + 
 \frac{\bar n^{\mu}}{2} \left(i  n \cdot \partial  + g  n \cdot A_n + g n \cdot A_{us}\right)
 +  \left(i  \partial^{\mu}_{ \perp}  + g  A^{\mu}_{n, \,\perp}\right)
 \; .
\end{equation}
The ultrasoft gluons only enter in the small component of the
covariant derivative. This fact can be exploited to decouple ultrasoft and
ultracollinear modes
in the leading-order Lagrangian
through a field redefinition reminiscent of the
collinear-ultrasoft decoupling in SCET \cite{Bauer:2001yt}
\cite{Fleming:2007qr}. The ultracollinear-ultrasoft decoupling is an
essential ingredient for the factorization of the decay rate.

Therefore, the appropriate EFT to
calculate the decay rate is a
combination of pNRQCD, for the bottomonium, and two copies of bHQET,
with fields collinear to the $n$ and $\bar n$ directions, 
for the $D$ and $\bar D$ mesons, symbolically written as $\eftii \equiv \text{pNRQCD} +
\text{bHQET}$.

As we mentioned earlier, we plan to describe the bottomonium structure
with
a two-step scheme 
$\text{QCD} \to \text{NRQCD} \to \text{pNRQCD}$. However, at the intermediate
stage, where we first integrate out the hard scale $2m_b$ and arrive
at the scale $m_bw$, the $D$
meson cannot yet be described in bHQET. This is because the interactions
relevant at the intermediate scale $m_bw$ can change the
$c$-quark velocity and leave the $D$ meson off-shell
of order
$\sim (m_bw)^2 \sim m^2_c \gg \lQ^2$. 
Highly energetic $c$ and $\bar c$ 
travelling in opposite directions can be described properly by SCET with mass.
Thus, at the scale $\mu = 2m_b$, we
match QCD onto an intermediate EFT, $\efti \equiv \text{NRQCD} +
\text{SCET}$, in which the EFT expansion is organized by $\lambda$ and
$w$. The degrees of freedom of $\efti$ are tabulated in Tab. \ref{Tab:1}.
 
\begin{table}[t]
 \begin{tabular}{c |c|c|c||c|c|c}
 & NRQCD                   & field                                                          &\,  momentum    &
 SCET			 & field				                          &\,  momentum   \\
 \hline 
quark \; & $b$, $\bar b$\;          &\, $\psi_b$,\, $\chi_{\bar b}$\;                                & $(m_b w^2, m_b w)$    & 
$c$, $\bar c$            &\, $ \xi^{c}_{\bar n}$, \, $ \xi^{\bar
  c}_{n}$ \;               & $2 m_b(1,\lambda^2,\lambda)$, \; $2 m_b(\lambda^2,1,\lambda)$ \\ 
gluon \; & potential         &\, $A^{\mu}$                                                    & $(m_b w^2, m_b w)$    &
collinear          &\, $A^{\mu}_{\bar n}$, \, $A^{\mu}_{n}$ \;                      & $2m_b(1,\lambda^2,\lambda)$, \; 													    $2m_b(\lambda^2,1,\lambda)$ \\ 
 &soft               &\, $A^{\mu}$							  & $(m_b w, m_b w)$      &
soft                &\, $A^{\mu}_s        $ \;                                       & $2m_b(\lambda,\lambda,\lambda)$  \\
& usoft              &\, $A^{\mu}$							  & $(m_b w^2, m_b w^2)$      &
usoft              &\, $A^{\mu}_{us}     $ \;               & $2m_b(\lambda^2,\lambda^2,\lambda^2)$\\ 
\end{tabular}
\caption{Degrees of freedom in $\efti (\text{NRQCD} + \text{SCET})$. $w$ is
  the $b\, \bar b$ relative velocity in the bottomonium rest frame,
  while $\lambda  \sim  m_c/2m_b$ is the SCET expansion parameter.
We assume $m_b w \sim m_c$ (or, equivalently, $w \sim \lambda$)
and $m_b w^2 \sim m_b \lambda^2 \sim \lQ$.
}\label{Tab:1}
\end{table}

Then, we integrate out
$m_c$ and $m_bw$ at the same time, matching $\efti$
onto $\eftii$ at the scale
$\mu^{\prime} = m_c$. In
 $\eftii$, the low-energy
approximation
is organized by $\lQ/m_c$ and $w$. The degrees of freedom of
$\eftii$ are summarized in Tab. \ref{Tab:2}. When no subscript is
specified in the rest of this paper, any reference to EFT applies to
both $\efti$ and $\eftii$. To facilitate the power counting, we adopt $w \sim
\lambda \sim \lQ/m_c$. As a first study, we will perform in this paper the
leading-order calculation of the 
bottomonium decay rates.

\begin{table}[t]
 \begin{tabular}{c|c|c|c||c|c|c }
&\, pNRQCD\,                   & field                                                        &\,  momentum 
&\, bHQET \,                   & field                                                        &\,  momentum   \\
 \hline 
quark \; & $b$, $\bar b$\;               &\, $\psi_b$,\, $\chi_{\bar b}$\;                   & $(m_b w^2, m_b w)$  
	 & $c$, $\bar c$                 &\, $ h^{c}_{\bar n}$, \, $ h^{\bar c}_{n}$ \;      & $ Q   \, (1,\lambda^2,\lambda)$, \; $Q \, (\lambda^2,1,\lambda)$  \\ 
         &                               &                                                   &
         & $u$, $d$         &\, $\xi_{\bar n}$, \, $\xi_{n} $                   & $  Q \, (1,\lambda^2,\lambda)$, \; $Q \, (\lambda^2,1,\lambda)$  \\
gluon    & usoft        &\, $A^{\mu}     $ \;               & $(m_b w^2, m_b w^2)$ 
         & usoft        &\, $A^{\mu}_{us}     $ \;               & $ Q (\lambda,\lambda,\lambda)$\\ 
    &				 & 						     &
         & ucollinear              &\, $A^{\mu}_{\bar n}$, \, $A^{\mu}_{n}$  \;        &  $ Q \, (1,\lambda^2,\lambda)$, \; $ Q \, (\lambda^2,1,\lambda)$                    \\ 
\end{tabular}
\caption{Degrees of freedom in $\eftii (\text{pNRQCD} +
\text{bHQET})$. The scale $Q$
in bHQET is $Q = n \cdot v^{\prime} \lQ$ for the $\bar n$-collinear
sector
and $Q = \bar n \cdot v \lQ$ for the $n$-collinear sector. $n \cdot
v^{\prime}$ and $\bar n \cdot v$ are the large light-cone components
of the
$D$-meson velocities in the
bottomonium rest frame, $n \cdot
v^{\prime} \sim \bar n \cdot v \sim 2m_b/m_c$. $\lambda$  and  $w$ are
defined as in Tab. \ref{Tab:1}.}\label{Tab:2}
\end{table}

\section{$ \text{NRQCD} + \text{SCET}$}\label{SCET}
\subsection{Matching}\label{SCETmatch}
\begin{figure}[t]
\includegraphics[width = 5cm]{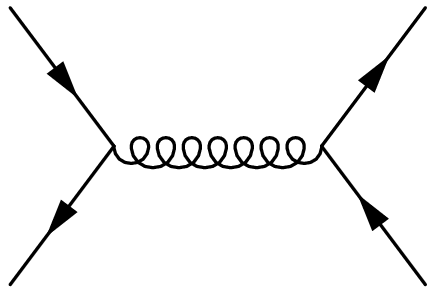} \hspace{0.5cm} \hspace{1cm} \hspace{0.5cm}
\includegraphics[width = 2.7cm]{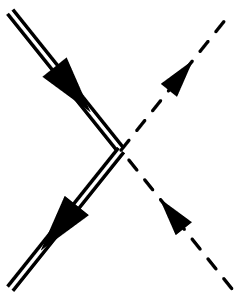}
\caption{Matching QCD onto
$\efti$.
On the r.h.s., the double
  lines represent the
non-relativistic $b$ ($\bar b$) (anti)quark, while the
  dashed lines represent the collinear $c$ ($\bar
  c$) (anti)quark.} \label{Fig:1}
\end{figure}

In the first step, we integrate out the dynamics
related to the hard
scale $2 m_b$ by matching the QCD diagrams for the production of a
$c\, \bar c$ pair in the annihilation
of a $b \, \bar b$ pair onto
their $\efti$ counterparts.
The tree-level
diagrams for the process
are shown in Fig. \ref{Fig:1}.
The gluon propagator in the QCD diagram has off-shellness of order
$q^2 = (2m_b)^2$ and it is not resolved in 
$\efti$, giving rise to a point-like interaction.

We calculate the diagrams on shell, finding
\begin{equation}\label{eq:M.1}
i J_{\rm{QCD}}  = i C(\mu) J_{\efti}(\mu) \; ,
\end{equation}
with, at tree level,
\begin{equation}\label{eq:M.2}
J_{\efti} = \chi_{\bar b}^{\dagger} \sigma^{\mu}_{\perp} t^a \psi_b \;
\bar{\chi}^{c}_{\bar n} \, S^{\dagger}_{\bar n} \, \gamma_{\mu \, \perp}
t^a \,  S_n\, \chi^{\bar c}_{n} \quad \text{and} \quad C(\mu= 2m_b) =
\frac{\alpha_s(2m_b)\pi}{m^2_b} \; ,
\end{equation}
where $t^a$ are color matrices and the symbol $\sigma^{\mu}$ denotes
the four matrices $\sigma^{\mu} = (1, \vec{\sigma})$, with
$\vec{\sigma}$ the Pauli matrices. The subscript $\perp$ refers to the
components orthogonal to the light-cone vectors $n^{\mu}$ and $\bar
n^{\mu}$.
The fields $\psi_b$ and $\chi^{\dagger}_{\bar b}$ are two-component spinors
that annihilate respectively a $b$ quark and a $\bar b$ antiquark.
$\chi^{\bar
  c}_{n,\, \bar n \cdot p}$ and $\chi^{c}_{\bar n,\, n \cdot p}$ are
collinear gauge-invariant fermion fields:
\be
\chi^{\bar c}_{n, \, \bar n \cdot p}  \equiv ( W_n^{\dagger} \xi^{\bar
  c}_n)_{\bar n \cdot p}\; , \qquad
\chi^c_{\bar n,\, n \cdot p} \equiv (W_{\bar n}^{\dagger} \xi^c_{\bar
  n})_{n \cdot p} \; ,
\ee
where $W_{n}$ is defined as
\begin{equation}\label{eq:w}
 W_{n} \equiv
\sum_{\rm{perms}} \exp \left( - \frac{g}{\bar n \cdot
     \mathcal P} \bar n \cdot A_n\right) \; .
\end{equation}
$W_{\bar n}$ has an analogous definition
with $n \rightarrow \bar n$. Collinear fields are labelled by the
large component of their momentum.
Note, however, we omit in
Eq. \eqref{eq:M.2} the
subscripts $n \cdot p$ and $\bar{n} \cdot p$ of the collinear fermion fields, in order
to simplify the notation.
The operator $\bar n \cdot \mathcal P$ in the definition \eqref{eq:w}
is a label operator that extracts the large component of the momentum
of a collinear field, $\bar n \cdot \mathcal P\, \phi_{n,\, \bar n \cdot
  p} = \bar n \cdot p \, \phi_{n,\, \bar n \cdot p}$, where $\phi_{n,\,
  \bar n \cdot p}$ is a generic collinear field.
$S_{n (\bar n)}$ is a soft Wilson line,
\begin{equation}\label{eq:s}
S_n
\equiv \sum_{\rm{perms}} \left[\exp\left( -\frac{g }{n \cdot \mathcal P} n \cdot A_s\right)\right],
\end{equation}
where the operator  $n \cdot \mathcal P$ acts on soft fields, $n \cdot
\mathcal P \, \phi_s = n \cdot k \, \phi_s$.

Since in SCET different gluon modes are represented by different
fields, we have to guarantee the gauge invariance of the operator
$J_{\efti}$ under separate soft and collinear gauge
transformations. A soft transformation is defined by $V_s(x) = \exp{(i
  \beta^a_s t^a)}$, with $\partial_{\mu} V \sim 2m_b( \lambda,
\lambda,\lambda)$, while a gauge transformation $U(x)$ is $n$-collinear if
$U(x) = \exp{(i \alpha^a(x) t^a)}$ and $\partial_{\mu} U(x) \sim 2m_b
(\lambda^2,
1,\lambda)$.
It has been shown in Ref.~\cite{Bauer:2001yt}
that collinear fields do not transform under a soft transformation and
that the combination $W_n^{\dagger} \xi_n$ is gauge invariant under a
collinear transformation. Soft fields do not transform under collinear
transformations
but they do under soft transformations. For
example, the NRQCD quark and antiquark fields
transform as $\psi_b  \rightarrow V_s(x) \psi_b$. The soft Wilson
line has the
same transformation, $S_n \rightarrow V_s(x) S_n$.
Therefore, $\chi_{\bar b}^{\dagger} \sigma^{\mu}_{\perp} t^a \psi_b$
transforms
as an octet under soft gauge transformations. Since
$\bar{\chi}^{c}_{\bar n} \, S^{\dagger}_{\bar n} \, \gamma_{\mu \,
  \perp} t^a \,  S_n\, \chi^{\bar c}_{n}$ behaves like an octet as
well,
$J_{\efti}$ is invariant.
It is worth noting that the soft Wilson lines are necessary to
guarantee the gauge invariance of $J_{\efti}$.
We have explicitly checked their appearance at
one gluon by matching QCD
diagrams like the one in Fig. \ref{Fig:1}, with all the possible
attachments of an extra soft or collinear gluon, onto four-fermion
operators in $\efti$.

\subsection{Running}\label{SCETrun}
The matching coefficient $C$ and the effective operator $J_{\efti}$
depend on the renormalization scale $\mu$.
Since the effective operator is sensitive to the
low-energy scales in 
$\efti$, logarithms that would appear in the evaluation of
$J_{\efti}$ are minimized by the choice $\mu \sim m_c$. On the
other hand, since the coefficient encodes the high-energy dynamics of
the scale $2 m_b$, such a choice would induce large logarithms of
$m_c/2m_b$ in the matching coefficient. These logarithms can
be resummed using
RGEs in $\text{NRQCD} + \text{SCET}$. 

\begin{figure}[htb]
\center
\includegraphics[width=2.5cm]{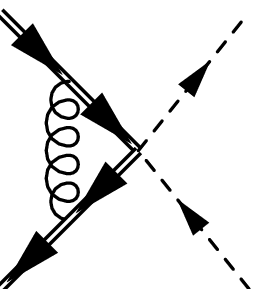}
\hspace{1cm}
\includegraphics[width=2.5cm]{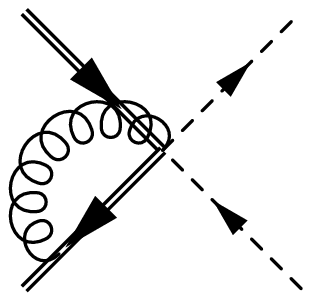}
\hspace{1cm}
\includegraphics[width=2.5cm]{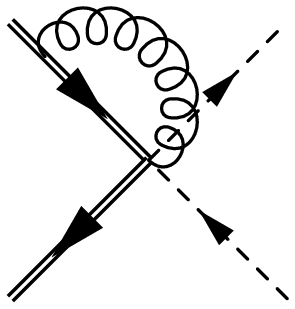}
\hspace{1cm}	
\includegraphics[width=2.5cm]{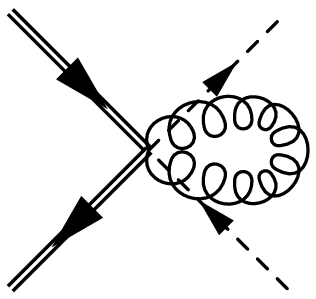} 
\caption{Soft diagrams at one loop.}\label{Fig:2}
\end{figure}

\begin{figure}[htb]
\centering
\includegraphics[width=2.5cm]{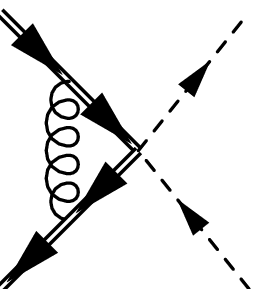} \hspace{0.5cm}
\includegraphics[width=2.5cm]{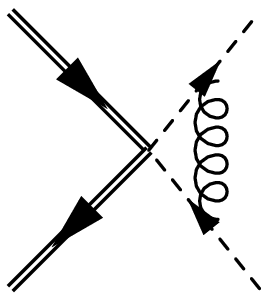} \hspace{0.5cm}
\includegraphics[width=2.5cm]{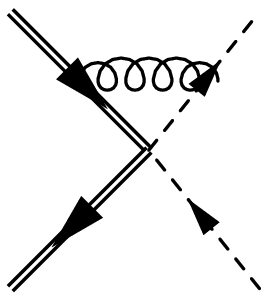}

\includegraphics[width=2.5cm]{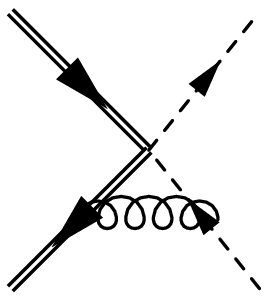} \hspace{0.5cm}
\includegraphics[width=2.5cm]{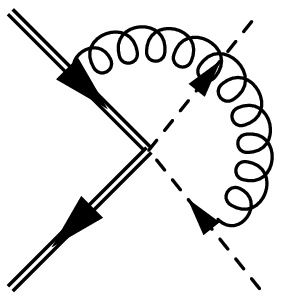} \hspace{0.5cm}
\includegraphics[width=2.5cm]{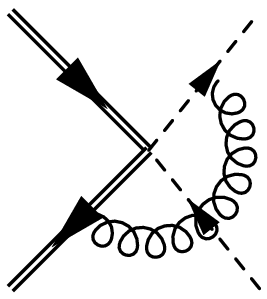} 
\caption{Ultrasoft diagrams at one loop.}\label{Fig:3}
\end{figure}

\begin{figure}[htb]
\center
\includegraphics[width=2.5cm]{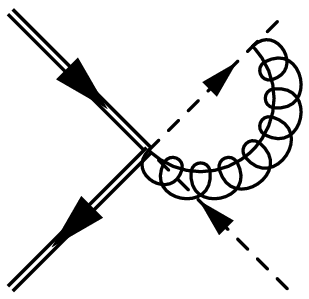}
\hspace{1cm}
\includegraphics[width=2.5cm]{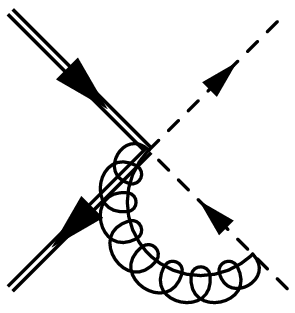}
\caption{Collinear diagrams at one loop. }\label{Fig:4}
\end{figure}

The $\mu$ dependence of
$J_{\efti}$ is governed by an equation of
the following form \cite{Manohar},
\begin{equation}\label{eq:run.1}
 \frac{d}{d \ln \mu} J_{\efti}(\mu) = - \gamma_{\efti}(\mu)
 J_{\efti}(\mu) \; ,
\end{equation}
where the anomalous dimension $\gamma_{\efti}$ is given by
\begin{equation}\label{eq:run.2}
 \gamma_{\efti} = Z^{-1}_{\efti}
 \frac{d}{ d \ln\mu} Z_{\efti}
\end{equation}
and $Z_{\efti}$ is the counterterm that relates the bare operator $J^{(0)}_{\efti}$ to the renormalized one,
$ J^{(0)}_{\efti} = Z_{\efti}(\mu) J_{\efti}(\mu)$.
Since the l.h.s. of Eq. \eqref{eq:M.1} is independent of the scale
$\mu$, the RGE \eqref{eq:run.1} can be recast as an equation for the
matching coefficient $C(\mu)$,
\begin{equation}\label{eq:run.3}
\begin{split}
 & \frac{d}{d \ln\mu} C(\mu) = \gamma_{\efti}(\mu) C(\mu) \; .
\end{split}
\end{equation}
The counterterm $Z_{\efti}$ cancels the divergences that appear in
Green functions with the insertion of the operator $J_{\efti}$. We
calculate $Z_{\efti}$ in the $\overline{\rm{MS}}$ scheme
by evaluating the divergent part of the four-point Green function at one
loop, given by the diagrams in Figs. \ref{Fig:2} - \ref{Fig:4}.

Since in NRQCD we do not introduce different
gluon fields for different
momentum modes,
``soft'' and ``ultrasoft''
in Fig. \ref{Fig:2} and Fig. \ref{Fig:3} refer to the convention that
we impose soft or ultrasoft scaling to the corresponding loop momentum. The
potential region,
which should be considered in the diagrams
of Fig. \ref{Fig:2}, does not give any divergent contribution.

The integrals are evaluated in dimensional regularization, with $d = 4
- 2\varepsilon$.
We regulate the infrared divergences by keeping the non-relativistic
$b$ and $\bar
b$ and the collinear $c$ and $\bar c$ off-shell:
 $E_{b, \, \bar b}  - \vec{p}^{\; 2}_{b,\, \bar b }/2 m_b =
 \Delta_{b}$, $p^2_c - m^2_c = \Delta^2$ and $p^2_{\bar c} - m^2_c =
 \bar{\Delta}^2$.
We power count the $c$-quark off-shellness as $\Delta^2 \sim \bar{\Delta}^2 \sim
m_b^2 \lambda^2$ and the $b$-quark off-shellness
as $\Delta_b \sim m_b w^2$. We also assume
$\Delta^2, \, \bar \Delta^2 >0$. To avoid double
counting, we define the
one-loop integrals with the $0$-bin subtraction \cite{Manohar:2006nz}.

Even with an off-shellness, the soft diagrams in Fig. \ref{Fig:2} do not contain any scale
and they are completely cancelled by their 0-bin.

The divergent part of the ultrasoft diagrams 
in Fig. \ref{Fig:3} is
\begin{equation}\label{eq:softren.1}
\begin{split}
  i \mathcal{M}_{usoft} & = -i \frac{\alpha_s}{4\pi} \, \left\{ 2
    C_F \, \left[\frac{1}{\varepsilon^2} - \frac{1}{\varepsilon} \ln
      \left(\frac{\Delta^2 \, \bar{\Delta}^2}{n \cdot p^{}_c \bar n
          \cdot p^{}_{\bar c} \, \mu^2}\right)\right]
+ \frac{1}{N_c}\frac{1}{\varepsilon} \ln (-1 - i 0)
- \frac{1}{N_c} \,\frac{1}{\varepsilon} \right\}  J_{\efti} \; ,
\end{split}
\end{equation}
where $C_F = (N^2_c-1)/2N_c$ and $\mu$ is the $\overline{\rm{MS}}$
unit mass, $\mu^2 = 4 \pi \mu^2_{\rm{MS}} \exp{(- \gamma_{E})}$.
The first term in the curly brackets of Eq. \eqref{eq:softren.1}
corresponds to the sum of
the divergences in the second diagram in Fig. \ref{Fig:3}, where an
ultrasoft gluon is exchanged
between the $c$ and $\bar c$ quarks collinear
in back-to-back directions, and those in the last four diagrams of
the same figure,
which contain ultrasoft interactions between the
initial and final states.
The second term is an extra imaginary piece generated by the second
diagram in Fig. \ref{Fig:3}. The $- i 0$ prescription in the argument of the logarithm, where $0$ is a positive infinitesimal quantity,
follows from the prescriptions in the
quark propagators and from the choice $\Delta^2, \bar \Delta^2 > 0$.
The divergences arising from the ultrasoft exchanges between the $b \,
\bar b$ pair in the first diagram in Fig. \ref{Fig:3} are encoded in the last term in Eq. \eqref{eq:softren.1}.

The initial and final
states cannot interact by exchanging collinear gluons
because the emission or absorption of a collinear gluon would give the $b$ quark
an off-shellness of order $m^2_b$, which
cannot appear in the effective theory. For the same reason, the $c$
and $\bar c$ cannot exchange $n$ or $\bar n$-collinear gluons. The
only collinear loop diagrams consist of the emission of a $n$($\bar
n$)-collinear gluon from the Wilson line $W_{n\, (\bar n)}$ in
$J_{\efti}$ and its absorption by the $\bar c$($c$) quark, as
shown in Fig. \ref{Fig:4}.
The divergent part of the sum of the two collinear diagrams is 
\begin{equation}\label{eq:softren.5}
 i \mathcal M_{\rm{coll}} = i  \frac{\alpha_s}{4\pi}2C_F
 \left[\frac{2}{\varepsilon^2} + \frac{1}{\varepsilon} \left(2-
     \ln\left( \frac{\Delta^2 \bar \Delta^2}{{\mu}^2  \mu^2}\right)
   \right)\right] J_{\efti} \; .
\end{equation} 
The collinear diagrams are calculated with a 0-bin subtraction
\cite{Manohar:2006nz}, that is, we subtract from the naive collinear
integrals the same integrals in the limit in which the loop momentum
is ultrasoft. In this way we avoid double counting between the
diagrams in Figs. \ref{Fig:3} and \ref{Fig:4}.

Summing Eqs. \eqref{eq:softren.1} and \eqref{eq:softren.5} and adding
factors of $Z^{1/2}_{\psi}$ for each field,
\begin{equation*}
 Z_{\psi_b} = Z_{\chi_b}  = 1 + \frac{1}{\varepsilon}
 \frac{\alpha_s}{2 \pi}C_F \; , \qquad Z_{\xi_n} = Z_{\xi_{\bar n}} = 1 -
 \frac{1}{\varepsilon} \frac{\alpha_s}{4\pi} C_F \; ,
\end{equation*}
the divergent piece becomes
\begin{equation}\label{eq:softren.5prime}
 i \mathcal M_{\rm{div}} = i  \frac{\alpha_s}{4\pi}\left \{
 C_F \left[\frac{2}{\varepsilon^2} + \frac{2}{\varepsilon}
   \left(\frac{3}{2} - \ln \left( \frac{n \cdot p_c \bar n \cdot
         p_{\bar c}}{{\mu}^2}\right)\right)\right] +
 \frac{1}{\varepsilon} N_c  + \frac{i \pi}{\varepsilon} \frac{1}{N_c}
\right\}J_{\efti} \; .
\end{equation}
The counterterm $Z_{\efti}$ is chosen so as to cancel the
divergence in Eq. \eqref{eq:softren.5prime},
\begin{equation}\label{eq:softren.5second}
Z_{\efti} = \frac{\alpha_s}{4\pi}\left \{
 C_F \left[\frac{2}{\varepsilon^2} + \frac{2}{\varepsilon}
   \left(\frac{3}{2} - \ln \left( \frac{n \cdot p_c \bar n \cdot
         p_{\bar c}}{{\mu}^2}\right)\right)\right] +
 \frac{1}{\varepsilon} N_c + \frac{i \pi}{\varepsilon}  \frac{1}{N_c}
\right\} \; .
\end{equation}
From the definition \eqref{eq:run.2}, Eq. \eqref{eq:softren.5second},
and recalling that $d\alpha_{s} / d \ln \mu  = - 2 \varepsilon \alpha_{s} + \mathcal O(\alpha^2_{s})$, the
anomalous dimension at one loop is
\begin{equation}
 \label{eq:softren.6}
\gamma_{\efti} = - 2 \frac{\alpha_s(\mu)}{4 \pi} \left\{ 3 C_F  +
  N_c + 4C_F  \ln \left({\frac{\mu}{ \sqrt{n \cdot p_c \bar n \cdot
          p_{\bar c }}}}\right)  + i \pi  \frac{1}{N_c}\right\}\; . 
\end{equation}

An important feature of the anomalous dimension
\eqref{eq:softren.6} is the presence of a term proportional to
$\ln \mu$. Because of this term, the RGE \eqref{eq:run.3}
can be used to resum
Sudakov double logarithms. As we will show shortly, the general
solution of Eq. \eqref{eq:run.3} can be written in the following form:
\begin{equation}\label{eq:sol.1}
C(\mu) = C( \mu_0 ) \left(\frac{\mu_0}{\sqrt{n \cdot p_c \bar n \cdot
      p_{\bar c}}}\right)^{g(\mu_0, \, \mu)} \exp U(\mu_0, \mu) \; ,
\end{equation}
where $g$ and $U$ depend on the initial scale $\mu_0$ and the final
scale $\mu$ that we run down to. For an anomalous dimension of the form
\eqref{eq:softren.6}, $U$ can be expanded as a series,
\begin{equation}\label{eq:sol.2}
U(\mu_0,\mu) = \sum_{n = 1}^{\infty} \alpha^n_{{s}}(\mu_0) \sum_{L =
  0}^{n+1} u_{n, L} \ln^{n - L +1} \frac{\mu}{\mu_0} \; .
\end{equation} 
If $\mu/\mu_0 \ll 1$, the most relevant terms in the expansion
\eqref{eq:sol.2} are those with $L=0$, which we call ``leading logs''
(LL).
Terms with higher $L$ are subleading; we call the
terms with
$L=1$ ``next-to-leading logs'' (NLL), those with $L=2$
``next-to-next-leading logs'' (NNLL), and, if
$L = m$, we denote them with $\mathrm{N}^m\mathrm{LL}$.
The RGE \eqref{eq:run.3} determines the coefficients in the expansion \eqref{eq:sol.2}.
With the anomalous dimensions written as
\begin{equation}\label{eq:softren.8}
 \gamma_{\efti} = - 2  \left\{ \gamma(\alpha_s) + \Gamma_{}(\alpha_s)
   \ln \left(\frac{ \mu}{ \sqrt{n \cdot p_c \bar n \cdot p_{\bar
           c}}}\right)  \right\} \; ,
\end{equation}
where $\gamma(\alpha_s)$ and $\Gamma_{}(\alpha_s)$ are series in
powers of $\alpha_s$,
\begin{equation*}
 \gamma(\alpha_s) = \frac{\alpha_s}{4\pi} \gamma^{(0)} +
 \left(\frac{\alpha_s}{4\pi}\right)^2 \gamma^{(1)} + \ldots \; , \qquad
 \Gamma_{} (\alpha_s) = \frac{\alpha_s}{4\pi} \Gamma^{(0)}_{}  +
 \left(\frac{\alpha_s}{4\pi}\right)^2 \Gamma^{(1)}_{} + \ldots \; , 
\end{equation*}
it can be proved that the coefficients of the LL, $u_{n 0}$, are
determined by the knowledge of $\Gamma^{(0)}$ and of
the QCD $\beta$ function at one loop. The NLL coefficients $u_{n 1}$ are instead
completely determined if $\Gamma$ and $\beta$ are known at two loops
and $\gamma(\alpha_s)$ at one loop.

In the case we are studying, the ratio of the scales $ \mu/\mu_0 \sim
m_c/ 2 m_b$ is not extremely small. Indeed, as to be seen shortly,
the numerical contributions of the LL and NLL terms in the
series \eqref{eq:sol.2} are of the same size. It is
therefore important to work at NLL accuracy, which requires the
calculation of the coefficient of  $\ln \mu$ to two loops. The factors
of $\ln \mu$  are induced by cusp angles involving light-like Wilson
lines and their coefficients are universal $\Gamma(\alpha_s)  \propto
\Gamma_{\rm{cusp}}(\alpha_s)$ \cite{cusp}. The cusp anomalous dimension
$\Gamma_{\rm{cusp}}(\alpha_s)$ is known at two loops \cite{cusp},

\begin{equation}\label{eq:sol.8}
\Gamma_{\textrm{cusp}} (\alpha_s)  = \frac{\alpha_s}{4\pi}
\Gamma^{(0)}_{\textrm{cusp}}  + \left(\frac{\alpha_s}{4\pi}\right)^2
\Gamma^{(1)}_{\textrm{cusp}} \; ,\end{equation}
with
\begin{equation}
\Gamma^{(0)}_{\textrm{cusp}}  = 4 C_F \; ,  \qquad \qquad 
\Gamma^{(1)}_{\textrm{cusp}}  = 4 C_F \left[ \left(\frac{67}{9} -
    \frac{\pi^2}{3} \right) N_c - \frac{10}{9} n_f  \right] \; ,
\end{equation}
while the  constant of proportionality between $\Gamma(\alpha_s)$ 
and $\Gamma_{\rm{cusp}}(\alpha_s)$ is fixed by the
one-loop calculation. Since we have determined $\gamma^{(0)}$, 
\begin{equation}\label{eq:softren.9}
 \gamma^{(0)} =  3C_F + N_c +   i  \frac{\pi}{N_c} \; ,
\end{equation}
and the $\beta$ function is known, we have all the ingredients to
provide the NLL approximation for $U(\mu_0,\mu)$ and $g(\mu_0,\mu)$.
Taking into account the tree-level initial condition in
Eq. \eqref{eq:M.2}, Eq. \eqref{eq:sol.1} determines the leading-order
matching coefficient, with NLL resummation. 

The solution \eqref{eq:sol.1} can be derived by writing
Eq. \eqref{eq:run.3}  as
\begin{equation}
d \ln C = - 2 \frac{d\alpha}{\beta(\alpha)} \left\{ \gamma(\alpha) +
  \Gamma_{\rm{cusp}}(\alpha)
\left[  \ln \left(\frac{ \mu_0}{ \sqrt{n
          \cdot p_c \bar n \cdot p_{\bar c}}} \right)
    +\int_{\alpha(\mu_0)}^{\alpha} \frac{d
      \alpha^{\prime}}{\beta(\alpha^{\prime})}
\right] \right\} \; ,
\end{equation}
where we have used the definition of the $\beta$ function, 
$\beta(\alpha) = d\alpha/d \ln\mu$, to write $\ln \mu$ and $d\ln\mu$ in
terms of $\alpha$. Integrating both sides from $\mu_0$ to $\mu$ and
exponentiating the result
we find the form given in Eq. \eqref{eq:sol.1}, with
\begin{equation}
 \begin{split}
   U(\mu_0,\mu) & = -2 \int_{\alpha_s(\mu_0)}^{\alpha_s(\mu)}
 \frac{d\alpha}{\beta(\alpha)} \left\{ \gamma(\alpha) +
   \Gamma_{\rm{cusp}}(\alpha) \int_{\alpha(\mu_0)}^{\alpha} \frac{d
     \alpha^{\prime}}{\beta(\alpha^{\prime})}  \right\} \; ,\\
g(\mu_0,\mu) & = -2 \int_{\alpha_s(\mu_0)}^{\alpha_s(\mu)}
 \frac{d\alpha}{\beta(\alpha)}  \Gamma_{\rm{cusp}}(\alpha) \; . 
 \end{split}
\end{equation}
At NLL, we find
\begin{equation}\label{eq:softren.13}
\begin{split}
U(\mu_b, \mu)  & = \frac{2 \pi
  \Gamma^{(0)}_{\textrm{cusp}}}{\beta_0^2} 
\left[  \frac{r -1 - r
    \ln r}{\alpha_s(\mu)}
+ \frac{\beta_0 \gamma^{(0)}_{\rm{Re}}}{2\pi \Gamma^{(0)}_{\textrm{cusp}}}\ln r 
 +
 \left(\frac{\Gamma^{(1)}_{\textrm{cusp}}}{\Gamma^{(0)}_{\textrm{cusp}}}
   - \frac{\beta_1}{\beta_0}\right) \frac{1 - r + \ln r}{4\pi}
\right. \\    & \hspace{65pt} +  \left.    \frac{\beta_1}{8 \pi
    \beta_0} \ln^2 r
\right] + \frac{\gamma^{(0)}_{\rm{Im}}}{\beta_0}\ln r \; , 
\end{split}
\end{equation}
and
\begin{equation}\label{eq:softren.14}
g(\mu_b,\mu)     =  \frac{\Gamma^{(0)}_{\rm{cusp}}}{\beta_0}
\left[ \ln r +
  \left(\frac{\Gamma^{(1)}_{\rm{cusp}}}{\Gamma^{(0)}_{\rm{cusp}}} -
    \frac{\beta_1}{\beta_0}\right) \frac{\alpha_{\rm s}(\mu_b)}{4\pi}
  (r-1)
\right] \; ,
\end{equation}
where $r = \alpha_s(\mu)/\alpha_s(\mu_b)$ and we have renamed the initial
scale $\mu_b$, to denote its connection to the scale $2m_b$. 
In Eqs. \eqref{eq:softren.13} and \eqref{eq:softren.14} 
we have used the two-loop beta function,
\begin{equation}\label{eq:softren.11}
\beta(\alpha_s)  = - 2 \alpha_s \left( \frac{\alpha_s}{4\pi}
   \beta_0 + \left(\frac{\alpha_s}{4\pi}\right)^2 \beta_1 \right) \; ,
\end{equation}
with
\begin{equation}
\beta_0   = 11 - \frac{2}{3} n_f \; , \qquad \qquad
 \beta_1  = \frac{34}{3} N^2_c - \frac{10}{3} N_c n_f - 2 C_F n_f \; .\end{equation}
In Eq. \eqref{eq:softren.13} we have kept the contributions of the real and imaginary part of 
$\gamma^{(0)}$ separated. 
The imaginary part of $\gamma^{(0)}$ changes the phase of the matching
coefficient $C(\mu)$, but this phase is irrelevant for the calculation
of physical observables like the decay rate,
which
depend on the
square
modulus of $C(\mu)$.
In Sec. \ref{rate} the factor $U(\mu_b,\mu)$
will be evaluated between the scales $\mu_b = 2m_b$ and $\mu = m_c$,
with $n_f = 4$ active quark flavors. The numerical evaluation shows
that the LL term, represented by the first term in the
brackets in Eq. \eqref{eq:softren.13}, is slightly smaller than and have the
opposite sign of the term proportional to
$\gamma^{(0)}_{\textrm{Re}}$, which dominates the NLL
contribution. This observation confirms, \textit{a posteriori}, the
necessity to work at NLL accuracy in the resummation of logarithms of
$m_c/2m_b$.

The RGE \eqref{eq:run.3} and its solution \eqref{eq:sol.1} thus
allow us to rewrite Eq. \eqref{eq:M.1} as
\begin{equation*}
 J_{\rm{QCD}} = C(\mu) J_{\efti}(\mu) = C(\mu_b = 2 m_b) \exp U(2 m_b,
 m_c) J_{\efti}(\mu = m_c ) \; ,
\end{equation*}
which avoids the occurrence of any large
logarithm in the matching
coefficient or in the matrix element of the effective operator.

\section{$\text{pNRQCD} + \text{bHQET}$}\label{bHQET}
\subsection{Matching}\label{bHQETmatch}
In the second step, we integrate out the soft modes by matching
$\efti$ onto  $\eftii$.
In $\text{NRQCD} + \text{SCET}$, contributions to the exclusive
decay processes are
obtained by considering time-ordered products of
$J_{\efti}$
and the terms in the $\efti$
Lagrangian that contain
soft-gluon emissions. The soft
gluons have enough virtuality to produce a pair of light quarks
travelling in opposite directions with ultracollinear momentum
scaling. These light quarks bind
to the charm quarks to form
back-to-back $D$ mesons. The total momentum of two back-to-back
ultracollinear quarks is $2 m_b \lQ/m_c \, (1,1,\lambda)$ and the invariant
mass of the pair is $q^2 \sim (2m_b \lQ/m_c)^2 \sim m^2_c$:
in $\text{NRQCD} + \text{SCET}$, only soft gluons have enough energy to
produce them. 
The time-ordered products in $\text{NRQCD} + \text{SCET}$
are matched onto six-fermion
operators in $\text{pNRQCD} + \text{bHQET}$,
where fluctuations of order $m^2_c$ cannot be resolved.

\begin{figure}[htb]
 \includegraphics[width=2.5cm]{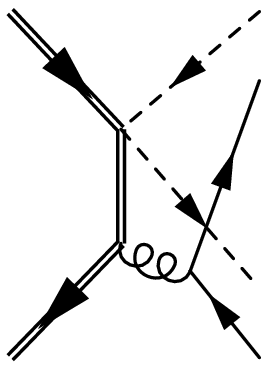} \hspace{0.5cm}
 \includegraphics[width=2.5cm]{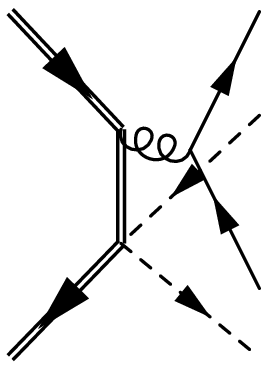} \hspace{1cm}
\hspace{1cm} \hspace{0.5cm}
\includegraphics[width=2.5cm]{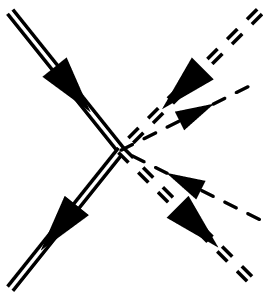}
\caption{Matching $\text{NRQCD} + \text{SCET}$ onto $\text{pNRQCD} + \text{bHQET}$.
On the r.h.s. the double solid lines represent heavy $b$ ($\bar{b}$)
(anti)quarks, the double dashed lines 
bHQET $c$ ($\bar c$) (anti)quarks, and the single dashed lines
collinear light quarks.}\label{Fig.5a}
\end{figure}

We consider the scale $\mu^{\prime} = m_c$ to be much bigger than
$\lQ$, so
the matching can be done in perturbation theory.
The Feynman diagrams contributing to the matching are shown in
Fig. \ref{Fig.5a}. 
The gluon
and the $b$-quark
propagators have off-shellness
of order $m^2_c$, so the two diagrams on the l.h.s. match onto
six-fermion operators on the r.h.s.

The amplitude for the decay of a bottomonium with quantum numbers $^{2S+1} L_J$ into two $D$
mesons has the following form:
\begin{equation}\label{eq:MM.1}
\begin{split}
 i \mathcal M = i C(\mu)  \int \frac{d\omega}{\omega}
 \frac{d\bar{\omega}}{\bar\omega} T(\omega, \bar{\omega}, \mu,
 \mu';^{2S+1}L_J) F^2(\mu^{\prime})  \langle D A, \, D B |\mathcal
 O^{^{2S+1}L_J}_{AB} (\omega, \bar{\omega}, \mu' ) | \bar b b
 (^{2S+1}L_J) \rangle.
\end{split}
\end{equation}
$A$ and $B$,
which label the final
states and the $\eftii$
operators $\mathcal O^{^{2S+1}L_J}_{AB}$,
denote the possible parity, spin, and polarization of the $D$ mesons,
$A, B = \{P, V_L, V_T \}$,
indicating respectively a pseudoscalar $D$ meson, a
longitudinally-polarized vector meson $D^*$, and a
transversely-polarized vector meson $D^*$. Unlike $J_{\efti}$, we have dropped the subscript
$\eftii$ in $\mathcal O^{^{2S+1}L_J}_{AB}$ in order to simplify the notation.

The
$\eftii$ operators that contribute to the decay of the $P$-wave states are
\begin{equation}\label{eq:MM.2}
\begin{split}
F^2(\mu^{\prime}) \, \mathcal
O^{^3P_J}_{PP}(\omega,\bar\omega,\mu^{\prime}) & =
\chi^{\dagger}_{\bar b} \, \vec{p}_b \cdot \vec{\sigma}_{\perp}
\psi_{b} \; \bar{\mathcal H}^{c}_{\bar n} \, \frac{\slashchar n}{2}
\gamma^5 \, \delta\left(-\bar{\omega} - n \cdot \mathcal P\right)
\chi^{\bar l}_{\bar n} \; \bar{\chi}^{l}_{n}  \delta\left(\omega- \bar
  n \cdot \mathcal P^{\dagger}\right)\, \frac{\slashchar {\bar n}}{2}
\gamma^5 \,  \mathcal H^{\bar c}_{n} \; , \\
F^2(\mu^{\prime}) \, \mathcal O^{^3P_J}_{V_L \,
  V_L}(\omega,\bar\omega,\mu^{\prime}) & =  \chi^{\dagger}_{\bar b} \,
\vec{p}_b \cdot \vec{\sigma}_{\perp} \psi_{b} \; \bar{\mathcal
  H}^{c}_{\bar n} \, \frac{\slashchar n}{2}  \,
\delta\left(-\bar{\omega} - n \cdot \mathcal P\right) \chi^{\bar
  l}_{\bar n} \; \bar{\chi}^{l}_{n}  \delta\left(\omega- \bar n \cdot
  \mathcal P^{\dagger}\right)\, \frac{\slashchar {\bar n}}{2}  \,
\mathcal H^{\bar c}_{n} \; , \\
F^2(\mu^{\prime}) \, \mathcal O^{^3P_J}_{V_T \,
  V_T}(\omega,\bar\omega,\mu^{\prime}) & =  \chi^{\dagger}_{\bar b} \,
{p}^{\,(\mu}_{b \, \perp}  \sigma^{\nu)}_{\perp} \psi_{b} \;
\bar{\mathcal H}^{c}_{\bar n} \, \frac{\slashchar n}{2} \gamma_{\mu \,
  \perp} \, \delta\left(-\bar{\omega} - n \cdot \mathcal P\right)
\chi^{\bar l}_{\bar n} \; \bar{\chi}^{l}_{n}  \delta\left(\omega- \bar
  n \cdot \mathcal P^{\dagger}\right)\, \frac{\slashchar {\bar n}}{2}
\gamma_{\nu \,\perp} \,  \mathcal H^{\bar c}_{n}, \\
\end{split}
\end{equation}
where ${p}^{\,(\mu}_{b \, \perp}  \sigma^{\nu)}_{\perp}$ is a
symmetric, traceless tensor,
\begin{equation*}
{p}^{\,(\mu}_{b \, \perp}  \sigma^{\nu)}_{\perp} = \frac{1}{2} \left(
  {p}^{\, \mu}_{b \, \perp}  \sigma^{\nu}_{\perp} + {p}^{\,\nu}_{b \,
    \perp}  \sigma^{\mu}_{\perp}-  g^{\mu \nu}_{\perp}  \vec{p}_b
  \cdot \vec{\sigma}_{\perp} \right) \; .
\end{equation*}
At leading order in the
$\eftii$ expansion, the $\eta_b$ can only
decay into a pseudoscalar and a vector meson, with an operator given by
\begin{equation}\label{eq:MM.2.a}
\begin{split}
F^2(\mu^{\prime}) \, \mathcal O^{^1S_0}_{P\,
  V_L}(\omega,\bar\omega,\mu^{\prime})   =  \chi^{\dagger}_{\bar b} \,
\psi_{b} \; &
\left[ \bar{\mathcal H}^{c}_{\bar n} \, \frac{\slashchar
    n}{2} \gamma^5 \, \delta\left(-\bar{\omega} - n \cdot \mathcal
    P\right) \chi^{\bar l}_{\bar n} \; \bar{\chi}^{l}_{n}
  \delta\left(\omega- \bar n \cdot \mathcal P^{\dagger}\right)\,
  \frac{\slashchar {\bar n}}{2}  \,  \mathcal H^{\bar c}_{n}
\right. \\  & + \left. \bar{\mathcal H}^{c}_{\bar n} \,
  \frac{\slashchar n}{2}  \, \delta\left(-\bar{\omega} - n \cdot
    \mathcal P\right) \chi^{\bar l}_{\bar n} \; \bar{\chi}^{l}_{n}
  \delta\left(\omega- \bar n \cdot \mathcal P^{\dagger}\right)\,
  \frac{\slashchar {\bar n}}{2} \gamma^5  \,  \mathcal H^{\bar c}_{n} 
\right].
\end{split}
\end{equation}
For later convenience, in the 
definition
 of the effective operators
\eqref{eq:MM.2} and \eqref{eq:MM.2.a} we have factored out the term
$F^2(\mu^{\prime})$, which is related to the $D$-meson decay
constant. 
The definition of $F^2(\mu^{\prime})$ will become clear
when we introduce the $D$-meson distribution amplitudes.
The fields $\chi^{l}_{n}$ and $\chi^{\bar l}_{\bar n}$ are
ultracollinear gauge-invariant light-quark fields, while  $\mathcal H^{c}_{\bar n} =
W^{\dagger}_{\bar n} h^{c}_{\bar n}$ and $\mathcal H^{\bar c}_{n} =
W^{\dagger}_n h^{\bar c}_{n}$ are bHQET heavy-quark fields, which are
invariant under an ultracollinear gauge transformation. The Wilson
lines $W_{n}$ and $W_{\bar n}$ have the same definition as in
Eq.~\eqref{eq:w}, with the restriction to ultracollinear gluons. 
Eqs. \eqref{eq:MM.2} and \eqref{eq:MM.2.a} allow us to interpret
$\omega$  as the component of the light-quark momentum along the
direction $n$. Similarly, $\bar \omega$ represents the component of
the light-antiquark momentum along $\bar n$. 
The minus
sign in the delta function $\delta (-\bar\omega- n \cdot \mathcal P)$
is chosen so that $\bar\omega$ is positive.

The tree-level matching coefficients are
\begin{equation}\label{eq:MM.3}
\begin{split}
 T(\omega, \bar{\omega}, \mu, \mu' = m_c;\, ^3P_J) & =
 \frac{C_F}{N_c^2} \frac{4 \pi \alpha_s(m_c)}{m_b}  \frac{1}{\omega +
   \bar{\omega} } \; ,\\
 T(\omega, \bar\omega,   \mu, \mu^{\prime} = m_c;\, ^1S_0) & =
 \frac{C_F}{N_c^2} \frac{4 \pi \alpha_s(m_c)}{m_b} \frac{1}{2}
 \frac{\omega - \bar\omega}{\omega + \bar{\omega} } \; . 
\end{split}
\end{equation}
Note that, at leading order in the
$\eftii$ expansion, the matching
coefficient $T(\omega,\bar\omega, \mu, \mu^{\prime};\, ^3P_J)$ is
independent of the spin and polarization of the final states,
or of the total angular momentum $J$ of the $\chi_b$. 

An important feature of bHQET is that the ultracollinear and ultrasoft
sectors can be decoupled at leading order in the power counting by a
field redefinition reminiscent of the collinear-usoft decoupling in
SCET \cite{Bauer:2001yt} \cite{Fleming:2007qr}. 
For bHQET in the $n$ direction, the decoupling is achieved by defining
$h^{\bar c}_n \rightarrow Y_n h^{\bar c}_n$
and $\bar \xi^{l}_n \rightarrow \bar \xi^l_n Y^{\dagger}_n$, where $Y_n$
is an ultrasoft Wilson line,
\begin{equation}\label{eq:y}
Y_n = \sum_{\rm{perms}} \left[\exp\left( -\frac{g }{n \cdot \mathcal
      P} n \cdot A_{us}\right)\right] \; .
\end{equation}
An analogous redefinition with $n \rightarrow \bar n$ decouples ultrasoft
from
$\bar n$-ultracollinear quarks and gluons. 
These redefinitions do not affect the operators in Eqs. \eqref{eq:MM.2} and \eqref{eq:MM.2.a}
because all the induced Wilson lines cancel out.
As a consequence, at leading order in the $\eftii$ power counting,
there is no interaction between the initial and the final
states, since the former can only emit and absorb ultrasoft 
gluons that do not couple to ultracollinear degrees of freedom. 
Furthermore, fields in the two copies of bHQET, boosted in opposite
directions, cannot interact with each other
because the interaction
with a
$\bar n$-ultracollinear gluon would give a
$n$-ultracollinear quark or
gluon a virtuality of order $m^2_c$,
which, however, cannot appear in $\eftii$. 
The matrix elements of the operators $\mathcal
O^{^{2S+1}L_J}_{AB}(\omega,\bar\omega,\mu)$, therefore,
factorize as
\begin{equation}\label{eq:MM.4}
\begin{split}
 F^2(\mu^{\prime}) \langle A B |\mathcal O^{^{2S+1}L_J}_{A\,B}
 (\omega, \bar{\omega}, \mu' ) | \bar b b \rangle  =
& \langle 0 | \chi^{\dagger}_{\bar b} \, T^{^{2S+1}L_J}_{AB} \psi_{b}
| \bar b b \rangle \, \langle A | \bar{\mathcal H}^{c}_{\bar n} \,
\frac{\slashchar n}{2} \Gamma_A \, \delta\left(- \bar{\omega} - n
  \cdot \mathcal P\right) \chi^{\bar l}_{\bar n} | 0\rangle  \\ & 
\langle B | \bar{\chi}^{l}_{n}  \delta\left(\omega- \bar n \cdot
  \mathcal P^{\dagger}\right)\, \frac{\slashchar {\bar n}}{2} \Gamma_B
\, \mathcal H^{\bar c}_{n} | 0\rangle \; ,
\end{split}
\end{equation}
where $\Gamma_A = \{\gamma_5, 1, \gamma^{\mu}_{\perp} \}$ and
$T^{^{2S+1}L_J}_{AB} = \{1, \, \vec{p}_b \cdot \vec{\sigma}_{\perp},\,
{p}^{\,(\mu}_{b \, \perp}  \sigma^{\nu)}_{\perp}\}$. The
charge-conjugated contribution is understood in the $\eta_b$ case. 

The quarkonium state and the $D$ mesons in Eq. \eqref{eq:MM.4} have
respectively non-relativistic
and HQET normalization:
\begin{equation*}
 \langle \chi_{bJ}(E^{\prime},\vec p^{\, \prime})| \chi_{bJ}(E,\vec p)
 \rangle = (2\pi)^3 \delta^{(3)}(\vec p - \vec{p}^{\, \prime}) \; ,
\quad 
  \langle D(v^{\prime}, k^{\prime})| D(v, k) \rangle =  2 v^0
  \delta_{v,v^{\prime}}(2\pi)^3 \delta^{(3)}(\vec k -
  \vec{k}^{\prime}) \; ,
\end{equation*}
where $v^0$ is the 0th component of the 4-velocity $v^{\mu}$.

The $D$-meson matrix elements can be expressed in terms of the
$D$-meson light-cone distribution amplitudes:
\begin{eqnarray}
 & \langle P | \bar\chi^{l}_{n} \, \frac{\slashchar{\bar n}}{2}
 \gamma^5 \, \delta\left({\omega} - \bar n \cdot \mathcal
   P^{\dagger}\right) {\mathcal H}^{\bar c}_{ n}  | 0\rangle  & =    i
 F_P(\mu^{\prime})  \frac{\bar n \cdot v}{2} \phi_{P} ({\omega},
 \mu^{\prime}) \; ,
 \label{eq:MM.6}\\ 
 & \langle V_L | \bar\chi^{l}_{n} \, \frac{\slashchar{\bar n}}{2}  \,
 \delta\left({\omega} - \bar n \cdot \mathcal P^{\dagger}\right)
 {\mathcal H}^{\bar c}_{ n}  | 0\rangle  & =    F_{V_L}(\mu^{\prime})
 \frac{\bar n \cdot v}{2} \phi_{V_L} ({\omega}, \mu^{\prime}) \; ,
 \label{eq:MM.6.1}
 \\ 
& \langle V_T | \bar\chi^{l}_{n} \, \frac{\slashchar{\bar n}}{2}
\gamma^{\mu}_{\perp} \, \delta\left({\omega} - \bar n \cdot \mathcal
  P^{\dagger}\right) {\mathcal H}^{\bar c}_{ n}  | 0\rangle  & =
F_{V_T}(\mu^{\prime})  \frac{\bar n \cdot v}{2} \varepsilon^{\mu}_{\perp}
\phi_{V_T} ({\omega}, \mu^{\prime}) \; , \label{eq:MM.6.2} 
\end{eqnarray}
where $\varepsilon^{\mu}_{\perp}$ is the transverse polarization of
the vector meson. 
The constants 
$F_A(\mu^{\prime})$, with $A = \{P, V_L, V_T\}$, are related to the matrix elements of the local
heavy-light currents in coordinate space. In the heavy-quark limit,
where $D$ and $D^{*}$ are degenerate, $F_A $ is the same for all the
three states: $F \equiv F_P = F_{V_L} = F_{V_T} $. In this limit,
\begin{equation}\label{eq:MM.6b}
 \langle 0 | \bar{\xi}^{\, \bar l}_{n}  \, \frac{\slashchar{\bar n}}{2} \gamma^5   {h}^{c}_{n} (0) |  P \rangle
 =   - i F(\mu^{\prime})  \frac{\bar n \cdot v^{\prime}}{2} \; .
\end{equation}
At tree level, the matrix element is proportional to the $D$-meson
decay constant  $f_D = 205.8 \pm 8.5 \pm 2.5$ MeV \cite{:2008sq}. More
precisely,  $F(\mu^{\prime}) = f_D \sqrt{m_D} $,
where the factor $\sqrt{m_D}$ is due to
HQET normalization.
The scale dependence of $F$ is determined by the renormalization of
heavy-light HQET currents.
At one loop, Ref. \cite{Manohar} showed that
\begin{equation}\label{eq:defgammaf}
 \frac{d}{ d \ln \mu^{\prime}} F(\mu^{\prime}) = - \gamma_F
 F(\mu^{\prime}) = 3 C_F\frac{\alpha_s}{4 \pi}  F(\mu^{\prime}) \; .
\end{equation}

The pNRQCD matrix elements can be expressed in terms of the heavy quarkonium wavefunctions.
The operator $\chi^{\dagger}_{\bar b} \vec p_{b} \cdot \vec{\sigma}_{\perp}
\psi_b$ contains a component with
$J=0$ and a component with $J=2$ and $J_z = 0$,
so its matrix element has
non-vanishing overlap with both $\chi_{b0}$ and $\chi_{b2}$. 
The operator $\chi^{\dagger}_{\bar b}  p^{\, (\mu}_{b}  {\sigma}^{\,
  \nu)}_{\perp} \psi_b$ instead has only contributions with $J=2$ and
$J_{z} = \pm 2$ and therefore it only overlaps with $\chi_{b2}$. 
In terms of the bottomonium wavefunctions, the pNRQCD matrix elements
are expressed as
\begin{eqnarray}
&\langle 0| \chi^{\dagger}_{\bar b} \vec p_{b} \cdot \vec{\sigma}_{\perp}
\psi_b | \chi_{b0} \rangle & = \frac{2}{\sqrt{3}} \,
\sqrt{\frac{3N_c}{2\pi}} R^{\prime}_{\chi_{b0}}(0,\mu^{\prime}) \; , 
\label{eq:quarkwave1}\\
&\langle 0| \chi^{\dagger}_{\bar b} \vec p_{b} \cdot \vec{\sigma}_{\perp}
\psi_b | \chi_{b2} \rangle & = -\sqrt{\frac{2}{15}} \,
\sqrt{\frac{3N_c}{2\pi}} R^{\prime}_{\chi_{b2}}(0,\mu^{\prime}) \; ,
\label{eq:quarkwave2} \\
&\langle 0| \chi^{\dagger}_{\bar b}  p^{\, (\mu}_{b}  {\sigma}^{\,
  \nu)}_{\perp} \psi_b | \chi_{b2} \rangle & =
(\varepsilon^{(2)}_{\mu\nu} + \varepsilon^{(-2)}_{\mu\nu} ) \,
\sqrt{\frac{3N_c}{2\pi}}
R^{\prime}_{\chi_{b2}}(0,\mu^{\prime}) \; , \label{eq:quarkwave3}
\end{eqnarray}
where $R^{\prime}_{\chi_{bJ}}(0)$ is the derivative of the radial
wavefunction of the  $\chi_{bJ}$ evaluated at the origin. At
leading order, the pNRQCD Hamiltonian does not depend on
$J$, so, up to corrections of order $w^2$,
$R^{\prime}_{\chi_{b2}}(0) = R^{\prime}_{\chi_{b0}}(0)$. The numerical
pre-factors in Eqs. \eqref{eq:quarkwave1} and \eqref{eq:quarkwave2}
follow from decomposing $\vec p_{b}\cdot \vec{\sigma}_{\perp}$
into components with definite $J_z$.  $\varepsilon^{(j)}_{\mu \nu}$ is the
polarization tensor of the $\chi_{b2}$ state, and
Eq. \eqref{eq:quarkwave3} states that, at leading order in the $w^2$
expansion, only the particles with polarization
$J_z=\pm 2$ contribute to
$\chi_{b2}$ decay into two transversely-polarized vector mesons.
Similarly, one finds
\begin{equation}
\langle 0 | \chi^{\dagger}_{\bar b}  \psi_b | \eta_b \rangle =
\sqrt{\frac{N_c}{2\pi}} R_{\eta_b} (0,\mu^{\prime}) \; .
\end{equation} 

The factorization of the matrix elements \eqref{eq:MM.4} implies that
the decay rate also factorizes.
For the
decays of
$\chi_{b0}$ and $\chi_{b2}$ into two
pseudoscalar mesons or two longitudinally-polarized vector mesons, we find
\begin{equation}\label{eq:MM.5}
\begin{split}
 \Gamma\left(\chi_{b 0} \rightarrow A A \right) = & \frac{4}{3} \frac{
   m^2_{D} \sqrt{m^2_{\chi_{b 0}}-4 m^2_D}}{8 \pi m_{\chi_{b0}} }
 \frac{3N_c}{2\pi} \, |C\left(\mu\right)|^2 \,
 |R^{\prime}_{\chi_{b0}}(0,\mu^{\prime}) |^2 \\ &
\left[ F^2(\mu^{\prime})  \frac{n \cdot v^{\prime}}{2} \frac{\bar n
    \cdot v}{2} \int \frac{d\omega}{\omega}
  \frac{d\bar{\omega}}{\bar\omega} T\left(\omega, \bar{\omega}, \mu,
    \mu^{\prime};\, ^3P_J\right) \phi_{A} (\bar{\omega},\mu^{\prime})
  \phi_{A}(\omega,\mu^{\prime})\right]^2 
\end{split}
\end{equation}
and
\begin{equation}\label{eq:MM.5b}
\begin{split}
 \Gamma\left(\chi_{b 2} \rightarrow A A \right) = & \frac{2}{15}
 \frac{ m^2_{D} \sqrt{m^2_{\chi_{b 2}}-4 m^2_D}}{8 \pi m_{\chi_{b2}} }
 \frac{3N_c}{2\pi} |C\left(\mu\right)|^2 \,
 |R^{\prime}_{\chi_{b2}}(0,\mu^{\prime}) |^2 \\ &
\left[ F^2(\mu^{\prime})  \frac{n \cdot v^{\prime}}{2} \frac{\bar n
    \cdot v}{2} \int \frac{d\omega}{\omega}
  \frac{d\bar{\omega}}{\bar\omega} T\left(\omega, \bar{\omega}, \mu,
    \mu^{\prime};\, ^3P_J\right) \phi_{A} (\bar{\omega},\mu^{\prime})
  \phi_{A}(\omega,\mu^{\prime})\right]^2 \; , 
\end{split}
\end{equation}
where $A = P, V_L$.
For the decay of
$\chi_{b2}$ into two transversely-polarized vector mesons, one finds
the decay rate by summing over the possible transverse polarizations:
\begin{equation}\label{eq:MM.6c}
\begin{split}
 \Gamma\left(\chi_{b 2} \rightarrow V_T V_T \right) = &
 \frac{2}{5} \frac{ m^2_{D} \sqrt{m^2_{\chi_{b 2}}-4 m^2_D}}{8 \pi
   m_{\chi_{b2}} } \frac{3N_c}{2\pi}  |C\left(\mu\right)|^2  \, \,
 |R^{\prime}_{\chi_{b2}}(0,\mu^{\prime}) |^2  \\ &
\left[ F^2(\mu^{\prime})  \frac{n \cdot v^{\prime}}{2} \frac{\bar n
    \cdot v}{2} \int \frac{d\omega}{\omega}
  \frac{d\bar{\omega}}{\bar\omega} T\left(\omega, \bar{\omega}, \mu,
    \mu^{\prime};\, ^3P_J\right) \phi_{V_T}
  (\bar{\omega},\mu^{\prime})
  \phi_{V_T}(\omega,\mu^{\prime})\right]^2 \; . 
\end{split}
\end{equation}
In the case of
$\eta_b$ decay into a pseudoscalar and a longitudinally-polarized vector meson, we find
\begin{equation}\label{eq:MM.6d}
\begin{split}
  \Gamma\left(\eta_{b} \rightarrow P V_L + \rm{c.c.} \right)  =   \frac{ m^2_{D}
   \sqrt{m^2_{\eta_{b}}-4 m^2_D}}{8 \pi m_{\eta_{b}} }
 \frac{N_c}{2\pi}  |C\left(\mu\right)|^2  \, \,
 |R_{\eta_{b}}(0,\mu^{\prime}) |^2 \,  \frac{1}{2} \left[ F^2(\mu^{\prime})
   \frac{n \cdot v^{\prime}}{2} \frac{\bar n \cdot v}{2} 
  \right. & \\  \left. \int \frac{d\omega}{\omega}\frac{d\bar{\omega}}{\bar\omega}
  T\left(\omega, \bar{\omega}, \mu, \mu^{\prime};\, ^1S_0\right) 
   (\phi_{V_L} (\bar{\omega},\mu^{\prime})
   \phi_{P}(\omega,\mu^{\prime}) - \phi_{V_L} (\omega,\mu^{\prime})
   \phi_{P}(\bar\omega,\mu^{\prime}) ) \right]^2 & . 
\end{split}
\end{equation}
Note that we are working in the limit $m_c \to \infty$,
where the $m_{D^*}- m_D$ mass splitting vanishes.

The factorized formulas Eqs. \eqref{eq:MM.4} and \eqref{eq:MM.5} -
\eqref{eq:MM.6d} are the main
results of this paper.
Each decay rate of \eqref{eq:MM.5} - \eqref{eq:MM.6d} depends on two
calculable matching coefficients, $C$ and $T$, and three
non-perturbative, process-independent matrix elements, namely,
two $D$-meson distribution amplitudes and the bottomonium wavefunction.
In Sec.~\ref{rate} we will provide a model-dependent estimate of the
decay rates \eqref{eq:MM.5} - \eqref{eq:MM.6d} and will discuss the
phenomenological implications.
We conclude this section by
observing that all the non-perturbative matrix elements cancel out in
the ratios $\Gamma(\chi_{b0} \rightarrow PP)/\Gamma(\chi_{b2}
\rightarrow PP)$ and $\Gamma(\chi_{b0} \rightarrow V_L
V_L)/\Gamma(\chi_{b2} \rightarrow V_L V_L)$, since the spin symmetry of pNRQCD guarantees $R^{\prime}_{\chi_{b0}}(0) = R^{\prime}_{\chi_{b 2}}(0)$, at leading order in $\eftii$.
Neglecting the $\chi_{b0}$ - $\chi_{b2}$ mass difference, we find, up to
corrections of order $w^2$, 
\begin{equation}\label{ratio}
 \Gamma(\chi_{b0} \rightarrow A A)/\Gamma(\chi_{b2} \rightarrow A A)=
 \frac{4}{3}\frac{15}{2}  = 10 \; ,
\end{equation}
with $A = P, V_L$.

\subsection{Running}\label{bHQETrun}
\begin{figure}[htb]
\center
 \includegraphics[width=2.5cm]{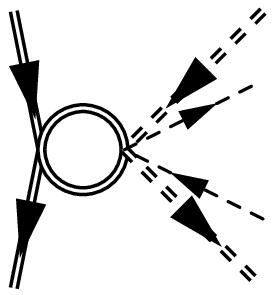}
\hspace{2cm}
 \includegraphics[width=2.5cm]{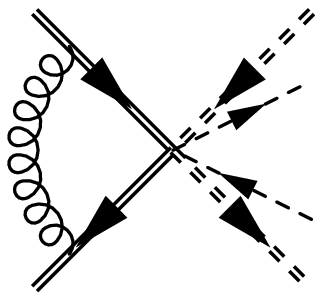} 
\caption{One-loop diagrams in pNRQCD.  The first diagram contains
  insertions of quark-antiquark potentials. In the second diagram the
  gluon is ultrasoft.}\label{Fig.5}
\end{figure}

\begin{figure}[htb]
\center
 \includegraphics[width=2.5cm]{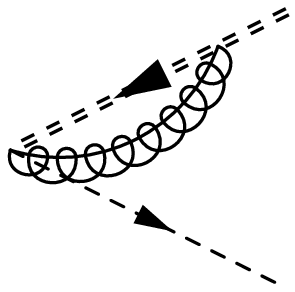} 
\hspace{1cm}
 \includegraphics[width=2.5cm]{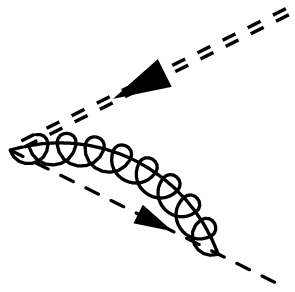}
\hspace{1cm}
 \includegraphics[width=2.5cm]{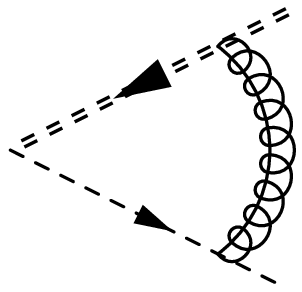}
\caption{One-loop diagrams in bHQET. There are three analogous diagrams for the other copy of bHQET.}\label{Fig.6}
\end{figure}

The dependence of the matching coefficient $T(\omega,\bar\omega,
\mu,\mu^{\prime};\, ^{2S+1}L_J)$ and of the operators in
Eqs. \eqref{eq:MM.5} - \eqref{eq:MM.6d} on the scale $\mu^{\prime}$ is
driven by
a RGE that can be obtained by renormalizing the
$\eftii$ operators. 
The RGE for the $\eftii$ operators, which
also defines the anomalous dimension $\gamma_{\eftii}$, 
is similar to Eq. \eqref{eq:run.1},
\be\label{eq:rgeeftii}
\begin{split}
 \frac{d}{d \ln  \mu^{\prime}} & \left[ F^2(\mu^{\prime})  \mathcal
   O^{^{2S+1}L_J}_{AB}  \left(\omega, \bar{\omega}, \mu^{\prime}\right)^{}\right] = \\  
 & \hspace{2.5cm} -\int d\omega^{\prime} \int d\bar{\omega}^{\prime}
\gamma_{\eftii}(\omega, \omega^{\prime};  \bar{\omega},
\bar{\omega}^{\prime};
\mu^{\prime}) F^2(\mu^{\prime}) \mathcal
O^{^{2S+1}L_J}_{AB}(\omega^{\prime}, \bar{\omega}^{\prime},
\mu^{\prime}) \; .
\end{split}
\ee
To calculate the anomalous dimension at one loop, we 
compute the divergent part of the diagrams in Figs. \ref{Fig.5} and \ref{Fig.6}.
As mentioned in Sec.~\ref{Eft}, the pNRQCD Lagrangian has the
following structure,
\begin{equation*}
  L_{\rm{pNRQCD}} =  \int d^3x \, \mathcal L^{usoft}_{\rm{NRQCD}} +
  L_{\rm{pot}} \; ,
\end{equation*}
where the superscript $usoft$ indicates that the gluons in the NRQCD
Lagrangian are purely ultrasoft ($m_b w^2, m_b w^2$), while $
L_{\rm{pot}}$ contains four-fermions operators, which are non-local in space,
\begin{equation*}
L_{\rm{pot}} = \int d^3 x_1 d^3 x_2 \psi^{\dagger}_{\alpha}(t, \vec x_1)
\chi_{\beta}(t, \vec x_2) V_{\alpha \beta, \gamma \delta}(\vec r\,)
\chi^{\dagger}_{\gamma}(t,\vec x_2) \psi_{\delta}(t,\vec x_1) \; .
\end{equation*}
At leading order in $\alpha_s(m_b w)$ and $r$, $V$ is the Coulomb potential 
\begin{equation*}
V_{\alpha \beta, \gamma \delta} = \frac{\alpha_s(m_b w)}{r}
t^a_{\alpha \delta} t^a_{\gamma \beta} \; .
\end{equation*}
For the explicit form of higher-order potentials, see, for example,
Refs. \cite{Brambilla:2004jw} \cite{Hoang:2002yy}.
Vertices from $L_{\rm{pot}}$ generate one-loop diagrams as the
first diagram in Fig. \ref{Fig.5}. However, these diagrams do not
give any contribution to the anomalous dimension at one loop.
Indeed, the insertion of the Coulomb potential
$1/r$ in
Fig. \ref{Fig.5} does not produce UV divergences.
Insertions of the $1/m_b$ potentials yield divergences
but the coefficient of the $1/m_b$ potential is proportional to
$\alpha_s^2(m_b w)$,
so it is not relevant if we are content with a NLL
resummation. Insertions of $1/m^2_b$ potentials give
divergences proportional to subleading operators,
which can be neglected.
The second diagram in Fig. \ref{Fig.5} yields a result completely
analogous to the last term in Eq. \eqref{eq:softren.1}, with
the only difference of a color pre-factor,
\begin{equation}
 i \mathcal M_{\rm{pNRQCD}} = - i   \frac{\alpha_s}{2 \pi}
 \,\frac{C_F}{\varepsilon} \mathcal
 O^{^{2S+1}L_J}_{AB}(\omega,\bar\omega, \mu) \; .
\end{equation}
This divergence is completely cancelled by the $b$-quark field
renormalization constant $Z_b$, and hence the pNRQCD diagrams in Fig. \ref{Fig.5}
do not contribute to the anomalous dimension at one loop.


On the bHQET side, the third diagram in Fig. \ref{Fig.6} is
convergent, and hence it does not contribute to the anomalous dimension.
The first two diagrams give
\begin{equation}\label{eq:RR.1} i \mathcal M_{\rm{bHQET}, \bar n}  =  i \int d\omega^{\prime}
 d\bar\omega^{\prime}  \Delta(\omega,\omega^{\prime}, \, \bar\omega,
 \bar\omega^{\prime}) \mathcal O^{^{2S+1}L_J}_{AB} (\omega^{\prime},
 \bar\omega^{\prime}, \mu) \; , 
\end{equation}
with
\begin{equation}\label{eq:RR.1.2}
\begin{split}
\Delta(\omega,\omega^{\prime}, \, \bar\omega, \bar\omega^{\prime})  =
&  \frac{\alpha_s}{2\pi} C_F \delta(\omega - \omega^{\prime}) \left\{
  \delta\left(\bar{\omega} -\bar{\omega}^{\prime}\right) \left[
    -\frac{1}{2\varepsilon^2} - \frac{1}{\varepsilon} \ln \left(
      \frac{\mu^{\prime} n \cdot v^{\prime}}{\bar{\omega}^{\prime}} \right) +
    \frac{1}{\varepsilon}\right] \right. \\  & + \left.
  \frac{1}{\varepsilon}\left[
    \theta\left(\bar{\omega}-\bar{\omega}^{\prime}\right)\left(\frac{1}{\bar{\omega}-
        \bar{\omega}^{\prime}}\right)_+
    +\theta\left(\bar{\omega}^{\prime}-\bar{\omega}\right)
    \theta\left(\bar{\omega}\right)
    \frac{\bar{\omega}}{\bar{\omega}^{\prime}}\left(\frac{1}{\bar{\omega}^{\prime}-\bar{\omega}}\right)_+
  \right]  \right\} \; .
\end{split} 
\end{equation}
The diagrams for the bHQET copy in the $n$-direction give a result
analogous to Eqs. \eqref{eq:RR.1} and \eqref{eq:RR.1.2}, with $\bar\omega \rightarrow
\omega$,
$\bar\omega^{\prime} \rightarrow \omega^{\prime}$, and $n
\cdot v^{\, \prime} \rightarrow \bar n \cdot v$.
Extracting $\gamma_{\eftii}$ from the divergence is again
standard, just as we did in the case of $\gamma_{\efti}$.
After adding to Eq. \eqref{eq:RR.1.2} the bHQET field renormalization
constants $Z_{h}$ and $Z_{\xi}$ for heavy and light quarks
\begin{equation*}
Z_{h} = 1 + \frac{1}{\varepsilon}\frac{\alpha_s }{2\pi } C_F \; ,
\qquad Z_{\xi} = 1 - \frac{1}{\varepsilon} \frac{\alpha_s}{4\pi} C_F
\; ,
\end{equation*}
we find
\begin{equation}\label{eq:RR.2.0}
\begin{split}
\gamma_{\eftii}(\omega, \omega^{\prime};  \bar{\omega}, \bar{\omega}^{\prime};
\mu^{\prime})  =     
 2 \gamma_F\, \delta\left(\omega - \omega^{\prime}\right)
 \delta\left(\bar\omega-\bar\omega^{\prime}\right)  + \gamma_{\mathcal
   O}(\omega, \omega^{\prime};  \bar{\omega}, \bar{\omega}^{\prime};
 \mu^{\prime}) \; ,
\end{split}
\end{equation}
with
\begin{equation}\label{eq:RR.2}
\begin{split}
& \gamma_{\mathcal
   O}(\omega, \omega^{\prime};  \bar{\omega}, \bar{\omega}^{\prime};
 \mu^{\prime}) \\
& = \frac{\alpha_s}{4\pi}\, 4 C_F
\delta\left(\omega-\omega'\right)
\delta\left(\bar{\omega}-\bar{\omega}^{\prime}\right)\left[-1 +
  \ln\left(\frac{{\mu}^{\prime} n \cdot v^{
        \prime}}{\bar{\omega}^{\prime}}  \right) +
  \ln\left(\frac{{\mu}^{\prime} \bar n \cdot v^{ }}{{\omega}^{\prime}}
  \right)\right]   \\ & -  \frac{\alpha_s}{4\pi} 4 C_F  \delta\left(\omega-\omega'\right)\left[\theta\left(\bar{\omega}-\bar{\omega}^{\prime}\right) \left(\frac{1}{\bar{\omega}-\bar{\omega}^{\prime}}\right)_{+}+ \theta\left(\bar{\omega}^{\prime} - \bar{\omega}\right)\theta\left(\bar{\omega}\right) \frac{\bar{\omega}}{\bar{\omega}^{\prime}}\left(\frac{1}{\bar{\omega}^{\prime}-\bar{\omega}}\right)_{+}\right] 
 \\ &
-  \frac{\alpha_s}{4\pi} 4 C_F
\delta\left(\bar\omega-\bar\omega^{\prime}\right)\left[\theta\left({\omega}-{\omega}^{\prime}\right)
  \left(\frac{1}{{\omega}-{\omega}^{\prime}}\right)_{+}+
  \theta\left({\omega}^{\prime} -
    {\omega}\right)\theta\left({\omega}\right)
  \frac{{\omega}}{{\omega}^{\prime}}\left(\frac{1}{{\omega}^{\prime}-{\omega}}\right)_{+}\right]
\; .
\end{split}
\end{equation}
The term proportional to $\gamma_F$ in Eq. \eqref{eq:RR.2.0} reproduces the
running of $F^2(\mu^{\prime})$ \eqref{eq:defgammaf}.
$\gamma_\mathcal{O}$ is responsible for the running of
the $D$-meson distribution amplitudes and it agrees with the result found
in Ref. \cite{Lange:2003ff}. Also, in
Eq. \eqref{eq:RR.2} the coefficient of
$\ln \mu^{\prime}$ is proportional to $\Gamma_{\rm{cusp}}(\alpha_s)$.  
Note that, since the bHQET Lagrangian is spin-independent, the
anomalous dimension does not depend on the spin
or on the polarization of
the $D$ meson in the final state,
at leading order in the power counting.

Using Eqs. 
\eqref{eq:rgeeftii} and
\eqref{eq:RR.2.0} we find the following
integro-differential RGE for the operator $\mathcal O(\omega,
\bar{\omega}, \mu^{\prime})$:
\begin{equation}\label{eq:RGE.1}
 \frac{d}{d \ln  \mu^{\prime}} \mathcal O(\omega, \bar{\omega}, \mu^{\prime}) = 
-\int d\omega^{\prime} \int d\bar{\omega}^{\prime} \gamma_{\mathcal
  O}(\omega, \omega^{\prime};  \bar{\omega}, \bar{\omega}^{\prime};
\mu^{\prime}) \mathcal O(\omega^{\prime}, \bar{\omega}^{\prime},
\mu^{\prime}) \; ,
\end{equation}
where we have dropped both the subscripts $A$, $B$, and the superscript
$^{2S+1}L_J$, since $\gamma_{\mathcal O}$ does not depend on the
quantum numbers of the initial
or final
state.
Using the fact that the convolution of $ F^2(\mu^{\prime})\,  T(\omega,
\bar{\omega},\mu,\mu^{\prime};\, ^{2S+1}L_J)$ and the operator
$\mathcal O^{^{2S+1}L_J}_{AB}(\omega,\bar{\omega},\mu^{\prime})$ is
$\mu^{\prime}$-independent, we can write an equation for the
coefficient,
 \begin{equation}\label{eq:RGE.2}
\begin{split}
 \frac{d}{d \ln  \mu^{\prime}} \left[ F^2(\mu^{\prime}) \, T(\omega,
   \bar{\omega}, \mu, \mu^{\prime}) \right] & = 
\int d\omega^{\prime} \int d\bar{\omega}^{\prime}
\frac{\omega}{\omega^{\prime}} \frac{\bar\omega}{\bar\omega^{\prime}}
F^{2}(\mu^{\prime})\, T(\omega^{\prime}, \bar{\omega}^{\prime},
\mu,\mu^{\prime}) \gamma_{\mathcal O}(\omega^{\prime}, \omega;
\bar{\omega}^{\prime}, \bar{\omega};
\mu^{\prime}) 
 \\
& = \int d\omega^{\prime} \int d\bar{\omega}^{\prime}
F^{2}(\mu^{\prime}) \, T(\omega^{\prime}, \bar{\omega}^{\prime}, \mu,
\mu^{\prime})   \gamma_{\mathcal O}(\omega, \omega^{\prime};
\bar{\omega}, \bar{\omega}^{\prime};\mu^{\prime})\; ,
\end{split}
\end{equation}
where the last line follows from the property of
$\gamma_\mathcal{O}$ at one loop,
\begin{equation*}
 \frac{\omega}{\omega^{\prime}}
 \frac{\bar{\omega}}{\bar{\omega}^{\prime}} \gamma_{\mathcal
   O}(\omega^{\prime}, \omega;   \bar{\omega}^{\prime}, \bar{\omega};
\mu^{\prime}) =  \gamma_{\mathcal O}(\omega, \omega^{\prime};
\bar{\omega}, \bar{\omega}^{\prime};\mu^{\prime}) \; ,
\end{equation*}
as can be explicitly verified from the expression in Eq. \eqref{eq:RR.2}.

Eq. \eqref{eq:RGE.2} can be solved following the methods described in
Ref. \cite{Lange:2003ff}. We discuss the details of
the solution
in App. \ref{AppA},
where we derive the analytic expressions for $T(\omega, \bar\omega, \mu, \mu^{\prime}; \,^3P_J)$ and 
$T(\omega, \bar\omega, \mu, \mu^{\prime}; \,^1S_0)$, with the initial
conditions at the scale $\mu^{\prime}_c = m_c$ expressed in
Eq.~\eqref{eq:MM.3}.

\section{Decay Rates and Phenomenology}\label{rate}
In Sec.~\ref{bHQETmatch} we gave the factorized
expressions for the decay rates \eqref{eq:MM.5} - \eqref{eq:MM.6d}:
$\Gamma(\chi_{b0, \, 2} \rightarrow PP)$, $\Gamma(\chi_{b0, \, 2}
\rightarrow V_L V_L)$, $\Gamma(\chi_{b 2} \rightarrow V_T V_T)$, and
$\Gamma(\eta_{b} \rightarrow P V_L + \rm{c.c.})$.
In Secs.~\ref{SCETrun} and \ref{bHQETrun} we exploited the RGEs
\eqref{eq:run.3} and \eqref{eq:RGE.2} to run the scales $\mu$ and
$\mu^{\prime}$, respectively, from the matching scales $\mu = 2m_b$ and $\mu^{\prime}
= m_c$ to the natural scales that contribute to the matrix elements,
$\mu = m_c$ and $\mu^{\prime} \sim 1$ GeV, resumming in this way
Sudakov logarithms of the ratios
$m_c / 2m_b $ and $m_c/ 1$ GeV.

We proceed now to estimate the
decay rates
\eqref{eq:MM.5} - \eqref{eq:MM.6d}. In order to do so, we
need to evaluate the following ingredients: the light-cone distribution
amplitudes of the $D$ meson and of the longitudinally- and
transversely-polarized $D^*$ mesons, and the wavefunctions of the
states $\eta_b$ and $\chi_{bJ}$.
In principle, these non-perturbative objects could be extracted
from other $\eta_b$, $\chi_b$, and $D$-meson observables.
In the case of the $\eta_b$, the value of the wavefunction 
at the origin can be obtained from a measurement of the inclusive
hadronic width or of the decay rate for
the electromagnetic process $\eta_b \rightarrow \gamma \gamma$,
since they are both proportional to
$|R_{\eta_b}(0)|^2$. Unfortunately, at the moment there are not
sufficient data on $\eta_b$ decays. Another 
way to proceed is to use the spin symmetry of the leading-order pNRQCD
Hamiltonian, which implies $R_{\eta_b}(0) = R_{\Upsilon}(0)$, and to
extract the Upsilon wavefunction from $\Gamma(\Upsilon \rightarrow
e^{+}e^{-}) = 1.28 \pm 0.07$ KeV \cite{Amsler:2008zz}.
Using the leading-order expression for $\Gamma(\Upsilon \rightarrow
e^{+}e^{-})$ \cite{Barbieri:1975ki},
one finds $|R_{\Upsilon}(0)|^2 = 6.92 \pm 0.38 \, \rm{GeV}^3$,
where the error only includes the experimental uncertainty. The above value
is in good agreement with the lattice evaluation
by Bodwin, Sinclair, and Kim \cite{Bodwin:2001mk} and it falls within the
range of values obtained with four different potential models,
as listed in Ref. \cite{Eichten:1995ch}.  

$|R^{\prime}_{\chi_{b0,\,2}}(0)|^2 $ can be obtained from the
electromagnetic decay $\chi_{b 0,2} \rightarrow \gamma
\gamma$. Unfortunately, such decay rates have not been measured yet.
The values listed in Ref.~\cite{Eichten:1995ch} range from a minimum of
$|R^{\prime}_{\chi_{bJ}}(0)|^2 = 1.417$ GeV$^5$, obtained with the
Buchmuller-Tye potential \cite{Buchmuller:1980su}, to a maximum of
$|R^{\prime}_{\chi_{bJ}}(0)|^2 = 2.067$ GeV$^5$, obtained with a
Coulomb-plus-linear potential.
The lattice value is roughly of the same size,
$|R^{\prime}_{\chi_{bJ}}(0)|^2 = 2.3 \; \rm{GeV}^5$,
with an uncertainty of about 15\% \cite{Bodwin:2001mk}. We use this value in our estimate.

For the pseudoscalar $D$-meson distribution amplitude we use two model
functions widely adopted in the study of $B$ physics.
A first possible choice, suggested for example in Ref. \cite{Lange:2003ff},
is a simple exponential decay:
\begin{equation}\label{eq:DR.4}
 \phi^{\rm{Exp}}_{P, 0} (\omega, \mu^{\prime} = 1 \rm{GeV}) = \theta(\omega)
 \frac{\omega}{\lambda^2_D} \exp\left(- \frac{\omega}{\lambda_D}
 \right) \; .
\end{equation}
Another form,  suggested in Ref. \cite{Braun:2003wx}, is
\begin{equation}\label{eq:DR.5}
 \phi^{\textrm{Braun}}_{P, \,0}(\omega, \mu^{\prime} = 1\, \textrm{GeV}) = \theta(\tilde\omega)
 \frac{4}{\lambda_D \pi}  \frac{\tilde{\omega}}{1 + \tilde{\omega}^2}
 \left[\frac{1}{1 + \tilde\omega^2} - \frac{2(\sigma_D -1 )}{\pi^2} \ln
\tilde\omega \right] \; ,
\end{equation}
where
$\tilde \omega = \omega/\mu^{\prime}$. The theta function in Eqs. \eqref{eq:DR.4} and \eqref{eq:DR.5} reflects the fact that the distribution amplitudes $\phi_{A}(\omega, \mu^{\prime})$, with $A = \{ P, V_L, V_T\}$, have support on $\omega  > 0$ \cite{Grozin:1996pq}.
 
The subscript $0$ indicates that these functional forms are
valid in the $D$-meson rest frame, with a HQET velocity-label $v_0
= (1,0,0,0)$.
With the definition we adopt in Eq. \eqref{eq:MM.6}, the distribution
amplitude is not boost-invariant and
in the bottomonium rest frame, in which the $D$ meson has a velocity
$(n \cdot v, \bar n \cdot v,0) \sim (m_c/2m_b, 2m_b/m_c,0)$, it becomes
\begin{equation}\label{eq:DR.5.b}
 \phi_{P}(\omega, \mu^{\prime}) = \frac{1}{\bar n \cdot v} \phi_{P,\,
   0}\left(\frac{\omega}{\bar n \cdot v}, \mu^{\prime} \right) \; ,
\end{equation}
as shown in App. \ref{AppC}.
$\lambda_D$ and $\sigma_D$ in Eqs. \eqref{eq:DR.4} and \eqref{eq:DR.5}
are, respectively, the first inverse moment and the first logarithmic moment of the
$D$-meson distribution amplitude in the $D$-meson rest frame,
\begin{equation*}
 \lambda_D^{-1}(\mu^{\prime}) = \int_0^{\infty}  \frac{d\omega}{\omega} \phi_{P,
   0}(\omega,\mu^{\prime}) \; ,
\end{equation*}
\begin{equation*}
 \sigma_D(\mu^{\prime}) \lambda_D^{-1}(\mu^{\prime}) = - \int_0^{\infty}  \frac{d\omega}{\omega}
 \ln \left(\frac{\omega}{\mu^{\prime}} \right)\phi_{P, 0}(\omega,\mu^{\prime}) \; .
\end{equation*}
Furthermore we assume that the vector-meson distribution amplitudes
$\phi_{V_L}(\omega)$ and $\phi_{V_T}(\omega)$ have the same functional
form as $\phi_{P}(\omega)$,
but with different parameters
$\lambda_{D^*_L}$, $\sigma_{D^*_L}$ and $\lambda_{D^*_T}$,
$\sigma_{D^*_T}$.

The $D$-meson distribution amplitude and its moments have not been
intensively studied
unlike, for example, the $B$-meson distribution amplitude.
Therefore, we invoke
heavy-quark symmetry and
use the moments of the $B$-meson distribution amplitude in order to
estimate the decay rate.
However, the value of $\lambda_B$ is affected by a noticeable uncertainty.  
Using QCD sum rules, Braun \textit{et al.} estimated \cite{Braun:2003wx}
$\lambda_B (\mu^{\prime} = 1\,\rm{GeV}) =  0.460 \pm 0.110 \, \, \textrm{GeV}$,
where the uncertainty is about $25 \%$.
Other authors \cite{Beneke:1999br}  \cite{Lee:2005gza}
\cite{Pilipp:2007sb} give slightly different central values and
comparable uncertainties, so that $\lambda_B$  falls in the range
$0.350 \;\textrm{GeV} < \lambda_B < 0.600$ GeV. The first logarithmic
moment $\sigma_D$ is given in Ref. \cite{Braun:2003wx}, $\sigma_D =
\sigma_B (\mu^{\prime} = 1\, \rm{GeV})= 1.4 \pm 0.4$.
We assume that the moments of the $D^*$-meson distribution amplitudes fall
in the same range as the moments of $\phi_{P}(\omega)$.

We evaluate numerically the convolution integrals in Eqs. \eqref{eq:MM.5} -
\eqref{eq:MM.6d}.
 We choose the matching scales $\mu_b$
and $\mu_c^{\prime}$ to be
$2 m_b$ and $m_c$ respectively. Using the RGEs we run the matching coefficients down to the scales $\mu
= m_c$ and $\mu^{\prime} = 1$ GeV.
For the $b$ and $c$ quark masses we adopt the 1S mass definition \cite{Hoang:1998ng},
\begin{equation}\label{eq:DR.3}
\begin{split} 
 m_b (1S)& = \frac{m_{\Upsilon}}{2} = 4730.15 \pm 0.13 \, \,
 \textrm{MeV} \; ,\\
 m_c (1S)&= \frac{m_{J/\psi}}{2} = 1548.46 \pm 0.01 \, \, \textrm{MeV}
 \; .
\end{split}
\end{equation}
The values of $\alpha_s$ at the relevant scales are \cite{Amsler:2008zz} 
$\alpha_s(2 m_b) = 0.178 \pm 0.005$, $\alpha_s (m_c) = 0.340 \pm
0.020$, and $\alpha_s (1\, \rm{GeV}) \sim 0.5$.
With these choices, the value of $g$ in Eq. \eqref{eq:RGE.7aa} is
$g(m_c, 1 \, \rm{GeV}) = -0.12 \pm 0.02$.

The decay rates $\Gamma(\chi_{bJ}\rightarrow A A)$ with $A = \{P,
V_L, V_T\}$, \eqref{eq:MM.5} - \eqref{eq:MM.6c}, depend on the masses of the $\chi_{b J}$ and of the $D$
mesons, whose most recent values are reported in Ref. \cite{Amsler:2008zz}.
Since the effects due to the mass splitting of the $\chi_{bJ}$ and $D$
multiplets are subleading in the EFT power counting, we use in the
evaluation the average mass of the $\chi_{bJ}$ multiplet and the
average mass of $D$ and $D^*$ mesons: $m_{\chi_{b J}}  = 9898.87 \pm
0.28 \pm 0.31$ MeV and $m_{D} = 1973.27 \pm 0.18$ MeV. Therefore, the
velocity of the $D$ mesons in
$\chi_{bJ}$ decay is
$\bar n \cdot v = n \cdot v^{\prime } = m_{\chi_{b J}}/m_{D}
 = 5.02$,
with negligible error.
The decay rate $\Gamma(\eta_b \rightarrow P V_{L} + \rm{c.c.})$ \eqref{eq:MM.6d}
depends on the mass of the $\eta_b$,
which has been recently measured:
$m_{\eta_b} = 9388.9 ^{+ 3.1}_{-2.3} \pm 2.7$ MeV \cite{:2008vj}. The velocity of
the $D$ meson in the $\eta_b$ decay is $ \bar n \cdot v = n \cdot
v^{\prime } = m_{\eta_b}/m_D = 4.76$, again with negligible error.

\begin{figure}[t]
\includegraphics[width=8cm]{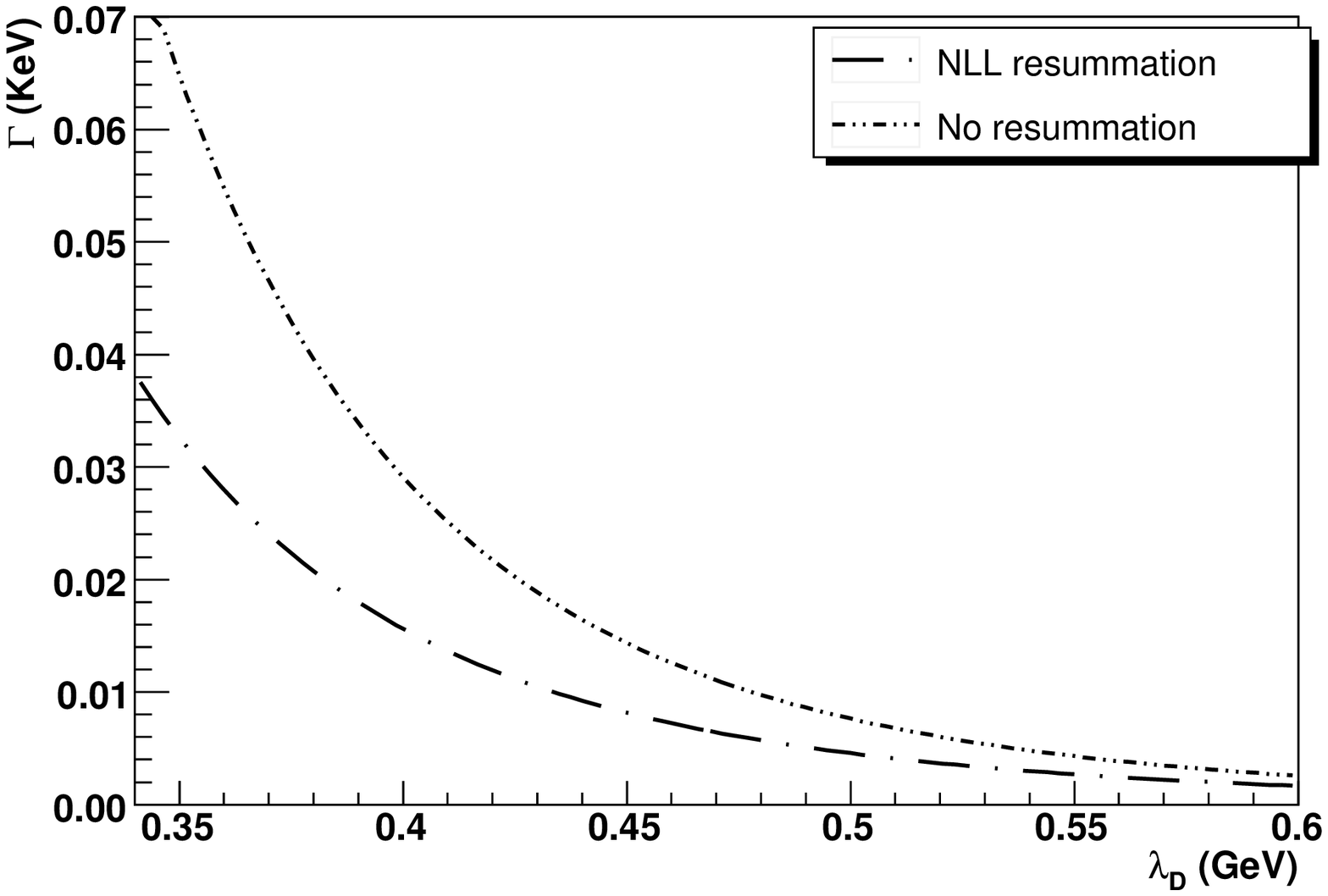}
\includegraphics[width=8cm]{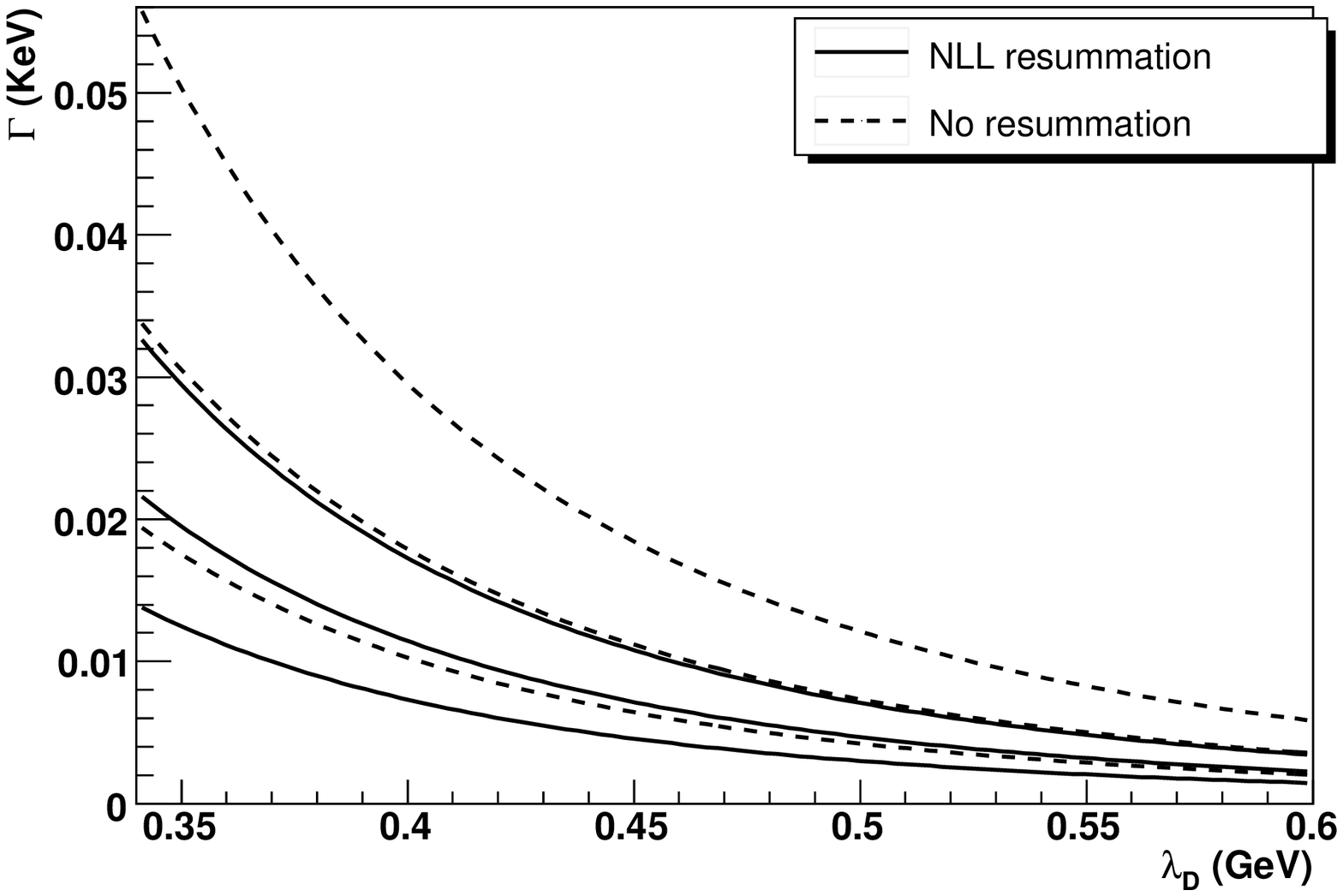}
\caption{$\Gamma(\chi_{b0}\rightarrow P P)$ as a function of
  $\lambda_D$, calculated with the distribution amplitudes
  $\phi^{\textrm{Exp}}$ (\textit{left}) and $\phi^{\textrm{Braun}}$
  (\textit{right}). The dash dotted and solid lines denote the
  NLL-resummed decay rate. For comparison,
the decay rate without resummation is also shown, denoted by
dash double-dotted (\textit{left}) and dashed (\textit{right}) lines. 
For $\phi^{\textrm{Braun}}$ we
vary the parameter $\sigma_D$ from
$\sigma_D = 1$ (lower curve) to $\sigma_D = 1.4$ (middle curve) to $\sigma_D = 1.8$ (upper
curve). 
}\label{Fig:graph.1}
\end{figure}

\begin{figure}[t]
\includegraphics[width=8cm]{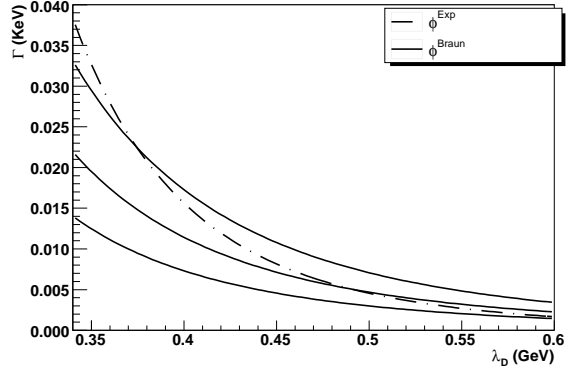}
\caption{$\Gamma(\chi_{b0}\rightarrow P P)$ as a function
  $\lambda_D$. The dash dotted line denotes the decay rate calculated
  with $\phi^{\textrm{Exp}}$, while the three solid lines with
  $\phi^{\textrm{Braun}}$. For $\phi^{\textrm{Braun}}$ we
vary the value of the parameter $\sigma_D$ from $\sigma_D = 1$ (lower curve) to $\sigma_D = 1.4$ (middle curve) to
  $\sigma_D = 1.8$ (upper curve).}\label{Fig:graph.3}
\end{figure}

The
decay rate $\Gamma(\chi_{b 0} \rightarrow P P)$ \eqref{eq:MM.5}, obtained
with $\phi^{\rm{Exp}}$ and $\phi^{\rm{Braun}}$ separately, is shown in
Fig. \ref{Fig:graph.1}. In order to see the impact of resumming
Sudakov logarithms, we show for both distribution amplitudes the
results with (\textit{i}) the LL and NLL resummations 
and (\textit{ii}) without any resummation at all.
In the plots, we call the resummed results NLL-resummed,
indicating that Sudakov logarithms are resummed up to NLL.
For both distribution amplitudes the resummation does have a
relevant effect on the decay rate. In the case of $\phi^{\rm{Exp}}$
the resummation decreases the decay rate by a factor of $ 2 - 1.5$ as
$\lambda_D$ goes from the lowest to the highest value under consideration.
In the case of $\phi^{\rm{Braun}}$  the
decay rate decreases too,
for example, by a factor 1.5
when $\sigma_D = 1.4$.
In Fig. \ref{Fig:graph.3} we compare the decay rates obtained with the
two distribution amplitudes. Over the range of $\lambda_D$ we are
considering the two decay rates are in rough agreement with each other.


Figs. \ref{Fig:graph.1} - \ref{Fig:graph.3} also describe the relation between the decay
rate $\Gamma(\chi_{b0}\rightarrow V_L V_L)$ and $\lambda_{D^*_L}$.
According to Eqs. \eqref{eq:MM.5b} and
\eqref{eq:MM.6c}, the processes $\chi_{b2} \rightarrow P P$,  $\chi_{b2} \rightarrow V_L
V_L$, and $\chi_{b2} \rightarrow V_T V_T$ show
an analogous dependence on the first inverse
moments of the light-cone
distribution
amplitudes, and they differ from Figs. \ref{Fig:graph.1} -
\ref{Fig:graph.3} by constant pre-factors. Therefore, we do not show
explicitly their plots.

Qualitatively, Figs. \ref{Fig:graph.1} - \ref{Fig:graph.3} show a
dramatic dependence of the decay rate on the inverse moment
$\lambda_D$. Using Eqs. \eqref{eq:MM.5}, \eqref{eq:DR.5.b} and \eqref{eq:RGE.18a}, one can
show that when $\phi^{\rm{Braun}}$ is used,
the decay rate is proportional to $\lambda_D^{-4}$, while it scales
as $\lambda_D^{-6 - 4 g}$ when we adopt $\phi^{\rm{Exp}}$, with $g$ defined in
Eq. \eqref{eq:RGE.7aa}.
As a
consequence, the decay rate 
drops by an order of magnitude when $\lambda_D$ goes from $0.350$ GeV
to $0.600$ GeV. 
The particular sensitivity of exclusive
bottomonium decays into two charmed
mesons
to the light-cone structure of the $D$ meson ---much stronger than
usually observed in $D$- and $B$-decay observables---
is due to the dependence of the amplitude on the product of two
distributions (one for each meson)
and to the non-trivial dependence
of the matching coefficient $T$ on the light-quark momentum labels $\omega$ and $\bar\omega$ at tree level.
On one hand, the strong dependence on a relatively
poorly known quantity prevents us from predicting the decay 
rate  $\Gamma(\chi_{b0} \rightarrow D^{} D^{})$.
On the other hand, however, it suggests that, if the decay rate is
measured, this channel could be used to better determine interesting
properties of the $D$-meson distribution amplitude,
such as $\lambda_D$ and $\sigma_D$.
The viability of this suggestion
relies on the control over the
theoretical error attached to the curves in Fig. \ref{Fig:graph.1}
and on the actual chances to observe the process $\chi_{b} \rightarrow
D D$ at current experiments.

The uncertainty of the decay rate stems mainly from three sources.
First, there are corrections coming from  subleading EFT operators.
In matching $\text{NRQCD} + \text{SCET}$ onto $\text{pNRQCD} +
\text{bHQET}$ (Sec. \ref{bHQETmatch}), we neglected the subleading
$\eftii$ operators 
that are 
suppressed
by powers of $\Lambda_{\rm{QCD}}/m_c$ and $w^2$, relative to the
leading $\eftii$ operators in Eqs. \eqref{eq:MM.2} and
\eqref{eq:MM.2.a}. In matching QCD onto $\text{NRQCD} + \text{SCET}$
(Sec. \ref{SCETmatch}), we kept only
$J_{\efti}$ \eqref{eq:M.2} and neglected subleading $\efti$ operators,
suppressed by powers of  
$\lambda$ and $w^2$. These subleading $\efti$ operators
would match onto subleading 
$\eftii$ operators, suppressed by powers of
$\Lambda_{\rm{QCD}}/m_c$ and $w^2$.
Using $w^2 \sim 0.1$ and $\Lambda_{\rm{QCD}}/m_c \sim 0.3$, we find a
conservative estimate for the non-perturbative corrections to be about
$30 \%$.

Second, there are perturbative corrections to the matching
coefficients $C$ and $T$. Since $\alpha_s(2 m_b) = 0.178$, we expect a
$20 \%$ correction from the one-loop contributions in matching QCD
onto
$\text{NRQCD} + \text{SCET}$.
In the second matching step, similarly, the one-loop corrections to
$T(\omega,\bar\omega, \mu, \mu^{\prime};\, ^{2S+1}L_J)$ would be proportional to
$\alpha_s (m_c) \sim 30 \%$.
We can get an idea of their relevance by estimating the
dependence of the decay rate
\eqref{eq:MM.5} on the matching
scales $\mu_b$ and $\mu^{\prime}_c$.
If the matching coefficients $C$ and $T$ and the anomalous dimensions
$\gamma_{\efti}$ and $\gamma_{\mathcal
  O}(\omega,\omega^{\prime},\bar\omega,\bar\omega^{\prime};\mu^{\prime})$ were
known at all orders, the decay rate would be independent of the
matching scales $\mu_b$ and $\mu^{\prime}_c$.
However, since we only know the first terms in the perturbative
expansions, the decay rate bears a residual renormalization-scale
dependence, whose size is determined by the first neglected terms.

\begin{figure}[t]
\includegraphics[width=8.1cm]{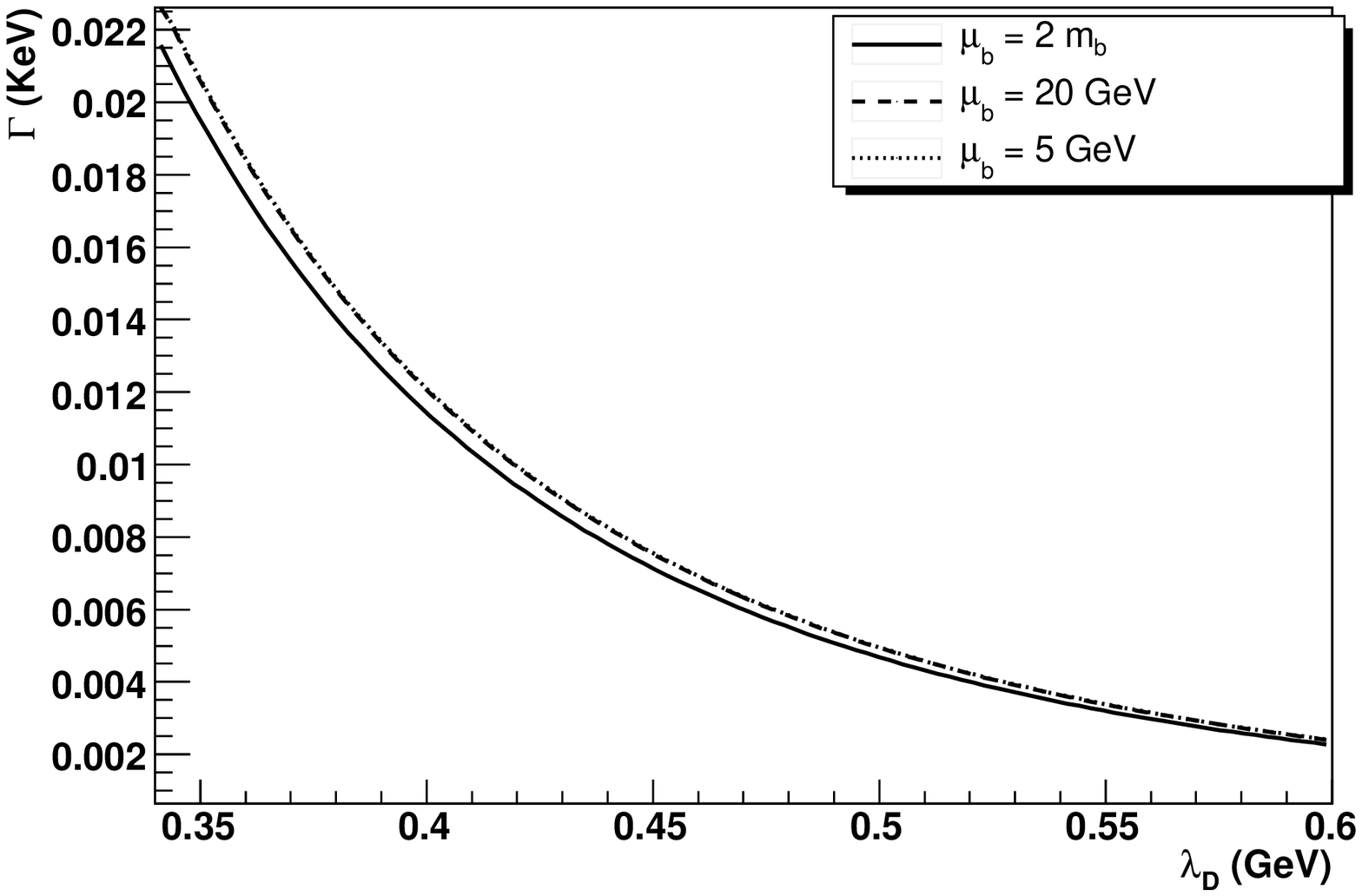}
\includegraphics[width=8.1cm]{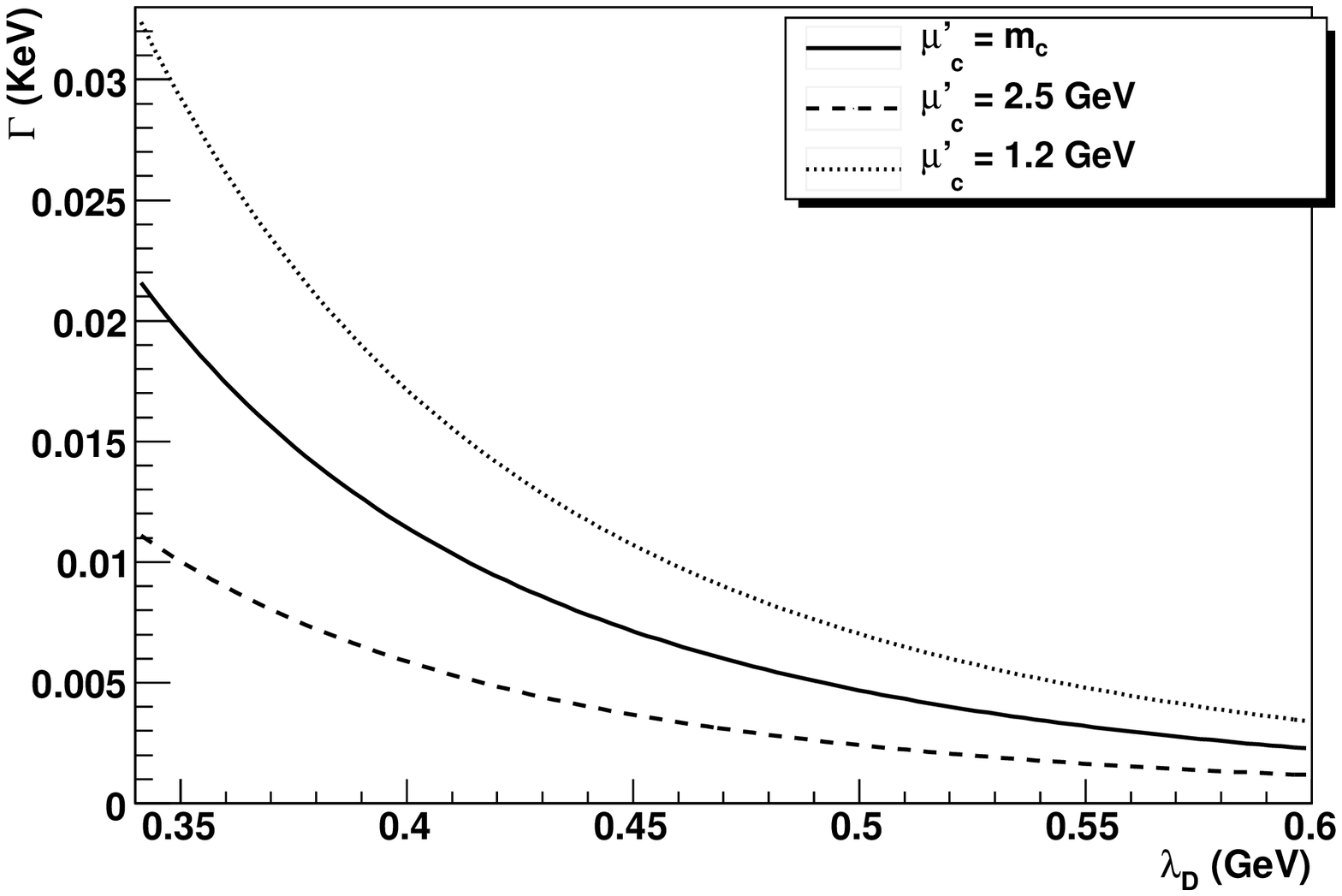}
\caption{\textit{Left}: 
Scale dependence of $\Gamma(\chi_{b0}\rightarrow P P)$ on the matching scale $\mu_b$.
We vary $\mu_b$ from a central value $\mu_b = 2 m_b$ (solid line) to a maximum
of $\mu_b = 20$ GeV (dashed line) and a minimum of $\mu_b = 5$ GeV
(dotted line). The dashed and dotted lines overlap almost perfectly.
\textit{Right}:
Scale dependence of 
$\Gamma(\chi_{b0}\rightarrow P P)$ on the matching scale
$\mu^{\prime}_c$. We varied $\mu^{\prime}_c$ from a central value of
$\mu^{\prime}_c =  m_c$ (solid line) to a maximum of $\mu^{\prime}_c =
2.5$ GeV (dashed line) and a minimum of $\mu^{\prime}_c = 1.2$ GeV
(dotted line).}\label{Fig:graph.4}
\end{figure}

In Fig. \ref{Fig:graph.4} we show the effect of varying
$\mu_b$ between $4 m_b \sim 20 $ GeV  and $m_b \sim 5$ GeV on the
decay rate, using $\phi^{\rm{Braun}}$. The solid line
represents the choice $\mu_b = 2 m_b$,
while the dashed and dotted lines,
which overlap almost perfectly, correspond respectively to
$\mu_b = 20$ GeV and $\mu_b = 5$ GeV. The dependence on $\mu_b$ is
mild, its effect being a variation of
about $5\%$. We obtain analogous results
for the decay rate computed with $\phi^{\rm{Exp}}$, which are
not shown here in order to avoid redundancy. 

On the other hand, even after the 
resummation, the decay rate strongly
depends on $\mu^{\prime}_c$. We vary this scale between $1.2$ GeV and
$2.5$ GeV and we observe an overall variation of about $50 \%$.
We expect the scale dependence to be compensated by the one-loop
corrections to the matching coefficient $T(\omega,\bar\omega, \mu,
\mu^{\prime};\, ^3P_J)$. This observation is reinforced by the fact that the
numerical values of the running factors $U(\mu_b,\mu)$ and
$V(\mu^{\prime}_c,\mu^{\prime})$ (defined respectively in Eqs. \eqref{eq:softren.13} and
\eqref{eq:RGE.7a})
at NLL accuracy are smaller than expected on the basis of naive
counting of the logarithms. As a consequence, the
next-to-leading-order corrections to the matching coefficient could be
as large as the effect of the
NLL
resummation.
In the light of Fig. \ref{Fig:graph.4},
the one-loop correction to $T(\omega,\bar\omega, \mu, \mu^{\prime};\, ^3P_J)$
seems to be an important ingredient for a reliable estimate of the decay rate.

A third source of error comes from the unknown functional form of the
$D$-meson distribution amplitude.
For the study of the $B$-meson shape
function, an expansion in a complete set of orthonormal functions has
recently been proposed and it has provided a systematic procedure to
control the uncertainties due to the unknown functional
form 
\cite{Ligeti:2008ac}. 
The same method should be generalized to the $B$- and $D$-meson
distribution amplitudes, in order to reduce the model dependence of
the decay rate. We leave such an analysis to future work.

To summarize, the calculation of the one-loop matching coefficients
and the inclusion of power corrections of order
$\Lambda_{\rm{QCD}}/m_c$ appear to be necessary to provide a decay
rate with an accuracy of $10 \%$, that would make
the decays $\chi_{bJ} \rightarrow D^{+} D^{-}$, $\chi_{bJ}
\rightarrow D^{0} \bar D^{0}$ competitive processes to improve the determination of
$\lambda_D$ and $\sigma_D$, if the experimental decay rate is observed
with comparable accuracy.

We estimate the decay rate $\Gamma(\eta_b \rightarrow P V_L + \rm{c.c.})$
\eqref{eq:MM.6d} using $\phi^{\textrm{Exp}}$ and $\phi^{\textrm{Braun}}$
for both $\phi_P$ and $\phi_{V_L}$.
In the limit
$m_c \to \infty$,
spin symmetry of the bHQET
Lagrangian would imply the equality of the pseudoscalar and vector
distribution amplitudes,
$\phi_{P} = \phi_{V_L}$, and hence
the vanishing of the decay rate $\Gamma(\eta_b \rightarrow P V_L + \rm{c.c.})$.
Assuming spin-symmetry violations,  the decay rate depends on (\textit{i}) the
two parameters  $\bar{\lambda}_D = (\lambda_D + \lambda_{D^*_L})/2$
and $\delta = (\lambda_{D^*_L} - \lambda_{D})/(\lambda_D +
\lambda_{D^*_L}) $, if
$\phi^{\textrm{Exp}}$ is
used, and on (\textit{ii}) three parameters $\bar{\lambda}_D$, $\delta$, and $|\sigma_{D^*_L} -
\sigma_{D}|$, if $\phi^{\textrm{Braun}}$ is used.

\begin{figure}[t]
\includegraphics[width=8cm]{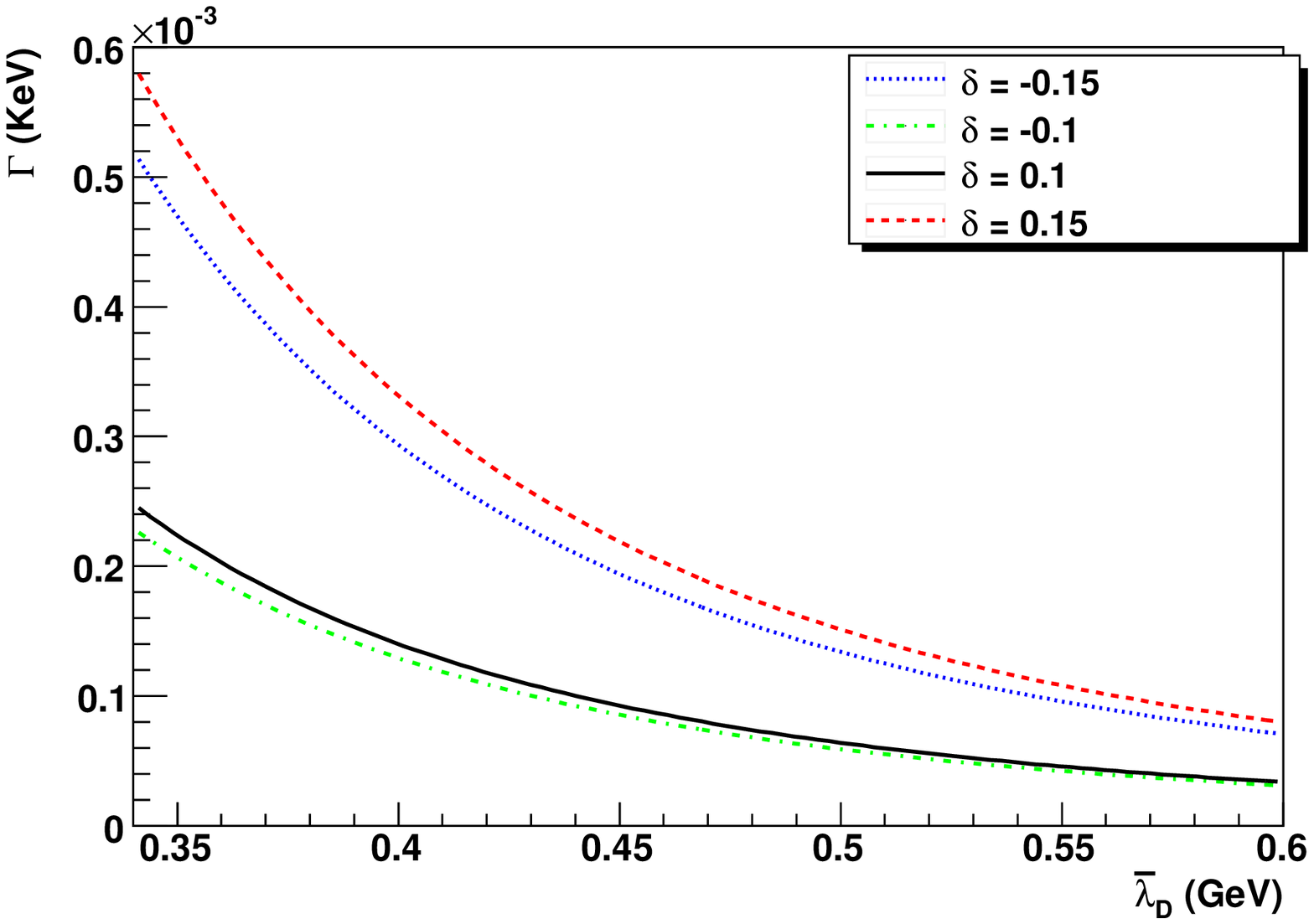}
\includegraphics[width=8cm]{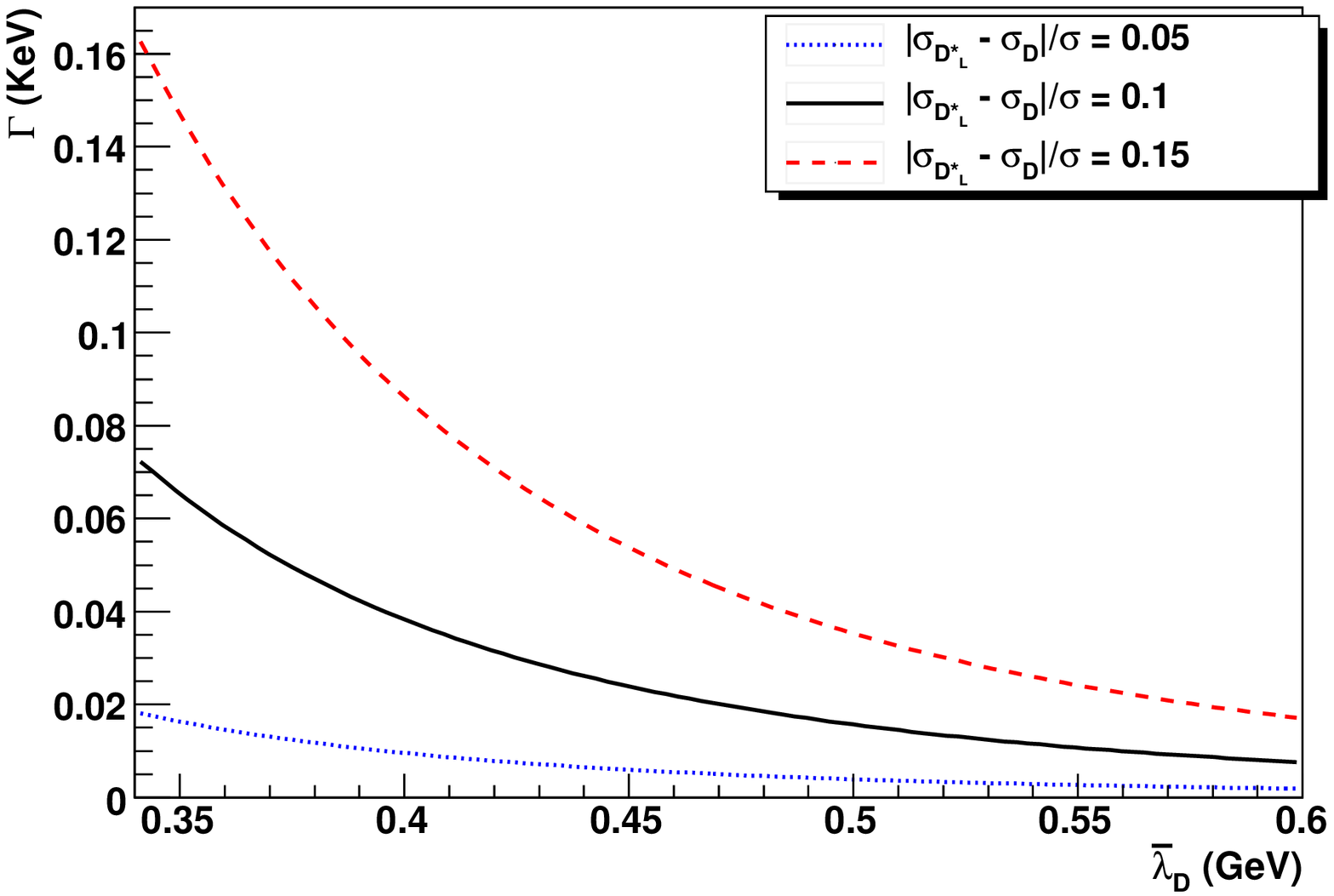}
\includegraphics[width=8cm]{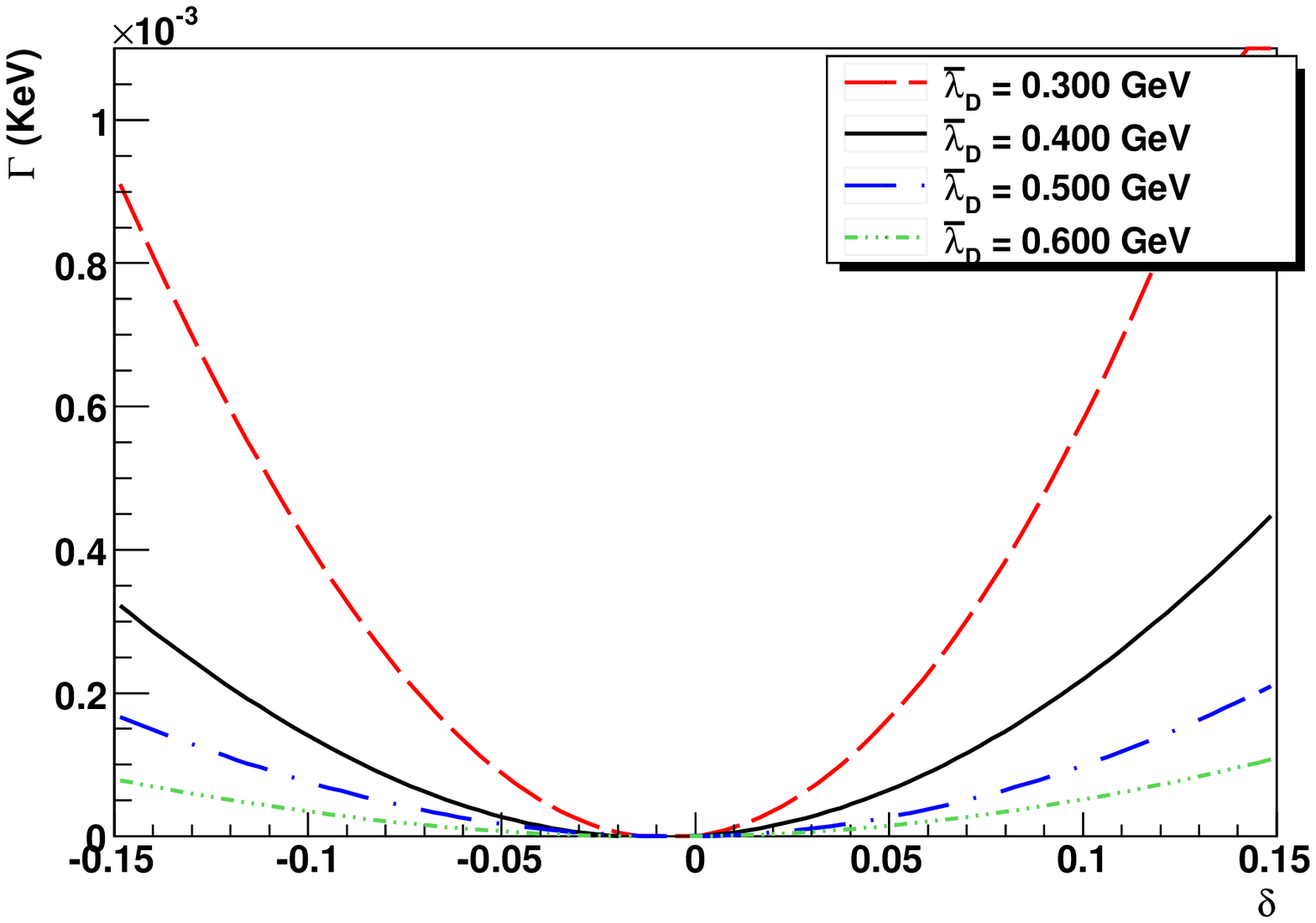}
\includegraphics[width=8cm]{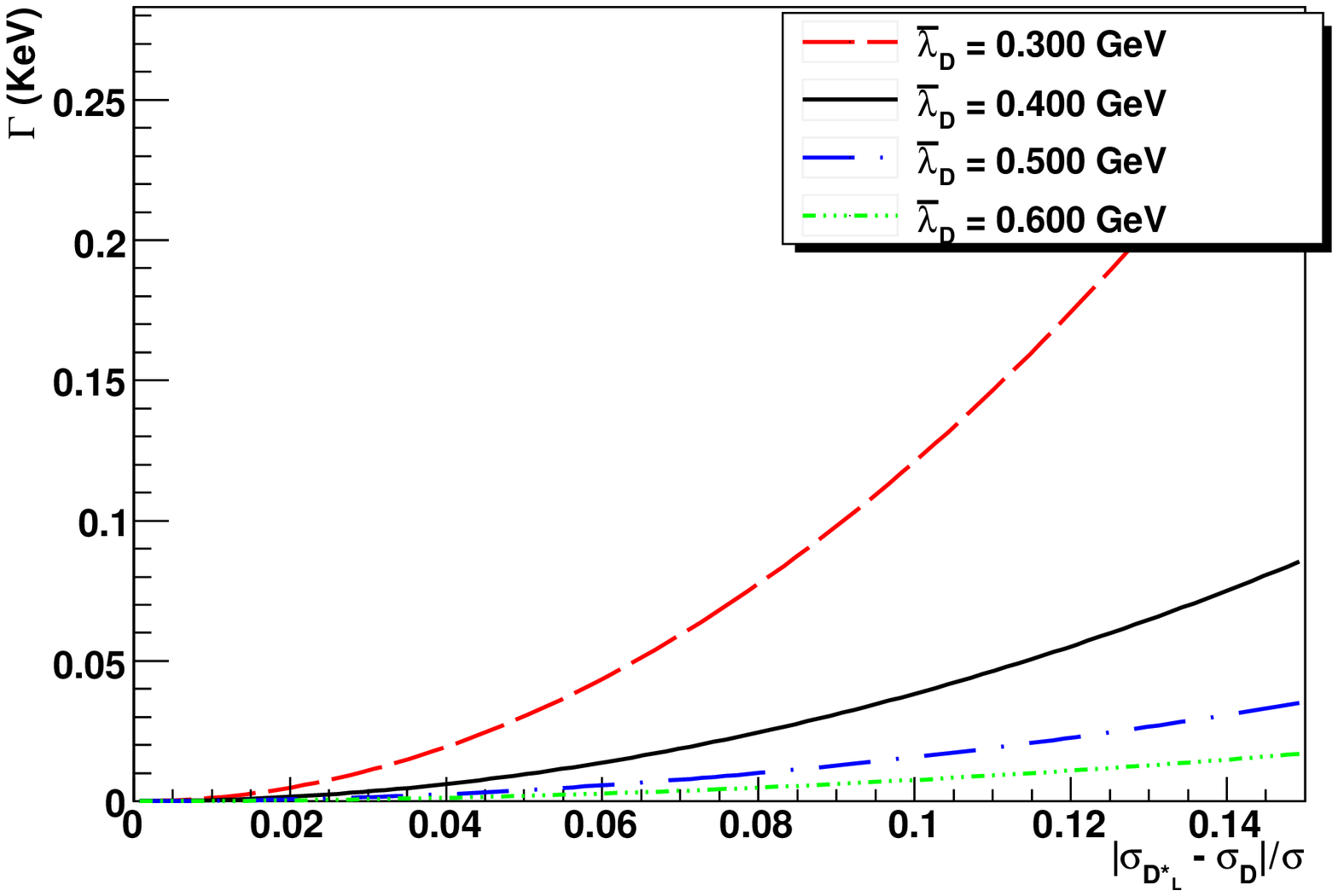}
\caption{\textit{Left}: $\Gamma(\eta_{b}\rightarrow P  V_L + \rm{c.c.})$ as a
function of $\lambda_D$ and $\delta$,
computed using exponential distribution amplitudes $\phi^{\rm{Exp}}_P$ and
$\phi^{\rm{Exp}}_{V_L}$. \textit{Right}: $\Gamma(\eta_{b}\rightarrow
P  V_L + \rm{c.c.})$ as a function of $\lambda_D$
and $|\sigma_{D^*_L} - \sigma_D|/\sigma$,
computed with the Braun
distribution amplitudes $\phi^{\rm{Braun}}_{P}$ and
$\phi^{\rm{Braun}}_{V_L}$.}\label{Fig:eta}
\end{figure}

The two plots in the left column of Fig. \ref{Fig:eta} show the decay
rate, computed with
$\phi^{\rm{Exp}}$, as a function of
$\bar{\lambda}_D$
with $\delta$ adopting various values,
and as a function of $\delta$ with $\bar{\lambda}_D$ now being the
parameter.
In the right column, the decay rate
computed with $\phi^{\rm{Braun}}$
is shown. Since in this case the decay
rate does not strongly depend on $\delta$, we
fix it at $\delta = 0$
and we show the dependence of the decay rate on $\bar{\lambda}_D$
and $|\sigma_{D^*_L} - \sigma_D|$.
We ``normalize'' the difference
between the first logarithmic moments by
dividing them by $\sigma = 2 \sigma_D$.

The most striking feature of Fig. \ref{Fig:eta}
is the huge sensitivity to the chosen functional form. Though a
precise comparison is difficult,
due to the
dependence
on different parameters, the decay rate
increases by two orders of magnitude when we switch from
$\phi^{\rm{Exp}}$ to $\phi^{\rm{Braun}}$. Once again, this effect
hinders our ability to predict $\Gamma(\eta_b \rightarrow P V_L + \rm{c.c.})$ but
it opens up the interesting possibility to discriminate between
different model distribution
amplitudes.

Using Eqs. \eqref{eq:MM.6d} and \eqref{eq:RGE.18as0}, we know that $\Gamma(\eta_b \rightarrow P
V_L + \rm{c.c.})$ goes like $\bar\lambda_D^{-4-4g}$ when
$\phi^{\textrm{Exp}}$ used
or $\bar\lambda_{D}^{-4}$
when $\phi^{\textrm{Braun}}$ used. Fig. \ref{Fig:eta} appears to
confirm this strong dependence on $\bar \lambda_D$.
The plots in the lower half of Fig. \ref{Fig:eta} reflect the fact
that the decay rate vanishes if one
assumes
$\phi_P(\omega) = \phi_{V_L}(\omega)$.

We conclude this section with
the determination of the branching ratios $\mathcal B(\chi_{b0} \rightarrow P P) =
\Gamma(\chi_{b0} \rightarrow P P )/\Gamma(\chi_{b 0} \rightarrow
\textrm{light hadrons})$ 
and $\mathcal B(\eta_b \rightarrow P V_L + \textrm{c.c.}) = \Gamma(\eta_{b}
\rightarrow P V_L + \textrm{c.c.})/\Gamma(\eta_{b} \rightarrow \textrm{light
  hadrons})$.
At leading order in pNRQCD, the only non-perturbative parameter
involved in the inclusive decay width of the $\eta_b$ is $|R_{\eta_b}(0)|^2$ \cite{Bodwin:1994jh},
\begin{equation}\label{eq:etaincl.1}
\Gamma(\eta_{b} \rightarrow
\text{light hadrons}) = \frac{2 \textrm{Im}
  f_1(^1S_0)}{m^2_b} \frac{N_c}{2\pi} |R_{\eta_b}(0)|^2 \; .
\end{equation}
Therefore, $\mathcal B(\eta_b \rightarrow P V_L + \rm{c.c.})$ does not depend
on the quarkonium wavefunction and the only non-perturbative
parameters in $\mathcal B(\eta_b \rightarrow P V_L + \rm{c.c.})$ are those
describing the $D$-meson distribution amplitudes.

For $P$-wave states, the inclusive decay rate was obtained in
Refs. \cite{Bodwin:1994jh} \cite{Bodwin:1992ye}, where the
contributions of the configurations in which the quark-antiquark pair
is in a color-octet $S$-wave state were first recognized.
In pNRQCD the inclusive decay rate is written as \cite{Brambilla:2001xy} \cite{Brambilla:2002nu}
\begin{equation}\label{eq:inclusive.1}
\Gamma(\chi_{b0}\rightarrow
\text{light hadrons}) =  \frac{1}{m_b^4} \frac{3N_c}{\pi}
|R^{\prime}_{\chi_b} (0)|^2 \left[  \textrm{Im} f_{1}(^3P_0) +
  \frac{1}{9 N_c^2} \textrm{Im} f_{8}(^3S_1) \mathcal E\right] \; ,
\end{equation}
where the color-octet matrix element has been expressed in terms of the heavy quarkonium wavefunction and
of the gluonic correlator $\mathcal E$, whose precise definition is
given in Ref. \cite{Brambilla:2001xy}.  
$\mathcal E$ is a universal parameter and is
completely independent of
any particular heavy quarkonium state under consideration.
Its value has been obtained by fitting to existing charmonium data and,
thanks to the universality, the same value can be used to predict
properties of bottomonium
decays.
It is found in Ref. \cite{Brambilla:2001xy} $\mathcal E = 5.3\,^{+3.5}_{-2.2}$.
The matching coefficients in Eqs. \eqref{eq:etaincl.1} and
\eqref{eq:inclusive.1} are 
known to one loop. For the updated value we refer to Ref. \cite{Vairo:2003gh}
and
references therein. For reference, the tree-level values of the
coefficients are as follows \cite{Bodwin:1994jh}:
\begin{equation}\label{eq:inclusive.2}
\textrm{Im} f_{1}(^1S_0) = \alpha^2_s(2m_b) \pi \frac{C_F}{2N_c} \; ,  \quad
 \textrm{Im} f_1(^3P_0) = 3 \alpha^2_s(2m_b) \pi \frac{C_F}{2N_c} \; , \quad 
 \textrm{Im} f_8(^3S_1) = \frac{n_f}{6} \alpha^2_s(2m_b) \pi \; . 
\end{equation}

\begin{figure}
\includegraphics[width=8cm]{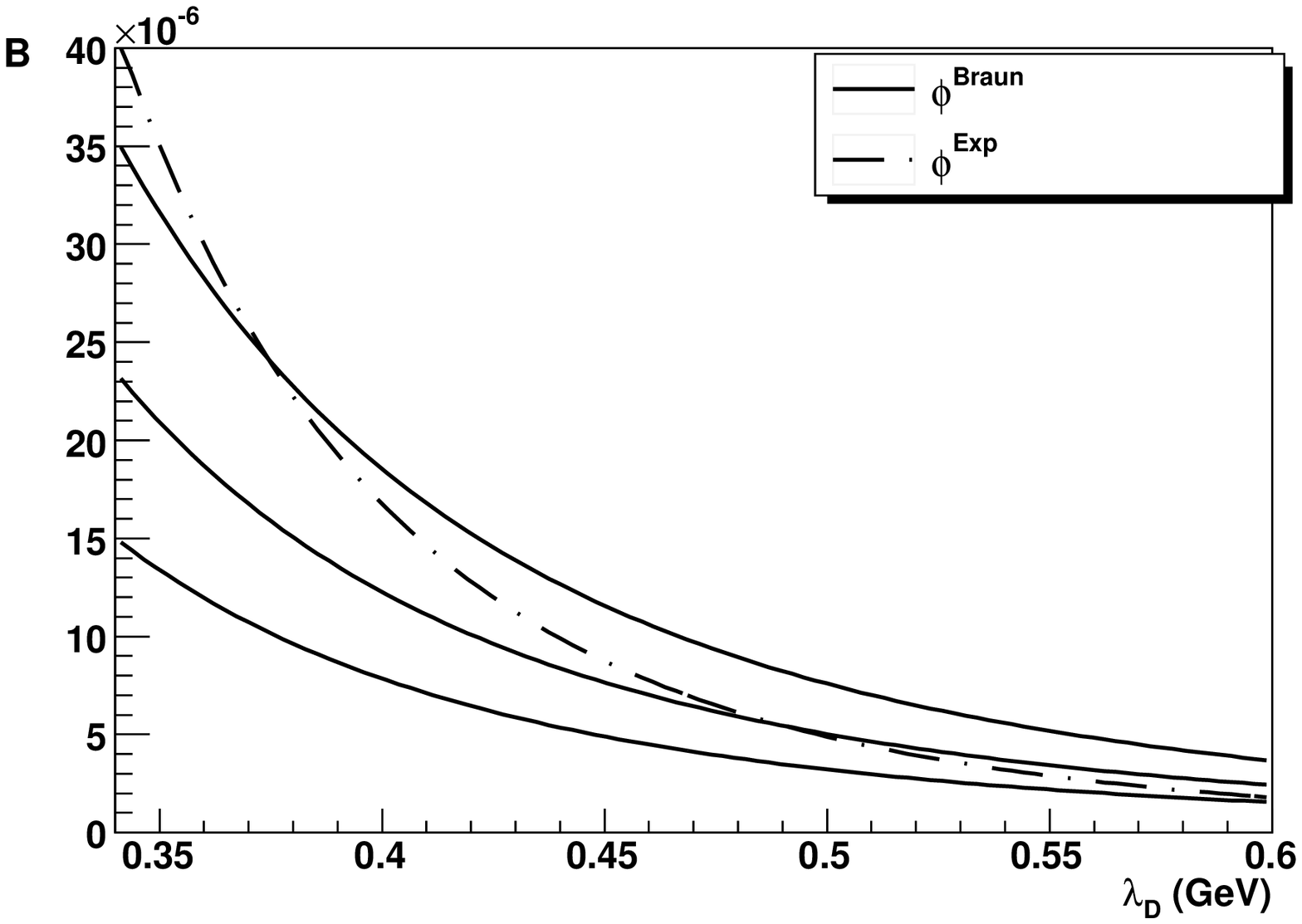}
\includegraphics[width=8cm]{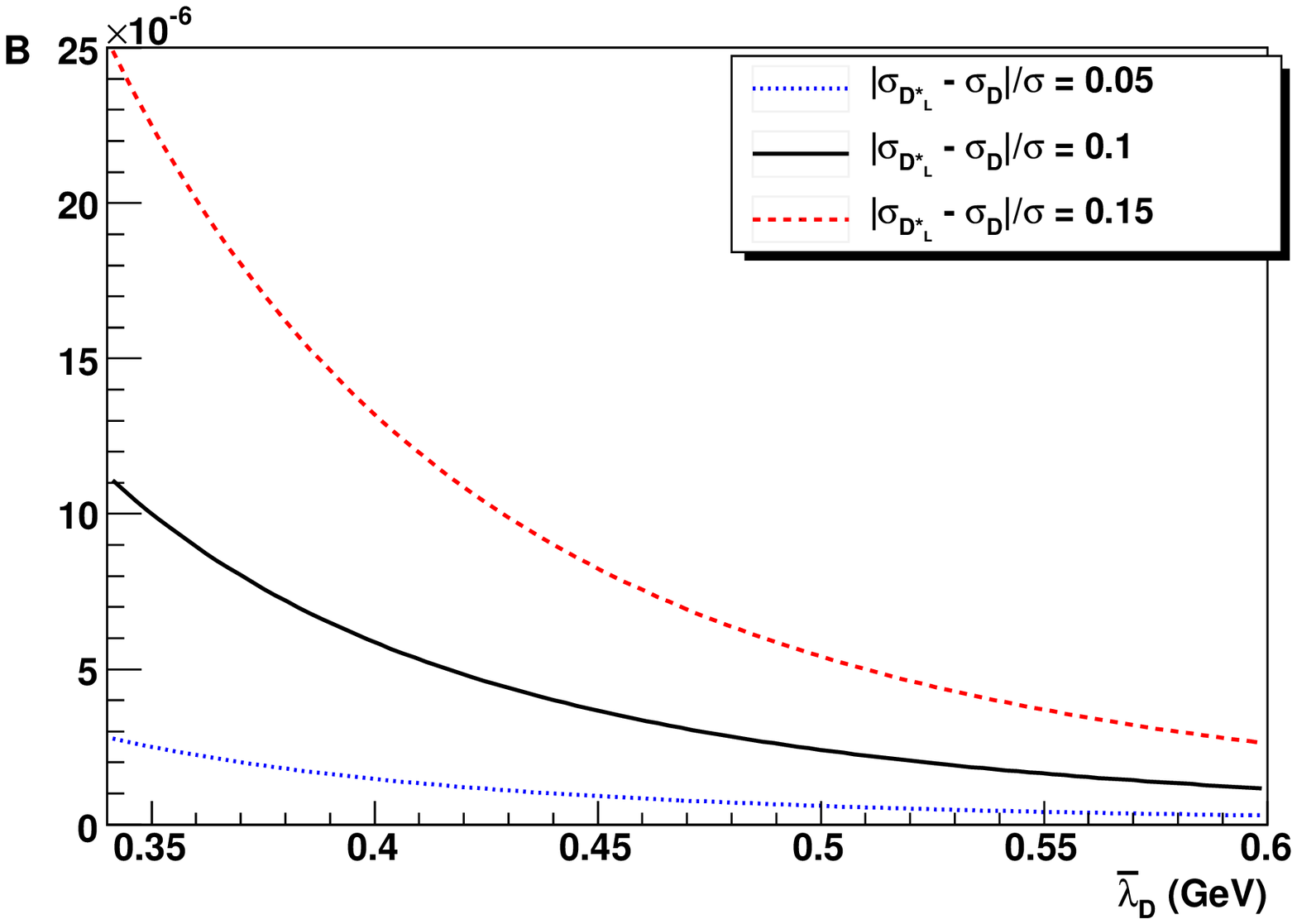}
\caption{Branching ratios $\mathcal B(\chi_{b0} \rightarrow P P)$
  (\textit{left}) and $\mathcal B(\eta_b \rightarrow P V_L + \rm{c.c.})$
  (\textit{right}). The latter
is computed using the distribution amplitude $\phi^{\rm{Braun}}$.}\label{Fig:branching}
\end{figure}

With the above parameters, we plot $\mathcal B(\chi_{b0} \rightarrow P P)$
and $\mathcal B(\eta_{b} \rightarrow P V_L + \rm{c.c.})$ as a function of
$\lambda_D$ and $\bar{\lambda}_D$, respectively, in Fig. \ref{Fig:branching}.
Over the range of $\lambda_D$ we are considering, $\mathcal
B(\chi_{b0} \rightarrow P P)$ varies between $4 \cdot 10^{-5}$ and
$4 \cdot 10^{-6}$;
it is approximately one or two orders of magnitude smaller than
the branching ratios observed in Ref. \cite{:2008sx} for $\chi_{bJ}$
decays into light hadrons. 
$\mathcal B(\eta_{b} \rightarrow P V_L + \rm{c.c.})$  depends on the
choice of the distribution amplitude. Choosing the parameterization
$\phi^{\rm{Braun}}$ \eqref{eq:DR.5}, it appears that, despite the suppression at
$|\sigma_{D^*_L} - \sigma_D| = 0$,
$\mathcal B(\eta_{b} \rightarrow P V_L + \rm{c.c.})$ assumes values comparable to
$\mathcal B(\chi_{b0} \rightarrow P P)$ even for a small deviation
from the spin-symmetry limit. If $\phi^{\rm{Exp}}$ is chosen, the
branching ratio is suppressed over a wide range of $|\sigma_{D^*_L} -
\sigma_{D}|$.
The branching ratio $\mathcal B(\eta_b \rightarrow P V_L + \rm{c.c.})$ was first estimated in \cite{Maltoni:2004hv}. 
The authors of \cite{Maltoni:2004hv} assumed that the exclusive decays into $D D^*$ dominate 
the inclusive decay into charm, $\Gamma(\eta_b \rightarrow P V_L + \textrm{c.c.}) \sim \Gamma(\eta_b \rightarrow c \bar c + X)$. 
With this  assumption, they estimated the branching ratio to be in the range $10^{-3} < \mathcal B(\eta_b \rightarrow P V_L + \rm{c.c.}) < 10^{-2}$.
Our analysis shows that such an assumption does not appear to be justified in the range of $\bar\lambda_{D}$  considered in Fig. \ref{Fig:branching}, 
while it would be appropriate for smaller values of the first inverse moments, for example for $\bar{\lambda}_D \sim 0.200$ GeV 
if the distribution amplitudes are described by $\phi^{\textrm{Braun}}$.

Our estimates indicate that
observing the exclusive processes
  $\eta_{b} \rightarrow D D^{*} + \textrm{c.c.} $ 
and $\chi_{b} \rightarrow D D$ would be extremely challenging. A preliminary analysis for $\chi_{b}
\rightarrow D^0 \bar D^0$ 
\cite{italianconnection}
suggests that the number of $\Upsilon (2S)$ produced at BABAR allows
for the measurement of a branching ratio
$\mathcal B(\chi_{b0} \rightarrow D^0 \bar D^0) \sim 10^{-3}$, which is two or three orders of magnitude bigger than the values in Fig. \ref{Fig:branching}. An even bigger branching ratio would be required for the smaller $\Upsilon(2S)$ sample of CLEO.
However, we stress once again the strong
dependence of the decay rates on the values of the first inverse
moments. In particular, our estimates
rely on the relation $\lambda_D = \lambda_B$,
which is valid in the
limit of $m_b, m_c \rightarrow \infty$; even small corrections to the
heavy flavor symmetry, if they had the effect of shifting the value of
$\lambda_{D}$ towards the range $0.250 - 0.350$ GeV,  could
considerably increase the  branching ratios.

\section{Conclusions}\label{concl}
In this paper we have analyzed the exclusive decays of the $C$-even bottomonia
into a pair of charmed mesons. We approached the problem using a
series of EFTs that
lead to the factorization formulas for the decay rates
(Eqs. \eqref{eq:MM.5} - \eqref{eq:MM.6d}),
valid at leading order in the EFT power counting and at all orders in $\alpha_s$.
We improved the perturbative results by resumming Sudakov logarithms of 
the ratios of the characteristic scales that are germane
to the dynamics of the processes.

The
decay rates \eqref{eq:MM.5} - \eqref{eq:MM.6d} receive both
perturbative and non-perturbative corrections. Perturbative
corrections come from loop corrections to the matching coefficients
$C$ and $T$,
which are
respectively
of order $\alpha_s(2m_b) \sim 0.2$
and $\alpha_s(m_c) \sim 0.3$. The largest non-perturbative
contribution could be as big as $\lQ/m_c$, which would amount
approximately to a $30 \%$ correction. Therefore, corrections to the
leading-order decay rates could be noticeable, as
the strong
dependence of the
decay rates on the renormalization-scale
$\mu^{\prime}_c$ suggests.
However, the EFT approach shown in this paper
allows for a systematic treatment of both perturbative corrections and
power-suppressed operators, so that, if the experimental data require,
it
is possible to extend the present analysis beyond the leading order.    
  
For simplicity, we have focused in this paper on the decays of $C$-even
bottomonia, in which cases the decays proceed via two intermediate
gluons and both the matching coefficients $C$ and $T$
are
non-trivial at tree level. The
same EFT approach can be
applied to the decays of $C$-odd states, in
particular, to the decays $\Upsilon \rightarrow D D$ and $\Upsilon
\rightarrow D^* D^*$, with the complication that the matching
coefficient $T$ arises only at one-loop level.
Moreover, the same EFT formalism  developed in this paper
can be applied to the study of the channels that have
vanishing decay rates at leading order in the power
counting, such as $\eta_b \rightarrow D^* D^*$, $\Upsilon \rightarrow
D D^* + \textrm{c.c.}$, and $\chi_{b2} \rightarrow D D^* + \textrm{c.c.}$.  
Experimental data for the charmonium system show that, for the decays
of charmonium into light hadrons, the expected suppression of the
subleading twist processes is not seen. It
is interesting to see whether such an effect appears in bottomonium
decays into two charmed mesons,
using
the EFT approach of this paper
to evaluate the power-suppressed decay rates.

Finally, in Sec.~\ref{rate} we used model distribution amplitudes to
estimate the decay rates. The most evident, qualitative feature of the
decay rates is the strong dependence
on the parameters of the $D$-meson distribution amplitude.
Even though this feature
may prevent us from giving reliable estimates of the
decay rates or of the branching ratios, it makes the channels analyzed
here ideal candidates for the extraction of important $D$-meson
parameters, when the branching ratios can be observed with
sufficient accuracy.

\acknowledgments
We would like to thank S. Fleming for proposing this problem and for
countless useful discussions, N. Brambilla and A. Vairo for
suggestions and  comments and R. Briere, V. M. Braun and S. Stracka
for helpful communications. BwL is grateful for hospitality to the
University of Arizona, where part of this work was finished.
This research was supported by the US Department of Energy
under grants DE-FG02-06ER41449 (RA and EM) and DE-FG02-04ER41338 (RA, BwL and EM).

\appendix
\section{Solution of the running equation in pNRQCD + bHQET}\label{AppA}
The RGE in Eq. \eqref{eq:RGE.2} can be solved by applying the methods
discussed in Ref. \cite{Lange:2003ff}  to find the evolution of the
$B$-meson distribution amplitude. We generalize this approach to the
specific case discussed here, where two distribution amplitudes are
present. 
Following Ref. \cite{Lange:2003ff}, we define
\begin{equation*}
 \omega \Gamma(\omega, \omega^{\prime}, \alpha_s) = - \frac{\alpha_s
   C_F}{\pi} \left[\theta\left({\omega}-{\omega}^{\prime}\right)
   \left(\frac{1}{{\omega}-{\omega}^{\,\prime}}\right)_{+}+
   \theta\left({\omega}^{\prime} -
     {\omega}\right)\theta\left({\omega}\right)
   \frac{{\omega}}{{\omega}^{\prime}}\left(\frac{1}{{\omega}^{\prime}-{\omega}}\right)_{+}\right]
 \; .
\end{equation*}
Lange and Neubert \cite{Lange:2003ff} prove that
\begin{equation}\label{eq:RGE.4}
 \int d\omega^{\prime}  \omega \Gamma(\omega, \omega^{\prime},
 \alpha_s) (\omega^{\prime})^a = \omega^a \mathcal F(a,\alpha_s) \; ,
\end{equation}
with 
\begin{equation*}
      \mathcal F(a,\alpha_s) = \frac{\alpha_s C_F}{\pi}
      \left[\psi(1+a) + \psi(1-a) + 2 \gamma_E\right] \; . 
\end{equation*}
$\psi$ is the digamma function and $\gamma_E$ the Euler
constant. Eq. \eqref{eq:RGE.4} is valid if $-1 <\textrm{Re}\, a < 1$.
Exploiting \eqref{eq:RGE.4}, a solution of the running equation
Eq. \eqref{eq:RGE.2} with initial condition $T(\omega, \bar{\omega},
\mu_0^{\prime}) = \left(\omega/\mu_0^{\prime}\right)^{\eta}
\left(\bar{\omega}/\mu_0^{\prime} \right)^{\xi}$ at a certain scale
$\mu_0^{\prime}$ is
\begin{equation}\label{eq:RGE.5}
F^2(\mu^{\prime}) T(\omega, \bar\omega, \mu^{\prime})  = F^2(\mu^{\prime}_0) f(\omega, \mu^{\prime},\mu_0^{\prime},
\eta) f(\bar{\omega}, \mu^{\prime}, \mu_0^{\prime}, \xi) \; ,
\end{equation}
with 
\begin{equation}\label{eq:RGE.6}
\begin{split}
 & f(\omega, \mu^{\prime}, \mu_0^{\prime}, \eta) = \left(\frac{\omega}{\mu_0^{\prime}}
 \right)^{\eta - g} \left(\bar n \cdot v \right)^g \exp
 U(\mu_0^{\prime},\mu^{\prime}, \eta) \; ,\\
& g \equiv g(\mu_0^{\prime}, \mu^{\prime}) =  \int_{\alpha_s(\mu_0^{\prime})}^{\alpha_s(\mu^{\prime})}
\frac{d\alpha}{\beta(\alpha)} \Gamma_{\textrm{cusp}}(\alpha)  \; , \\
& U(\mu_0^{\prime}, \mu^{\prime}, \eta) = \int_{\alpha_s(\mu_0^{\prime})}^{\alpha_s(\mu^{\prime})}
\frac{d\alpha}{\beta(\alpha)} \left[\Gamma_{\textrm{cusp}}(\alpha)
  \int_{\alpha_{s}(\mu_0^{\prime})}^{\alpha}
  \frac{d\alpha^{\prime}}{\beta(\alpha^{\prime})} + \gamma_1(\alpha) +
  \mathcal F(\eta - g,\alpha)\right] \; , \\
& \gamma_1(\alpha_s) = -2 \frac{\alpha_s C_F}{4\pi} \; .
 \end{split}
\end{equation}
The function $f(\bar \omega, \mu^{\prime}, \mu_0^{\prime}, \xi)$ has the same form as
$f(\omega, \mu^{\prime}, \mu_0^{\prime}, \eta)$ and
is obtained by replacing $\omega
\rightarrow \bar \omega$, $\eta \rightarrow \xi$, and $\bar n \cdot v
\rightarrow n \cdot v^{\prime}$ in Eq. \eqref{eq:RGE.6}.
The integrals over $\alpha$ can be performed explicitly using the beta
function in Eq. \eqref{eq:softren.11}.
The result is 
\begin{equation}\label{eq:RGE.7}
\begin{split}
  f(\omega, \mu^{\prime},\mu_0^{\prime}, \eta) f(\bar{\omega}, \mu^{\prime}, \mu_0^{\prime}, \xi)  = &
\left(\frac{\omega}{\mu_0^{\prime}}\right)^{\eta - g} \left(\frac{\bar{\omega}}{\mu_0^{\prime}} \right)^{\xi - g} \left(\bar n \cdot v \,n \cdot v^{\prime} \right)^g \exp\left[ V(\mu_0^{\prime},\mu^{\prime})\right] 
\\ & \frac{\Gamma(1- \eta + g) \Gamma(1+ \eta)}
{\Gamma(1+ \eta - g) \Gamma(1 - \eta)} 
\frac{\Gamma(1- \xi + g) \Gamma(1+ \xi)}
{\Gamma(1+ \xi - g) \Gamma(1 - \xi)} \; ,
\end{split}
\end{equation}
Where, at NLL,\begin{equation}\label{eq:RGE.7aa}
 g(\mu_0^{\prime},\mu^{\prime}) = - \frac{\Gamma^{(0)}_{\rm{cusp}}}{2\beta_0} \left\{ \ln r + \left(\frac{\Gamma^{(1)}_{\rm{cusp}}}{\Gamma_{\rm{cusp}}^{(0)}} -
\frac{\beta_1}{\beta_0} \right) \frac{\alpha_{\rm{s}}(\mu_0^{\prime})}{4\pi}
(r-1) \right\} \; ,
\end{equation}
and
\begin{equation}\label{eq:RGE.7a}
\begin{split}
  V(\mu_0^{\prime}, \mu^{\prime})  = & - \Gamma^{(0)}_{\textrm{cusp}} \frac{2
    \pi}{\beta_0^2} \left\{ \frac{r - 1 - r \ln r}{\alpha_s (\mu^{\prime})}
+
\left(\frac{\Gamma^{(1)}_{\textrm{cusp}}}{\Gamma^{(0)}_{\textrm{cusp}}}
  - \frac{\beta_1}{\beta_0}\right) \frac{1 - r + \ln r}{4\pi}   +
\frac{\beta_1}{8 \pi \beta_0} \ln^2 r 
\right\} \\ & + \frac{C_F}{\beta_0} \left(2  - 8 \gamma_E  \right) \ln
r \; ,
\end{split}
\end{equation}
with $r = \alpha_s(\mu^{\prime})/\alpha_s(\mu_0^{\prime})$.  Notice that in the
running from $\mu_0^{\prime} = m_c$ to $ \mu^{\prime} = 1$ GeV only three flavors are active, so in
the expressions for $\beta_0$, $\beta_1$, and
$\Gamma^{(1)}_{\textrm{cusp}}$ we  use $n_f = 3$.

Eq. \eqref{eq:RGE.7} is the solution for
the initial condition
$T(\omega, \bar{\omega}, \mu_0^{\prime}) =
\left(\omega/\mu_0^{\prime}\right)^{\eta}
\left(\bar{\omega}/\mu_0^{\prime} \right)^{\xi}$. To solve the RGE for
a generic initial condition,
we express $T$ as the Fourier transform with respect to $\ln
\omega/\mu_0^{\prime}$,
\begin{equation*}
\begin{split} T(\omega, \bar{\omega}, \mu_0^{\prime}) & = \frac{1}{(2 \pi)^2} \int_{-\infty}^{+\infty} dr ds \exp\left(- i r \ln\frac{\omega}{\mu_0^{\prime}}\right)
\exp\left(- i s \ln\frac{\bar\omega}{\mu_0^{\prime}}\right) F[T](r,s,\mu_0^{\prime}) \\ & =
\frac{1}{(2 \pi)^2} \int_{-\infty}^{+\infty} dr ds \left(\frac{\omega}{\mu_0^{\prime}}\right)^{-i r}
\left(\frac{\bar\omega}{\mu_0^{\prime}}\right)^{-i s} F[T](r,s,\mu_0^{\prime}) \; , 
\end{split}
\end{equation*}
where $F[T]$ denotes the Fourier transform of $T$.
From the solution \eqref{eq:RGE.5}-\eqref{eq:RGE.7} it follows that
\begin{equation}\label{eq:RGE.8}
\begin{split} 
F^2(\mu^{\prime})T(\omega, \bar{\omega}, \mu^{\prime})  = & 
\frac{F^2(\mu^{\prime}_0)}{(2 \pi)^2} \int_{-\infty}^{+\infty} dr ds \left(\frac{\omega}{\mu_0^{\prime}}\right)^{-i r - g}
\left(\frac{\bar\omega}{\mu_0^{\prime}}\right)^{-i s- g} \left( \bar n \cdot v\, n \cdot v^{\prime}\right)^g  F[T](r,s, \mu_0^{\prime})  \\
& \exp\left[ V(\mu_0^{\prime}, \mu^{\prime})\right] \frac{\Gamma(1 + ir  + g) \Gamma(1 - ir )}
{\Gamma(1 - ir - g) \Gamma(1 + i r)} 
\frac{\Gamma(1+ is + g) \Gamma(1- is)}
{\Gamma(1- is - g) \Gamma(1 + is)} \; .
\end{split}
\end{equation}

The Fourier transform of the matching coefficient in Eq. \eqref{eq:RGE.8}
has to be understood in the sense of distributions
\cite{Distribution}.
That is, we define the Fourier transform of $T$  as
the function of $r$ and $s$ that satisfies
\begin{equation}\label{eq:RGE.10}
\frac{1}{(2\pi)^2} \int dr ds\, F[T](r,s,\mu^{\prime}) \varphi_A(r,\mu^{\prime})
\varphi_B(s, \mu^{\prime}) = \int_0^{+ \infty} \frac{d\omega}{\omega}
\frac{d\bar\omega}{\bar\omega} T(\omega, \bar{\omega}, \mu^{\prime})
\phi_A(\omega, \mu^{\prime}) \phi_B(\bar{\omega}, \mu^{\prime}) \; ,
\end{equation}
or, more precisely, $F[T](r,s,\mu^{\prime})$ is
the linear functional that acts on the test functions $\varphi_A(r)$ and $\varphi_B(s)$ according to
\begin{equation}\label{eq:RGE.10a}
\frac{1}{(2\pi)^2}  ( F[T](r,s,\mu^{\prime}),\, \varphi_A(r,\mu^{\prime}) \varphi_B(s,
\mu^{\prime}) ) = \int_0^{+ \infty} \frac{d\omega}{\omega}
\frac{d\bar\omega}{\bar\omega} T(\omega, \bar{\omega}, \mu^{\prime})
\phi_A(\omega, \mu^{\prime}) \phi_B(\bar{\omega}, \mu^{\prime}) \; .
\end{equation}
The function $\varphi_A$ is the Fourier transform of the $D$-meson distribution amplitude,
\begin{equation}\label{eq:RGE.11}
\varphi_A(r, \mu^{\prime}) = \int_0^{\infty} \frac{d\omega}{\omega}
\left(\frac{\omega}{\mu^{\prime}}\right)^{i r} \phi_A(\omega, \mu^{\prime}) \; ,
\end{equation}
where the integral on the r.h.s.
should converge in the
ordinary sense
because of the regularity properties of the $D$-meson
distribution amplitude. As in Sec. \ref{bHQET}, the subscript $A$
denotes the spin and polarization of the $D$ meson.

In the distribution sense, the Fourier transform of the coefficient $1/(\omega + \bar\omega)$ is
\begin{equation}\label{eq:RGE.12}	
\begin{split}
F\left[\frac{1}{\omega + \bar\omega}\right](r,s, \mu_0^{\prime}) &= (2\pi)^2
\frac{1}{2\mu_0^{\prime}} \delta( r + s + i)\, \textrm{sech}\left[\frac{\pi}{2}
  (r-s)\right] \\& =
\frac{1}{2} (2\pi)^2 \frac{1}{2\mu_0^{\prime}} \delta( R + i)\,
\textrm{sech}\left[\frac{\pi}{2} S\right] \; ,
\end{split}
\end{equation}
where
$R = r + s$,
$S = r - s$, and the factor $\frac{1}{2}$ comes from the Jacobian of the change of
variables. The hyperbolic secant is defined as  $\rm{sech} =
1/\cosh$.
Similarly, we find
\begin{equation}\label{eq:RGE.12a}	
\begin{split}
F\left[\frac{\omega - \bar\omega}{\omega + \bar\omega}\right](R,S, \mu_0^{\prime}) = 
\frac{i}{2} (2\pi)^2  \delta( R)\,\left(
  \textrm{cosech}\left[\frac{\pi}{2} S + i\varepsilon\right] +
  \textrm{cosech}\left[\frac{\pi}{2} S - i\varepsilon\right] \right)
\; .
\end{split}
\end{equation}

The $\delta$ function in Eq. \eqref{eq:RGE.12} has complex
argument. The definition is analogous to the one in real space
\cite{Distribution},
\begin{equation}\label{eq:RGE.13}
(\delta(R+i), \varphi(R)) = \varphi(-i) \; . 
\end{equation}
Using
Eqs. \eqref{eq:RGE.12} and \eqref{eq:RGE.12a}, we can perform the
integral in Eq. \eqref{eq:RGE.8}, obtaining respectively $T(\omega, \bar\omega, \mu, \mu^{\prime}; \, ^3P_J)$ and $T(\omega, \bar\omega,\mu, \mu^{\prime}; \, ^1S_0)$.  
In order to give an explicit example, we proceed using Eq. \eqref{eq:RGE.12}.
Integrating the $\delta$ function we are left with
\begin{equation}\label{eq:RGE.16}
\begin{split} 
& F^{2}(\mu^{\prime}) T(\omega, \bar{\omega};\mu^{\prime})  =  
 F^{2}(\mu_0^{\prime})\exp \left[V(\mu_0^{\prime}, \mu^{\prime})\right] \frac{1}{\mu_0^{\prime}} \left(\frac{\mu_0^{\prime \, 2}}{\omega \bar{\omega}}\right)^{1/2+g} \left( \bar n \cdot v \, n \cdot v^{\prime}\right)^g \\ & \int_{-\infty}^{\infty} d S 
\exp\left[- i \frac{S}{2} \ln \frac{\omega}{\bar\omega} \right] 
\textrm{sech}\left[\frac{\pi}{2} S \right]
\frac{1}{1+ S^2} 
 \frac{\Gamma\left(\frac{3}{2} + g + \frac{i}{2} S\right)}{\Gamma\left(\frac{1}{2} - g - \frac{i}{2} S\right) }
 \frac{\Gamma\left(\frac{3}{2} + g - \frac{i}{2}
     S\right)}{\Gamma\left(\frac{1}{2} - g + \frac{i}{2} S\right) } \;
 .
\end{split}
\end{equation}

The integral \eqref{eq:RGE.16} can be done by contour.
The integrand has poles along the imaginary axis. In $S = \pm i$ there is a double pole, coming from the coincidence of one pole of the hyperbolic secant and the singularities in $1/(1+S^2)$. The $\Gamma$ functions in the numerator have poles respectively in $S = \pm i \left(2 n + 3 + 2g \right)$ with $n > 0$, while the other poles of sech are in $S = \pm i (2 n +1)$, with $n \ge 1$.
We close the contour in the  upper half plane for $\bar{\omega} >
\omega$ and
in the lower half plan
for $\omega > \bar{\omega}$, obtaining
\begin{equation}\label{eq:RGE.17}
\begin{split}
& F^2(\mu^{\prime}) T(\omega,\bar\omega,\mu^{\prime}) =  F^2(\mu_0^{\prime})  \exp \left[ V(\mu^{\prime}_0, \mu^{\prime}) \right] \,\theta(\bar\omega - \omega) \frac{1}{\bar\omega} \left(\frac{\mu_0^{\prime \, 2} \bar n \cdot v\, n \cdot v^{\prime}}{\omega \bar\omega}  \right)^g\\
& \qquad \left\{ 
\frac{\Gamma(1+g) \Gamma(2+g)}{\Gamma(1-g) \Gamma(-g)}\left[1 - \ln\frac{\omega}{\bar\omega} + \psi(1-g) - \psi(-g) + \psi(1+g) -\psi(2+g)\right]
\right. \\ & \left. 
+ \sum_{n = 1}^{\infty} (-)^{n+1} \left(\frac{\omega}{\bar\omega}\right)^{n}\frac{1}{n(n+1)} \frac{\Gamma(1-n+g) \Gamma(2+n+g)}{\Gamma(-n-g)\Gamma(1-g+n)}\right. \\ & \left. 
- \sum_{n = 1}^{\infty} \left(\frac{\omega}{\bar\omega}\right)^{n+g} \frac{\pi}{(n-1)!} \textrm{csc}(g\pi) \frac{1}{(n+g)(1+n+g)} \frac{\Gamma(2+n + 2g)}{\Gamma(1+n) \Gamma(-n-2g)}
\right\} 
 + \left( \omega \rightarrow \bar\omega \right) \; ,
\end{split}
\end{equation}
with csc$(g \pi) = 1/\sin(g\pi)$ and $\psi$ is the digamma function. 
More compactly, we can express Eq. \eqref{eq:RGE.17} using the hypergeometric functions $_4F_3$ and $_3F_2$, 
\begin{equation}\label{eq:RGE.18a}
 \begin{split}
& F^2(\mu^{\prime})  \,  T(\omega,\bar\omega,\mu, \mu^{\prime}; \,^3P_J) = F^2(\mu^{\prime}_c)   \frac{C_F}{N_c^2} \frac{4 \pi \alpha_s(\mu^{\prime}_c)}{m_b} \exp\left[ V(\mu^{\prime}_c, \mu^{\prime}) \right] \, \left(\frac{\mu^{\prime\, 2}_c \bar n \cdot v  n \cdot v^{\prime } }{\omega  \,\bar\omega} \right)^g  \\
& \frac{\theta(\bar\omega - \omega)}{\bar\omega} \left\{ 
\frac{\Gamma(1+g) \Gamma(2+g)}{\Gamma(1-g) \Gamma(-g)}\left[1 - \ln\frac{\omega}{\bar\omega} + \psi(1-g) - \psi(-g) + \psi(1+g) -\psi(2+g)\right]
\right. \\ & \left. 
+ \frac{1}{2}\frac{\omega}{\bar\omega} \,\frac{\Gamma(g+2) \Gamma(g+3)}{\Gamma(1-g)\Gamma(2-g)} \,_4F_3\left(1,1,g+2,g+3; 3, 1-g, 2-g; -\frac{\omega}{\bar\omega}\right)\right. \\ & \left. 
- \left(\frac{\omega}{\bar\omega}  \right)^{1+g} 4 \cos(g\pi) \frac{\Gamma(2 + 2g)^2}{g+2} \,_3F_2\left(g+1,2g+2,2g+3;2,g+3;-\frac{\omega}{\bar\omega}\right)
\right\}
 + \left( \omega \rightarrow \bar\omega \right) \; ,  
\end{split}
\end{equation}
where we have introduced the constants that appear in the initial condition in Eq. \eqref{eq:MM.3}. 
In the same way, we obtain
\begin{equation}\label{eq:RGE.18as0}
 \begin{split}
 F^2(\mu^{\prime})\, & T(\omega,\bar\omega,\mu, \mu^{\prime};\,^1S_0) = F^2(\mu^{\prime}_c)   \frac{C_F}{2 N_c^2} \frac{4 \pi \alpha_s(\mu^{\prime}_c)}{m_b} \exp \left[V(\mu^{\prime}_c, \mu^{\prime})\right] \,\theta(\bar\omega - \omega)  \left(\frac{\mu^{\prime\, 2}_c \bar n \cdot v  n \cdot v^{\prime } }{\omega  \,\bar\omega} \right)^g  \\
& \left\{ 
2 \frac{\Gamma(1+g) \Gamma(2+g)}{\Gamma(1-g) \Gamma(2-g)}\frac{\omega}{\bar\omega} \, _3F_2\left(1,g+1,g+2; 1-g,2-g;-\frac{\omega}{\bar \omega}\right)
+ \frac{\Gamma^2(1+g)}{\Gamma^2(1-g)}\right. \\ & \left. 
 - \left(\frac{\omega}{\bar\omega}  \right)^{1+g} 4 \cos(g\pi)
 \Gamma(1 + 2g)\Gamma(2g+2) \,_2F_1\left(2g+2,2g+1;2;- \frac{\omega}{\bar\omega}\right)
\right\} 
 - \left( \omega \rightarrow \bar\omega \right) \; .  
\end{split}
\end{equation}
In Eqs. \eqref{eq:RGE.18a} and \eqref{eq:RGE.18as0} we renamed the initial scale $\mu^{\prime}_0 = \mu^{\prime}_c$ to denote its connection to the scale $m_c$.
Setting $\mu^{\prime} = \mu^{\prime}_c$ or, equivalently, $g=0$, it can be explicitly verified that the solutions Eqs.
 \eqref{eq:RGE.18a} and \eqref{eq:RGE.18as0} satisfy the initial conditions Eq. \eqref{eq:MM.3}.

\section{Boost transformation of the $D$-meson distribution amplitude}\label{AppC}
We derive in this Appendix the relation between the distribution
amplitudes in the $D$-meson and in the bottomonium rest
frames, as given in Eq. \eqref{eq:DR.5.b}.
In the $D$-meson rest frame, characterized by the velocity label $v_0
=(1,0,0,0)$, the local heavy-light matrix element is defined as
\begin{equation}\label{eq:boost.1}
\langle 0 |\bar\xi^{\, \bar l}_n(0) \frac{\slashchar{\bar n}}{2}
\gamma_5\,   h^{c}_{n}(0) | D \rangle_{v_0} = - i F(\mu^{\prime}) \frac{\bar n \cdot
  v_0}{2} \; .
\end{equation}
The matrix element of the heavy- and light-quark fields
at a light-like separation $z^{\mu}_0 = n \cdot z_0 \, \bar n^{\mu}/2$
defines the light-cone distribution $\tilde{\phi}_0(n \cdot z_0,\mu^{\prime})$
in coordinate space: 
\begin{equation}\label{eq:boost.2}
 \langle 0 | \bar\chi^{\, \bar l}_{n}(n \cdot z_0) \,
 \frac{\slashchar{\bar n}}{2} \gamma_5 \, {\mathcal H}^{c}_{n}(0) | D
 \rangle_{v_0} = - i F(\mu^{\prime}) \frac{\bar n \cdot v_0}{2} \, \tilde
 \phi_{0}(n \cdot z_0,\mu^{\prime}) \; .
\end{equation}
Eqs. \eqref{eq:boost.1} and \eqref{eq:boost.2} imply $\tilde \phi_0(0, \mu^{\prime})
= 1$. In the definitions \eqref{eq:boost.1} and \eqref{eq:boost.2}
the subscript $0$ is used to denote quantities in the $D$-meson rest
frame. This convention is used in the rest of this Appendix.
In the bottomonium rest frame, where the velocity label in light-cone
coordinates is $v = (n\cdot v,\bar n \cdot v,
0)$ and the light-like separation is $z^{\mu} = n \cdot z \, \bar n^{\mu}/2$, we define  
\begin{equation}\label{eq:boost.3}
\langle 0 |  \bar\xi^{\, \bar l}_{n}(0) \frac{\slashchar{\bar n}}{2} \gamma_5 \,  h^{c}_{n}(0) | D \rangle_{v}
= - i F(\mu^{\prime}) \frac{\bar n \cdot v}{2}  
\end{equation}
and
\begin{equation}\label{eq:boost.4}
 \langle 0 |\bar\chi^{\, \bar l}_{n}( n \cdot z) \,
 \frac{\slashchar{\bar n}}{2} \gamma_5 \, { \mathcal H}^{c}_{ n}(0) |
 D \rangle_{v} = - i F(\mu^{\prime})  \frac{\bar n \cdot v}{2} \, \tilde \phi(
 n \cdot z,\mu^{\prime}) \; .
\end{equation}

Suppose that $\Lambda$ is some standardized boost that takes the $D$ meson from
$v$, its velocity in the bottomonium rest frame, to rest.
It is straightforward to find the relations between the $D$-meson
momenta in the two frames:
\begin{equation*}
\begin{split}
n \cdot p_0  = \bar n \cdot v^{  } n \cdot p \quad \text{and} \quad
\bar n \cdot p_0 = n \cdot v^{ } \bar n \cdot p \; .
\end{split}
\end{equation*}
There is a similar relation for the light-cone coordinates,
$$
n \cdot z_0 =\bar n \cdot v \, n \cdot z \; .
$$
With $U(\Lambda)$, the unitary operator that implements the boost
$\Lambda$, one can write
$$
U(\Lambda) | D \rangle_{v} = | D^{} \rangle_{v_0} \; .
$$
We choose $\Lambda$ such that, for the Dirac fields, 
\begin{equation*}
  U^{}(\Lambda) \xi^{\, \bar l}_{ n}(x) \, U^{-1}(\Lambda) =
  \Lambda_{{1}/{2}}^{-1}\, \xi^{\, \bar l} (\Lambda x) \quad
  \text{and} \quad
  U^{}(\Lambda) h^{c}_{n}(x) \, U^{-1}(\Lambda) =
  \Lambda_{{1}/{2}}^{-1}\, h^{c} (\Lambda x) \; ,
\end{equation*}
where
\begin{equation*}
 \Lambda_{{1}/{2}} = \cosh \frac{\alpha}{2} + \frac{{\slashchar{\bar
       n}} \slashchar n - \slashchar n {\slashchar {\bar n}}}{4} \sinh
 \frac{\alpha}{2} \; ,
\end{equation*}
with $\alpha$ related to $v$
by $ e^{\alpha} = \bar n \cdot v$
and $e^{-\alpha} = n \cdot v$.

Now we can write the matrix element in Eq. \eqref{eq:boost.3} as
\begin{equation}\label{eq:boost.5}
\begin{split}
\langle 0 | \, \bar\xi^{\, \bar l}_{ n} &\frac{\slashchar{\bar n}}{2} \gamma_5  h^{c}_{n}(0)  | D \rangle_{v}
\\ & =\, \langle 0 |U^{-1}(\Lambda)\left( U^{}(\Lambda)  \bar\xi^{\,
    \bar l}_n(0) U^{-1}(\Lambda) \right) \frac{\slashchar{\bar n}}{2}
\gamma_5 
\left( U^{}(\Lambda)  h^{c}_{n}(0) U^{-1}(\Lambda)\right) U^{}(\Lambda)| D \rangle_{v} 
\\ &
=\, \langle 0 | \bar \xi^{\, \bar l}(0) \Lambda_{1/2}  \frac{\slashchar{\bar n}}{2} \gamma_5 
 \Lambda_{1/2}^{-1}  h^{c}_{}(0) | D \rangle_{v_0}   = \bar n \cdot
 v\, \langle 0 | \bar\xi^{\, \bar l}(0)   \frac{\slashchar{\bar n}}{2}
 \gamma_5
  h^{ c}_{}(0)  | D \rangle_{v_0}  \\ & =  - i F(\mu^{\prime}) \frac{\bar n
    \cdot v}{2} \bar n \cdot v_0 = - i F(\mu^{\prime}) \frac{\bar n \cdot v}{2}
  \; .
 \end{split}
 \end{equation}
where, in the last step, we have used $\bar n \cdot v_0 =
1$. Eq. \eqref{eq:boost.5} is thus in agreement with the definition in
Eq. \eqref{eq:boost.3}.
Applying the same reasoning to Eq. \eqref{eq:boost.4}, one finds
\begin{equation}\label{eq:boost.8}
\begin{split}
 \langle 0 |\bar\chi^{\, \bar l}_{ n}( n \cdot z) \, \frac{\slashchar{\bar n}}{2} \gamma_5 \, { \mathcal H}^{c}_{ n}(0) | D \rangle_{v} 
& = \bar n \cdot v \,\langle 0 |\bar\chi^{\, \bar l}(\bar n \cdot v \, n \cdot z)  \, \frac{\slashchar{\bar n}}{2} \gamma_5 \, { \mathcal H}^{c}_{}(0)| D \rangle_{v_{0}} \\ & 
  = - i F(\mu^{\prime})  \frac{\bar n \cdot v}{2}  \, \tilde \phi_0(n \cdot
  z_0,\mu^{\prime}) \; .
 \end{split}
 \end{equation}
Comparing Eq. \eqref{eq:boost.8}
with \eqref{eq:boost.4},
we see that $\tilde \phi( n \cdot z, \mu^{\prime}) =  \tilde \phi_0 (\bar n \cdot v \, n \cdot z, \mu^{\prime})$. 
Note that in the bottomonium rest frame the normalization condition for the distribution
amplitude is also $\tilde \phi(0,\mu^{\prime}) = 1$.

In the
main text of this paper we have used the $D$-meson distribution
amplitudes in momentum space,
\begin{equation*}
\begin{split}
\phi_0 (\omega_0, \mu^{\prime}) & \equiv \frac{1}{2\pi}\int d  n \cdot z_0 \,
e^{i \omega_0  n \cdot z_0} \tilde \phi_0 ( n\cdot z_0, \mu^{\prime}) \; ,\\
   \phi(\omega,   \mu^{\prime}) & \equiv \frac{1}{2\pi}\int d  n \cdot z   \,
   e^{i \omega    n \cdot z}   \tilde \phi ( n \cdot z, \mu^{\prime}) \; .
\end{split}
\end{equation*}
Using
Eq. \eqref{eq:boost.8}, we can relate the two distributions:
\begin{equation*}\label{eq:boost.9}
\begin{split}
\phi( \omega, \mu^{\prime})    & = \frac{1}{2\pi} \int d  n \cdot z    \, e^{i
  \omega  n \cdot z} \tilde \phi ( n \cdot z, \mu^{\prime})
= \frac{1}{2\pi} \int d  n \cdot z    \, e^{i \omega  n \cdot z} \tilde \phi_0 ( \bar n \cdot v\,  n \cdot z, \mu^{\prime}) \\
& = \frac{1}{2\pi} \frac{1}{ \bar n \cdot  v} \int d  n \cdot z    \,
e^{i \frac{\omega}{\bar  n \cdot v}   n \cdot z} \tilde \phi_0 (  n
\cdot z, \mu^{\prime})
 = \frac{1}{\bar n \cdot v} \phi_0 \left(\frac{\omega}{\bar n\cdot v},
   \mu^{\prime}\right) \; ,
\end{split}
\end{equation*}
as stated in Eq. \eqref{eq:DR.5.b}. The $D$-meson light-cone
distribution is normalized to 1 in both frames,
\begin{equation*}
\int d \omega_0 \phi_0 (\omega_0, \mu^{\prime}) = \int d \omega \phi(\omega, \mu^{\prime}) = 1,
\end{equation*}
as can be easily proved using $\tilde \phi_0 (0,\mu^{\prime}) =\tilde \phi(0,\mu^{\prime}) = 1$.


\begin{thebibliography}{999}

\bibitem{Chernyak:1983ej}
  V.~L.~Chernyak and A.~R.~Zhitnitsky,
  Phys.\ Rept.\  {\bf 112}, 173 (1984).

\bibitem{Brodsky:1989pv}
  S.~J.~Brodsky and G.~P.~Lepage,
  Adv.\ Ser.\ Direct.\ High Energy Phys.\  {\bf 5}, 93 (1989).



\bibitem{Brambilla:2004wf}
  N.~Brambilla {\it et al.}  [Quarkonium Working Group],
  arXiv:hep-ph/0412158.


\bibitem{Bodwin:1994jh}
  G.~T.~Bodwin, E.~Braaten, and G.~P.~Lepage,
  Phys.\ Rev.\  D {\bf 51}, 1125 (1995)
  [Erratum-ibid.\  D {\bf 55}, 5853 (1997)]
  [arXiv:hep-ph/9407339].



  
\bibitem{Bauer:2000ew}
  C.~W.~Bauer, S.~Fleming, and M.~E.~Luke,
  Phys.\ Rev.\  D {\bf 63}, 014006 (2000)
  [arXiv:hep-ph/0005275].

\bibitem{Bauer:2000yr}
  C.~W.~Bauer, S.~Fleming, D.~Pirjol, and I.~W.~Stewart,
  Phys.\ Rev.\  D {\bf 63}, 114020 (2001)
  [arXiv:hep-ph/0011336].

\bibitem{Bauer:2001yt}
  C.~W.~Bauer, D.~Pirjol, and I.~W.~Stewart,
  Phys.\ Rev.\  D {\bf 65}, 054022 (2002)
  [arXiv:hep-ph/0109045].

\bibitem{Bauer:2002nz}
  C.~W.~Bauer, S.~Fleming, D.~Pirjol, I.~Z.~Rothstein, and I.~W.~Stewart,
  Phys.\ Rev.\  D {\bf 66}, 014017 (2002)
  [arXiv:hep-ph/0202088].

\bibitem{Leibovich:2003jd}
  A.~K.~Leibovich, Z.~Ligeti, and M.~B.~Wise,
  Phys.\ Lett.\  B {\bf 564}, 231 (2003)
  [arXiv:hep-ph/0303099].

\bibitem{Pineda:1997bj}
  A.~Pineda and J.~Soto,
  Nucl.\ Phys.\ Proc.\ Suppl.\  {\bf 64}, 428 (1998)
  [arXiv:hep-ph/9707481].

\bibitem{Brambilla:1999xf}
  N.~Brambilla, A.~Pineda, J.~Soto, and A.~Vairo,
  Nucl.\ Phys.\  B {\bf 566}, 275 (2000)
  [arXiv:hep-ph/9907240].


\bibitem{Brambilla:2004jw}
  N.~Brambilla, A.~Pineda, J.~Soto, and A.~Vairo,
  Rev.\ Mod.\ Phys.\  {\bf 77}, 1423 (2005)
  [arXiv:hep-ph/0410047].

\bibitem{Grinstein:1990mj}
  B.~Grinstein,
  Nucl.\ Phys.\  B {\bf 339}, 253 (1990).

\bibitem{Eichten:1989zv}
  E.~Eichten and B.~R.~Hill,
  Phys.\ Lett.\  B {\bf 234}, 511 (1990).

\bibitem{Georgi:1990um}
  H.~Georgi,
  Phys.\ Lett.\  B {\bf 240}, 447 (1990).

\bibitem{Beneke:2003xh}
  M.~Beneke, A.~P.~Chapovsky, A.~Signer, and G.~Zanderighi,
  Phys.\ Rev.\ Lett.\  {\bf 93}, 011602 (2004)
  [arXiv:hep-ph/0312331].

\bibitem{Beneke:2004km}
  M.~Beneke, A.~P.~Chapovsky, A.~Signer, and G.~Zanderighi,
  Nucl.\ Phys.\  B {\bf 686}, 205 (2004)
  [arXiv:hep-ph/0401002].


\bibitem{Fleming:2007qr}
  S.~Fleming, A.~H.~Hoang, S.~Mantry, and I.~W.~Stewart,
  Phys.\ Rev.\  D {\bf 77}, 074010 (2008)
  [arXiv:hep-ph/0703207].

  
\bibitem{Fleming:2007xt}
  S.~Fleming, A.~H.~Hoang, S.~Mantry, and I.~W.~Stewart,
  Phys.\ Rev.\  D {\bf 77}, 114003 (2008)
  [arXiv:0711.2079 [hep-ph]].




\bibitem{Bodwin:2007zf}
  G.~T.~Bodwin, E.~Braaten, D.~Kang, and J.~Lee,
  Phys.\ Rev.\  D {\bf 76}, 054001 (2007)
  [arXiv:0704.2599 [hep-ph]].

\bibitem{Kang:2007uv}
  D.~Kang, T.~Kim, J.~Lee, and C.~Yu,
  Phys.\ Rev.\  D {\bf 76}, 114018 (2007)
  [arXiv:0707.4056 [hep-ph]].

\bibitem{Braguta:2009df}
  V.~V.~Braguta, A.~K.~Likhoded, and A.~V.~Luchinsky,
  arXiv:0902.0459 [hep-ph].

\bibitem{Braguta:2009xu}
  V.~V.~Braguta and V.~G.~Kartvelishvili,
  arXiv:0907.2772 [hep-ph].



\bibitem{:2008vj}
  B.~Aubert {\it et al.}  [BABAR Collaboration],
  Phys.\ Rev.\ Lett.\  {\bf 101}, 071801 (2008)
  [Erratum-ibid.\  {\bf 102}, 029901 (2009)]
  [arXiv:0807.1086 [hep-ex]].


\bibitem{:2008sx}
  D.~M.~Asner {\it et al.},
  Phys.\ Rev.\  D {\bf 78}, 091103 (2008)
  [arXiv:0808.0933 [hep-ex]].

\bibitem{Briere:2008cv}
  R.~A.~Briere {\it et al.}  [CLEO Collaboration],
  Phys.\ Rev.\  D {\bf 78}, 092007 (2008)
  [arXiv:0807.3757 [hep-ex]].
  
  

\bibitem{Hill:2005ju}
  R.~J.~Hill,
  Phys.\ Rev.\  D {\bf 73}, 014012 (2006)
  [arXiv:hep-ph/0505129].

 
\bibitem{Luke:1996hj}
  M.~E.~Luke and A.~V.~Manohar,
  Phys.\ Rev.\  D {\bf 55}, 4129 (1997)
  [arXiv:hep-ph/9610534].

\bibitem{Luke:1999kz}
  M.~E.~Luke, A.~V.~Manohar, and I.~Z.~Rothstein,
  Phys.\ Rev.\  D {\bf 61}, 074025 (2000)
  [arXiv:hep-ph/9910209].


\bibitem{Manohar:1999xd}
  A.~V.~Manohar and I.~W.~Stewart,
  Phys.\ Rev.\  D {\bf 62}, 014033 (2000)
  [arXiv:hep-ph/9912226].


\bibitem{Hoang:2002yy}
  A.~H.~Hoang and I.~W.~Stewart,
  Phys.\ Rev.\  D {\bf 67}, 114020 (2003)
  [arXiv:hep-ph/0209340].



\bibitem{Manohar}
A.~Manohar and M.~Wise, \textit{Heavy Quark Physics}, Cambridge University Press, Cambridge, 2000.




\bibitem{Manohar:2006nz}
  A.~V.~Manohar and I.~W.~Stewart,
  Phys.\ Rev.\  D {\bf 76}, 074002 (2007)
  [arXiv:hep-ph/0605001].

  
\bibitem{cusp}
  G.~P.~Korchemsky and A.~V.~Radyushkin,
  Nucl.\ Phys.\  B {\bf 283}, 342 (1987).

  I.~A.~Korchemskaya and G.~P.~Korchemsky,
  Phys.\ Lett.\  B {\bf 287}, 169 (1992).

\bibitem{:2008sq}
  B.~I.~Eisenstein {\it et al.}  [CLEO Collaboration],
  arXiv:0806.2112 [hep-ex].

  
\bibitem{Lange:2003ff}
  B.~O.~Lange and M.~Neubert,
  Phys.\ Rev.\ Lett.\  {\bf 91}, 102001 (2003)
  [arXiv:hep-ph/0303082].

\bibitem{Amsler:2008zz}
  C.~Amsler {\it et al.}  [Particle Data Group],
  Phys.\ Lett.\  B {\bf 667}, 1 (2008).


\bibitem{Barbieri:1975ki}
  R.~Barbieri, R.~Gatto, R.~Kogerler, and Z.~Kunszt,
  Phys.\ Lett.\  B {\bf 57}, 455 (1975).

  W.~Celmaster,
  Phys.\ Rev.\  D {\bf 19}, 1517 (1979).

\bibitem{Bodwin:2001mk}
  G.~T.~Bodwin, D.~K.~Sinclair, and S.~Kim,
  Phys.\ Rev.\  D {\bf 65}, 054504 (2002)
  [arXiv:hep-lat/0107011].

\bibitem{Eichten:1995ch}
  E.~J.~Eichten and C.~Quigg,
  Phys.\ Rev.\  D {\bf 52}, 1726 (1995)
  [arXiv:hep-ph/9503356].





\bibitem{Buchmuller:1980su}
  W.~Buchmuller and S.~H.~H.~Tye,
  Phys.\ Rev.\  D {\bf 24}, 132 (1981).


  
\bibitem{Braun:2003wx}
  V.~M.~Braun, D.~Y.~Ivanov, and G.~P.~Korchemsky,
  Phys.\ Rev.\  D {\bf 69}, 034014 (2004)
  [arXiv:hep-ph/0309330].

\bibitem{Grozin:1996pq}
  A.~G.~Grozin and M.~Neubert,
  Phys.\ Rev.\  D {\bf 55}, 272 (1997)
  [arXiv:hep-ph/9607366].
  
\bibitem{Beneke:1999br}
  M.~Beneke, G.~Buchalla, M.~Neubert, and C.~T.~Sachrajda,
  Phys.\ Rev.\ Lett.\  {\bf 83}, 1914 (1999)
  [arXiv:hep-ph/9905312].

\bibitem{Lee:2005gza}
  S.~J.~Lee and M.~Neubert,
  Phys.\ Rev.\  D {\bf 72}, 094028 (2005)
  [arXiv:hep-ph/0509350].


  
\bibitem{Pilipp:2007sb}
  V.~Pilipp,
  arXiv:hep-ph/0703180.

\bibitem{Hoang:1998ng}
  A.~H.~Hoang, Z.~Ligeti, and A.~V.~Manohar,
  Phys.\ Rev.\ Lett.\  {\bf 82}, 277 (1999)
  [arXiv:hep-ph/9809423].







\bibitem{Ligeti:2008ac}
  Z.~Ligeti, I.~W.~Stewart, and F.~J.~Tackmann,
  Phys.\ Rev.\  D {\bf 78}, 114014 (2008)
  [arXiv:0807.1926 [hep-ph]].


\bibitem{Bodwin:1992ye}
  G.~T.~Bodwin, E.~Braaten, and G.~P.~Lepage,
  Phys.\ Rev.\  D {\bf 46}, R1914 (1992)
  [arXiv:hep-lat/9205006].


\bibitem{Brambilla:2001xy}
  N.~Brambilla, D.~Eiras, A.~Pineda, J.~Soto, and A.~Vairo,
  Phys.\ Rev.\ Lett.\  {\bf 88}, 012003 (2002)
  [arXiv:hep-ph/0109130].

\bibitem{Brambilla:2002nu}
  N.~Brambilla, D.~Eiras, A.~Pineda, J.~Soto, and A.~Vairo,
  Phys.\ Rev.\  D {\bf 67}, 034018 (2003)
  [arXiv:hep-ph/0208019].

\bibitem{Vairo:2003gh}
  A.~Vairo,
  Mod.\ Phys.\ Lett.\  A {\bf 19}, 253 (2004)
  [arXiv:hep-ph/0311303].

\bibitem{Maltoni:2004hv}
  F.~Maltoni and A.~D.~Polosa,
  Phys.\ Rev.\  D {\bf 70}, 054014 (2004)
  [arXiv:hep-ph/0405082].
  
\bibitem{italianconnection}
S.~Stracka,
private communication.

\bibitem{Distribution}
I.~M.~Gelfand and G.~E.~Shilov, \textit{Generalized Functions, Vol. 1}, Academic Press, New York, 1964.


\end{thebibliography}
\end{document}